\documentclass{aa}

\usepackage{graphicx}
\usepackage{xcolor}
\usepackage{txfonts}
\usepackage{siunitx}
\usepackage{xspace}
\usepackage{placeins}

\newcommand{\rc}{\ensuremath{r_\mathrm{c}}\xspace}
\usepackage[export]{adjustbox}

\definecolor{rd}{HTML}{E41A1C}
\definecolor{xlinkcolor}{cmyk}{1,1,0,0}

\usepackage[bookmarks=false,         % show bookmarks bar?
    pdfnewwindow=true,      % links in new window
    colorlinks=true,    % false: boxed links; true: colored links
    linkcolor=xlinkcolor,     % color of internal links
    citecolor=xlinkcolor,     % color of links to bibliography
    filecolor=xlinkcolor,  % color of file links
    urlcolor=xlinkcolor,      % color of external links
final=true
]{hyperref}

\usepackage{siunitx,booktabs}
\makeatletter
\renewcommand*\aa@pageof{, page \thepage{} of \pageref*{LastPage}}
\makeatother

\definecolor{MyDarkGreen}{rgb}{0.0,0.7,0.0}

\addto\extrasenglish{%

}

\begin{document}

\title{A large population study of protoplanetary disks}
\subtitle{Explaining the millimeter size-luminosity relation with or without substructure}

\author{
    Apostolos Zormpas\inst{1}
    \and
    Tilman Birnstiel\inst{1,2}
    \and
    Giovanni P. Rosotti\inst{3,4}
    \and
    Sean M. Andrews\inst{5}
}
\institute{
    University Observatory, Faculty of Physics, Ludwig-Maximilians-Universit\"{a}t M\"{u}nchen, Scheinerstr.~1, 81679 M\"{u}nchen, Germany \\
    \email{zormpas@usm.lmu.de}
    \and
    Exzellenzcluster ORIGINS, Boltzmannstr. 2, D-85748 Garching, Germany
    \and
    Department of Physics and Astronomy, University of Leicester, University Road, Leicester LE1 7RH, UK
    \and
    Leiden Observatory, Leiden University, PO Box 9513, NL-2300 RA Leiden, the Netherlands
    \and
    Center for Astrophysics \textbar\ Harvard \& Smithsonian, 60 Garden Street, Cambridge, MA 02138, USA
}

\date{Received \today; accepted XXXX}

% 5 {} token are mandatory
%%\abstract

\abstract{
Recent subarcsecond resolution surveys of the dust continuum emission from nearby protoplanetary disks show a strong correlation between the sizes and luminosities of the disks. We aim to explain the origin of the (sub-)millimeter size-luminosity relation (SLR) between the $\mathrm{68\%}$ effective radius ($\mathrm{r_{eff}}$) of disks with their continuum luminosity ($\mathrm{L_{mm}}$), with models of gas and dust evolution in a simple viscous accretion disk and radiative transfer calculations.

We use a large grid of models ($\mathrm{10^{5}}$ simulations) with and without planetary gaps, and vary the initial conditions of the key parameters. We calculate the disk continuum emission and the effective radius for all models as a function of time. By selecting those simulations that continuously follow the SLR, we can derive constraints on the input parameters of the models.

We confirm previous results that models of smooth disks in the radial drift regime are compatible with the observed SLR ($\mathrm{L_{mm}\propto r_{eff}^{2}}$), but only smooth disks cannot be the reality. We show that the SLR is more widely populated if planets are present. However, they tend to follow a different relation than smooth disks, potentially implying that a mixture of smooth and substructured disks are present in the observed sample. We derive a SLR ($\mathrm{L_{mm}\propto r_{eff}^{5/4}}$) for disks with strong substructure. To be compatible with the SLR, models need to have an initially high disk mass ($\mathrm{\geq 2.5 \cdot 10^{-2}M_{\star}}$) and low turbulence-parameter $\mathrm{\alpha}$ values ($\mathrm{\leq 10^{-3}}$). Furthermore, we find that the grain composition and porosity drastically affects the evolution of disks in the size-luminosity diagram where relatively compact grains that include amorphous carbon are favored. Moreover, a uniformly optically thick disk with high albedo ($\rm{0.9}$) that follows the SLR cannot be formed from an evolutionary procedure.
}


\maketitle
%
%-------------------------------------------------------------------

\section{Introduction}
\label{sec:intro}
Planet formation is a process far from having a complete, robust, and widely accepted theory. Multiple theories aim to explore the way planets form \citep[e.g.,][]{Benz2014prpl.conf..691B,Kratter2016ARA&A..54..271K,Johansen2017AREPS..45..359J}. To understand planet formation, high resolution observations and large surveys of the birth places of planets, the protoplanetary disks (PPDs), are essential. In recent years the Atacama Large Millimeter/Submillimeter Array (ALMA) has not only provided a sizable number of highly resolved observations of PPDs, but due to its high sensitivity it has also enabled several large intermediate resolution (on the order of \SI{100}{mas}) surveys of different star-forming regions \citep[for a review see][and references therein]{Andrews2020} providing crucial information on population properties such as distributions of disk sizes, fluxes, or spectral indices for disks across stars of different masses and ages and across star-forming regions in different environments.

Highly resolved observations and large intermediate resolution surveys are complementary and are both essential for understanding the connections between key properties of PPDs. One of the important diagnostics is the continuum luminosity ($\mathrm{L_{mm}}$) at (sub-)millimeter wavelengths since it can be a tracer of the disk mass that is produced by the solid grains \citep{Beckwith1990AJ.....99..924B} (i.e., the amount of material available to form planets). Assuming a constant dust-to-gas ratio (usually $\rm{0.01}$ based on observations of the interstellar medium, see \citealp{Bohlin1978}), the dust mass can be converted to the total disk mass (dust and gas). Surveys that  measure the disk dust mass ($\mathrm{M_{dust}}$) have been used to correlate this property with the mass of the host star ($\mathrm{M_{\star}}$), and a linear relation has been found between them \citep{Andrews2013} that appears to steepen with time \citep[][]{Pascucci2016ApJ...831..125P,Ansdell2017AJ....153..240A,Barenfeld2016ApJ...827..142B}. Furthermore,  a steeper than linear relationship has been observed between $\mathrm{L_{mm}}$ and $\mathrm{M_{\star}}$ \citep[][]{Andrews2013,Pascucci2016ApJ...831..125P,Ansdell2016ApJ...828...46A}. Recently, theoretical studies have started to explain these correlations  \citep[e.g.,][]{Pascucci2016ApJ...831..125P,stammler2019,pinilla2020A&A...635A.105P} using numerical dust evolution models that include grain growth, radial drift, fragmentation of dust particles, and particle traps \citep[e.g.,][]{Birnstiel2010,BKE2012,Krijt2016}.

These observations of dust thermal continuum emission are crucial for the characterization of disk evolution. However, the methods described above that relate dust emission to physical quantities like total disk masses carry uncertainty arising from the multiple assumptions such as the dust opacity and the disk temperature that may vary for every disk \citep[e.g.,][]{Hendler2017,Ballering2019}. The grain opacity depends on the unknown particle size distribution, composition, and particle structure \citep[e.g.,][and references within]{Birnstiel_DSHARP}, although considerable efforts have been undertaken from modeling \citep[e.g.,][]{Wada2008,Wada2009,Okuzumi2009,Seizinger2013a,Seizinger2013b} and experimental studies \citep[e.g.,][]{Blum2008,Guttler2010,Gundlach2015} of aggregation and of the computation of optical properties of aggregates \citep[e.g.,][]{Kataoka2014,Min2016,Tazaki2016}.

Another important property for the characterization of a disk population is the disk size. Viscous theory \citep{LBP1974} predicts that a fraction of the disk mass keeps moving outward, known as viscous spreading, suggesting the disk size should increase with time. In principle a measurement of disk size as a function of time could measure this evolution to test viscous theory and measure its efficiency. The most readily available tracer of the disk size is the continuum as gas tracers suffer from uncertain abundances (due to freeze-out and  dissociation, among others) and sensitivity constraints. Since the disk does not have a clear outer edge, we need to introduce an effective radius and express the size as a function of the total continuum emission. This size metric is called emission size or effective radius ($\mathrm{r_{eff}}$) \citep{tripathi2017millimeter}. 

However, the dust component does not behave in the same way as the gas mainly due to an effect termed radial drift \citep[e.g.,][]{Whipple1972,Weidenschilling1977MNRAS.180...57W, Takeuchi2002ApJ...581.1344T}. The dust particles interact with the sub-Keplerian gas disk via aerodynamic drag forces, leading them to migrate toward the star. As an observational implication, the dust emission is less extended than the gas emission \citep[e.g.,][]{Andrews2012,Isella2012ApJ...747..136I,Andrews2016ApJ...820L..40A,Cleeves2016ApJ...832..110C}  predicted by \citet{Birnstiel2014ApJ...780..153B} (but see \citealp{Trapman2020}). Radial drift is also heavily dependent on the grain size; therefore, grain growth \citep[e.g.,][]{BKE2012} has to be included in the numerical studies that aim to use the dust disk radius. \citet{rosotti2019time_evolution} studied theoretically how the evolution of the disk dust radius changes with time in a viscously evolving disk and addressed whether the evolution of the dust disk radius is set by viscous spreading or by the dust processes such as grain growth and radial drift. They found that viscous spreading influences the dust and leads to the dust disk expanding with time.

Many surveys have been performed to explore the relation between the two diagnostics \citep[e.g.,][]{ Andrews2010ApJ...723.1241A,Pietu2014A&A...564A..95P,Hendler2020ApJ...895..126H}. Recently, a subarcsecond resolution survey of 50 nearby protoplanetary disks, conducted with the Submillimeter Array (SMA) by \citet{tripathi2017millimeter}, showed a strong size-luminosity relation (SLR) between the observed population. The follow-up program in \citet{Andrews2018a}, a combined analysis of the \citet{tripathi2017millimeter} data and the ALMA data from the Lupus disk sample (105 disks in total), confirmed the scaling relations between the $\mathrm{r_{eff}}$ and the $\mathrm{L_{mm}}$, $\mathrm{M_{\star}}$. However, not all studied star-forming regions show the same correlation, but appear to vary with the age of the region \citep{Hendler2020ApJ...895..126H}.

In recent years, due to the unprecedented sensitivity and resolution, ALMA has provided a plethora of groundbreaking images of protoplanetary disks. Most of these disks do not show a smooth and monotonically decreasing surface density profile, but instead are composed of single or multiple symmetric annular substructures, for example  HL Tauri \citep{ALMA_partnership2015ApJ...808L...3A}, TW Hya \citep[][]{Andrews2016ApJ...820L..40A,Tsukagoshi2016ApJ...829L..35T}, HD 163296 \citep{Isella2016PhRvL.117y1101I}, HD 169142 \citep{Fedele2017A&A...600A..72F}, AS 209  \citep{Fedele2018A&A...610A..24F}, HD 142527 \citep{Casassus2013Natur.493..191C}, and many more in the recent DSHARP survey \citep{Andrews_DSHARP_2018ApJ...869L..41A}. Moreover, non-axisymmetric features like spiral arms \citep[e.g.,][]{Perez2016Sci...353.1519P,Huang2018III} and lopsided rings \citep[e.g.,][]{Nienke2013Sci...340.1199V} have been observed.

Many ideas for the origin of these ring-like substructures have been explored but one of the most favorable explanations is the formation of gaps in the gas surface density, due to the presence of planets. A massive planet ($\rm{\geq 0.1 M_{jup}}$) \citep{Zhang2018DSHARPApJ...869L..47Z} is able to open a gap in the surrounding gaseous disk, thereby generating a pressure maximum. Later on, the dust particles   migrate  toward the local pressure bump, due to radial drift \citep{Weidenschilling1977MNRAS.180...57W,Nakagawa1986Icar...67..375N}, and consequently lead to the annular shape  \citep[e.g.,][]{Rice2006MNRAS.373.1619R,Pinilla2012A&A...545A..81P}. These narrow rings may be optically thick or moderately optically thick, but  between these features the material is approximated as optically thin \citep{Dullemond2018DSHARP}. However, these rings can contain large amounts of dust that can increase the total luminosity of a disk and its position with respect to the SLR. 

The SLR might contain crucial information about disk evolution and planet formation theory. Our goal is to explore the physical origins of the SLR from \citet{tripathi2017millimeter} and \citet{Andrews2018a} by performing a large population study of models with gas and dust evolution. We aim to characterize the key properties of disks that reproduce the observational results. We explore the differences in the SLR of disks that have a smooth surface density profile and disks that contain weak and strong substructures. In \autoref{sec:methods}, we discuss the methods we used to carry out our computational models. The results of this analysis are presented in \autoref{sec:results}, where we explain the global effect of every parameter in the population of disks and we present the general properties that disks should have to follow the SLR. In \autoref{sec:discussion} we discuss the theoretical and observational implications of our results. We draw our conclusions in \autoref{sec:conclusions}.

%--------------------------------------------------------------------
\section{Methods}
\label{sec:methods}

We carry out  1D gas and dust evolution simulations using a slightly modified version of the two-population model (two-pop-py) by \citet[][]{BKE2012,Birnstiel2015ApJ...813L..14B}, while we also mimic the presence of planets. As a post-processing step, we calculate the intensity profile and the disk continuum emission. With the purpose of running a population study, we use a large grid of parameters (see \autoref{tabel:input_param}), so that we can explore the differences that occur due to the different initial conditions. In the next sections, we explain the procedure in more detail.

\subsection{Disk evolution}

The gas  follows the viscous evolution equation. For the disk evolution, we use the turbulent effective viscosity as in \citet{shakura1973black},
\begin{equation}
    \mathrm{\nu = \alpha_{gas}  \frac{c_{s}^{2}}{\Omega_{K}}}\,,
  \label{eq:visc}
\end{equation} 
and the dust diffusion coefficient as
\begin{equation}
    \mathrm{D = \alpha_{dust}  \frac{c_{s}^{2}}{\Omega_{K}}}\,,
  \label{eq:diff}
\end{equation} 
with $\mathrm{\alpha_{gas}}$ being the turbulence parameter, $\mathrm{c_{s}}$ the sound speed, and $\mathrm{\Omega_{K}}$ the Keplerian frequency. The above equation lacks the term $\rm{\frac{1}{1+St^{2}}}$, where $\rm{St}$ is the Stokes number, but we can ignore it since the Stokes number is always $\rm{<1}$ in our simulations \citep{Youdin2007Icar..192..588Y}. We differentiate the $\alpha$ parameter in two different values. One for the gas $\alpha_\mathrm{gas}$ and one for the dust $\alpha_\mathrm{dust}$, since we later mimic planetary gaps by locally varying the viscosity (see \autoref{subsec:planets}). In the smooth case $\alpha_\mathrm{dust} = \alpha_\mathrm{gas}$.

The dust is described by the two populations model of \citet{BKE2012} which evolves the dust surface density under the assumption that the small dust is tightly coupled to the gas while the large particles can decouple from the gas and drift inward. The initial dust growth phase is using the current dust-to-gas ratio instead of the initial value as in \citet{BKE2012}.

We set the initial gas surface density according to the self-similar solution of \citet{LBP1974},
\begin{equation}
    \mathrm{\Sigma_{g}(r)} =\mathrm{\Sigma_{0} \left(\frac{r}{r_{c}}\right) ^{-\gamma} Exp\left[-\left(\frac{r}{r_{c}}\right)^{2-\gamma}\right]}\,,
  \label{eq:surf_dens}
\end{equation} 
where $\mathrm{\Sigma_{0} = (2-\gamma) M_{d} / 2 \pi \rc^{2}}$ is the normalization parameter, which is set in every simulation by the disk mass $\mathrm{M_{d}}$. The other parameters are $\mathrm{\gamma}=1$, which is the viscosity exponent and it is not varied throughout our models and \rc which is the characteristic radius of the disk (see \autoref{tabel:input_param}). When r $\ll$ \rc, then $\mathrm{\Sigma_{g}}$ is a power law and when $\mathrm{r \geq \rc}$, $\mathrm{\Sigma_{g}}$ is dominated by the exponential factor.

The initial dust distribution follows the gas distribution with a constant dust-to-gas ratio of $\mathrm{\Sigma_{d}}$/$\mathrm{\Sigma_{g}}$=0.01. The initial grain size (= monomer grain size) is $a_\mathrm{min} = \SI{0.1}{\mu m}$. This monomer size stays constant in time and space, while the representative size for the large grains increases with time as the particles grow. The particle bulk density is $\rm{\rho_{s}}=\SI{1.7}{g/cm^3}$ for the standard opacity model from \citet{Ricci_2010} and $\rm{\rho_{s}}=\SI{1.675}{g/cm^3}$ for the DSHARP \citep{Birnstiel_DSHARP} opacity, but decreases for different values of porosity (see \autoref{sub:Observables}). We evolve the disks up to $\mathrm{\SI{10}{Myr}}$ to study the long-term evolution, but in the following analysis we only show results from $\rm{\SI{300}{kyr}}$ to $\SI{3}{Myr}$ (see \autoref{sub:survival_frequency}).

Our $\mathrm{1D}$ radial grid ranges from $\rm{0.05}$ to $\rm{\SI{2000}{au}}$ and the grid cells are spaced logarithmically. We  use adaptive temperature, which depends on the luminosity of every star. Since the stellar mass changes in our grid, the stellar luminosity changes as well. We follow the temperature profile,
\begin{equation}
    \mathrm{T = \left(\phi \frac{L_{\star}}{4\pi\sigma_{SB}r^{2} }+(\SI{10}{K})^{4}\right)^{1/4}}\,,
  \label{eq:Temp}
\end{equation} 
as in \citet{Kenyon_Hartmann_1996}.
In this equation $\mathrm{L_{\star}}$ is the stellar luminosity,  $\mathrm{\phi} =0.05$ is the flaring angle,  $\mathrm{\sigma_{SB}}$ is the Stefan-Boltzmann constant, and $\mathrm{r}$ is the radius. The term $\mathrm{10^{4}}$ is a lower limit so that we do not allow the disk temperature to drop below $\SI{10}{K}$ at the outer parts of the disk. We  use the evolutionary tracks of \citet{Siess_2000A&A...358..593S} to get the luminosity of a $\mathrm{\SI{1}{Myr}}$ old star of the given mass. The stellar luminosity and effective temperature is not evolved in our simulations. However, the luminosity of a \SI{1}{M_{\odot}} star would decrease from $\rm{\sim \SI{2.4}{L_{\odot}}}$ at \SI{1}{Myr} to $\rm{\sim \SI{1}{L_{\odot}}}$ at \SI{3}{Myr}. In \autoref{disc:limitations} we explore how a change in stellar luminosity affects our results.

\subsection{Population study}

We use an extended parameter grid, by varying the initial values of the turbulence parameter ($\mathrm{\alpha_{gas}}$), disk mass ($\mathrm{M_{d}}$), stellar mass ($\mathrm{M_{\star}}$), characteristic radius (\rc), and fragmentation velocity ($\mathrm{v_{frag}}$). For every parameter, we pick the  ten values specified in \autoref{tabel:tbl_params}, taking all the possible combinations between them leading to a total of $\mathrm{100.000}$ simulations.

\begin{table}
 \caption{Grid parameters of the model}
 \begin{center}
 \centering
 \begin{tabular}{l l l } 
 \toprule
 Parameter & Description & Value-Range \\ [0.3ex]
 \hline
 \hline
 $\mathrm{\Sigma_{d}/\Sigma_{g}}$  & initial dust-to-gas ratio & $\rm{0.01}$  \\ [0.3ex]
 $\mathrm{\rho_{s}}$ \hfill [$\rm{g/cm^{3}}$]& particle bulk density & $\rm{1.7, 1.675}$  \\ [0.3ex]
 $\mathrm{\gamma}$ & viscosity exponent & $\rm{1}$  \\ [0.3ex]
 $r$ \hfill [$\mathrm{au}$] & logarithmic grid extent & $\rm{0.05 - 2000}$   \\ [0.3ex]
 $\mathrm{n_{r}}$ \hfill [cells] & grid resolution & $\rm{400}$   \\ [0.3ex]
  $\mathrm{t}$ \hfill [years] & duration of each simulation & $\rm{10^{7}}$   \\ [0.3ex]
 
 \bottomrule
\end{tabular}
\end{center}
\label{tabel:input_param}
\end{table}

\begin{table}
 \caption{Variables of the model}
 \begin{center}
 \centering
 \begin{tabular}{l l l } 
 \toprule
 Parameter & Description & Values \\ [0.3ex]
 \hline
 \hline
 $\mathrm{\alpha}$ & viscosity parameter & $\mathrm{\{1,2.5,5,7.5\}\cdot10^{-4}}$ \\ [0.3ex]
 
 & & $\mathrm{\{1,2.5,5,7.5\}\cdot10^{-3}}$\\ [0.3ex]
 
 & & $\mathrm{\{1,2.5\}\cdot10^{-2}}$ \\ [0.3ex]
 
 $\mathrm{M_{d}}$ \hfill [M$_{\star}$] & initial disk mass & $\mathrm{\{1,2.5,5,7.5\}\cdot10^{-3}}$ \\ [0.3ex]
 & & $\mathrm{\{1,2.5,5,7.5\}\cdot10^{-2}}$\\ [0.3ex]
 & & $\mathrm{\{1,2.5\}\cdot10^{-1}}$ \\ [0.3ex]

 $\mathrm{M_\star}$  \hfill [M$_{\odot}$] & stellar mass  &  $\mathrm{0.2, 0.4, 0.6, 0.8, 1.0}$ \\ [0.3ex]
 & & $\mathrm{1.2, 1.4, 1.6, 1.8, 2.0}$\\ [0.3ex]
 
 \rc \hfill [$au$]  & characteristic radius  & $\mathrm{10, 30, 50, 80, 100}$ \\ [0.3ex]
 & & $\mathrm{130, 150, 180, 200, 230}$\\ [0.3ex]
 
 $\mathrm{v_{f}}$ \hfill [$cm/s$]& fragmentation velocity & $\mathrm{200, 400, 600, 800, 1000}$ \\ [0.3ex]
 & & $\mathrm{1200, 1400, 1600, 1800, 2000}$\\ [0.3ex]
 q & planet/star mass ratio & $3\cdot10^{-4}$, $10^{-3}$, $3\cdot10^{-3}$ \\ [0.3ex]
 r$_{p}$ \hfill [r$_{c}$]& planet position & $1/3$, $2/3$  \\ [0.3ex]
 \bottomrule
\end{tabular}
\end{center}
\label{tabel:tbl_params}
\end{table}

\subsection{Planets}
\label{subsec:planets}
A large planet embedded in a disk produces a co-orbital gap in the gas density. To mimic gap opening by planets in our simulations, we altered the $\mathrm{\alpha_{gas}}$ turbulence parameter. Since in steady state $\mathrm{\alpha_{gas}}$ $\mathrm{\cdot}$ $\mathrm{\Sigma_{g}}$ is constant, the $\mathrm{\alpha}$ parameter and the surface density $\mathrm{\Sigma_{g}}$ are inversely proportional quantities, so a bump in the $\mathrm{\alpha_{gas}}$ profile leads to a gap in the surface density profile. The reason for the change in $\mathrm{\alpha_{gas}}$ and not in $\mathrm{\Sigma_{g}}$ is that the surface density evolves according to  \autoref{eq:surf_dens}. By inserting the bump in the $\mathrm{\alpha_{gas}}$, the $\mathrm{\Sigma_{g}}$ still evolves viscously and at the same time produces a planetary gap shape.

Following the prescription from \citet{Kanagawa2016}, we mimic the effect of planets with different planet/star mass ratios $q$ (see \autoref{tabel:input_param}). For reference, $q=10^{-3}$ represents a Jupiter-mass planet around a solar-mass star. This way, we can study the effect of planetary gaps and rings in the observable properties of the disk and extract the key observables in a computationally efficient way, avoiding the need to run expensive hydrodynamic simulations for each combination of parameters.

Choosing the appropriate profile that mimics a planetary gap is tricky, so we performed hydrodynamical simulations using FARGO-3D \citep{FARG03D_2015ascl.soft09006B}, and we compared the effect on the observable quantities. The \citet{Kanagawa2016} profile is an analytical approximation of the gap depth and width, but does not necessarily represent the pressure bump that is caused by the bump. Therefore, we tested how strongly this assumption affects the properties of the dust in the trap by comparing them against proper hydrodynamical solutions and disk evolution. We found that the depth of the gap is not as important to the evolution of the disk on the SLR, but the width is the dominant factor. As long as the planet is massive enough to create a strong pressure maximum and thus stop the particles, the position of the pressure maximum is more important than a precise value of the gap depth. In summary, the gap depth is not what matters the most but the associated amplitude and location of the pressure maximum. We should mention that the precise amount of trapping in the bumps should still matter, for example in
planetesimal formation, but for our results this is less relevant. We provide comparison plots and more details in \autoref{app:gap_profiles}.

We define the position of the planets in the disk $\mathrm{r_{p}}$, as a function of the characteristic radius \rc (see \autoref{tabel:input_param}). We locate them either at $\mathrm{2/3}$ or at $\mathrm{1/3}$ of \rc. In our simulations we used zero, one, or two planets in these positions. We refer the reader to \autoref{subsusb:planet_tracks} for the effect of the planet location and mass in the simulations. 

\subsection{Observables}
\label{sub:Observables}

Since the disk size is not one of the parameters that we measure directly, using the characteristic radius \rc as a size metric is problematic \citep{rosotti2019time_evolution}. For this reason we define an observed disk radius using the calculated surface brightness profile. Following \citet{tripathi2017millimeter} we adopt their approach to defining an effective radius ($\mathrm{r_{eff}}$) as the radius that encloses a fixed fraction of the total flux, $\mathrm{f_{\nu}}$ ($\mathrm{r_{eff}}$) = $\mathrm{xF_{v}}$. We choose $\mathrm{x=68\%}$ of the total disk flux as a suitable intermediate value to define $\mathrm{r_{eff}}$ as it is comparable to a standard deviation in the approximation of a Gaussian profile. 

We calculate the mean intensity $\rm{J_{\nu}}$ profile by using the scattering solution from \citet{Miyake1993Icar..106...20M} of the radiative transfer equation 
\begin{equation}
\frac{J_{\nu}(\tau_{\nu})}{B_{\nu}(T(r))} = 1-b\left(e^{-\sqrt{3\epsilon_{\nu}^{eff}}\left(\frac{1}{2}\Delta\tau-\tau_{\nu}\right)}+e^{-\sqrt{3\epsilon_{\nu}^{eff}}\left(\frac{1}{2}\Delta\tau+\tau_{\nu}\right)}\right)\,,  
\end{equation}
with
\begin{equation}
b=\left[\left(1-\sqrt{\epsilon_{\nu}^{eff}}\right)e^{-\sqrt{3\epsilon_{\nu}^{eff}}\Delta\tau}+1+\sqrt{\epsilon_{\nu}^{eff}}\right]^{-1}
,\end{equation}
where $\mathrm{B_{\nu}}$ is the Planck function and 
\begin{equation}
\tau_{\nu}=\left(\kappa_{\nu}^{\rm abs}+\kappa_{\nu}^{\rm sca,eff}\right) \Sigma_{d}
\end{equation}
is the optical depth with $\mathrm{\kappa_{\nu}^{\rm abs}}$ the dust absorption opacity and $\mathrm{\kappa_{\nu}^{\rm sca,eff}}$ the effective scattering opacity, which is obtained from \citet{Ricci_2010} or \citet{Birnstiel_DSHARP} (see below). As effective scattering opacity we refer to
\begin{equation}
\kappa_{\nu}^{\rm sca,eff}=(1-g_{\nu})\kappa_{\nu}^{sca}
,\end{equation}
where $\rm{g_{\nu}}$ is the forward-scattering parameter,
$\rm{\Delta \tau}$ is
\begin{equation}
\rm{\Delta \tau=\Sigma_{d}\kappa_{\nu}^{\rm tot} \Delta z}    
,\end{equation}
while 
\begin{equation}
\epsilon_{\nu}^{eff}=\frac{\kappa_{\nu}^{abs}}{\kappa_{\nu}^{abs}+\kappa_{\nu}^{sca,eff}}
\end{equation}
is the effective absorption probability.
To calculate the intensity $\rm{I_{\nu}^{out}}$ we follow the modified Eddington-Barbier approximation as in \citet{Birnstiel_DSHARP},
\begin{equation}
\label{eq:intensity}
    I_{\nu}^{out}\simeq \left(1-e^{-\Delta \tau/\mu} \right)S_{\nu}\left(\left(\frac{1}{2}\Delta\tau-\tau_{\nu}\right)/\mu=2/3\right)
,\end{equation}
where $\rm{\mu=cos\theta}$ and 
\begin{equation}
    S_{\nu}(\tau_{\nu})=\epsilon_{\nu}^{eff}B_{\nu}(T_{d})+(1-\epsilon_{\nu}^{eff})J_{\nu}(\tau_{\nu})
\end{equation}
is the source function.

Two-pop-py evolves only the dust and gas surface densities and the maximum particle size,\footnote{In the rest of the manuscript we  refer to grain size or particle size rather than maximum grain size.} thereby implicitly assuming a particle size distribution. The grain size at each radius is set by either the maximum size possible in the fragmentation- or drift-limited regimes, whichever is lower \citep[see][for details]{BKE2012}. To compute the optical properties of the dust, we therefore considered a population of grains with a power-law size distribution, $\mathrm{n(a)}$ $\mathrm{\propto}$ $\mathrm{a^{-q}}$, with an exponent $\mathrm{q = 2.5}$, for $\mathrm{a_{min}}$ $\mathrm{\leq}$ $\mathrm{a}$ $\mathrm{\leq}$ $\mathrm{a_{max}}$. This choice follows \citet{BKE2012} where the size distribution is closer to $\rm{q=2.5}$ for disks that are in the drift limit, while for the fragmentation-limited ones, a choice of  $\rm{q=3.5}$  would be more suitable. Considering the disk mass is dominated by the large grains, the choice of a smaller exponent does not alter our results significantly, but it matters for the details. Since the smooth simulations are mostly drift-limited the choice of $\rm{q=2.5}$ fits  these disks better. Moreover, if a disk is fragmentation-limited  then it is so mostly in the inner part, but the  bulk of the disk that defines the luminosity is in the outer part. Therefore, the luminosity will still depend mainly on the drift-limited regime. The disks with substructures can be fragmentation-limited  farther out in the formed rings, but considering that these rings are mostly optically thick, the difference between exponents is much smaller than for the smooth disks.

The grain composition consists of $\mathrm{10\%}$ silicates, $\mathrm{30\%}$ carbonaceous materials, and $\mathrm{60\%}$ water ice by volume. For a direct comparison with observations \citep[][]{tripathi2017millimeter,Andrews2018a}, we calculate the opacity in band 7 (i.e., at 850 $\mathrm{\mu m}$). Afterward, we use the absorption opacity to calculate the continuum intensity profile.

\begin{figure}
    \centering
    \includegraphics[width=\linewidth]{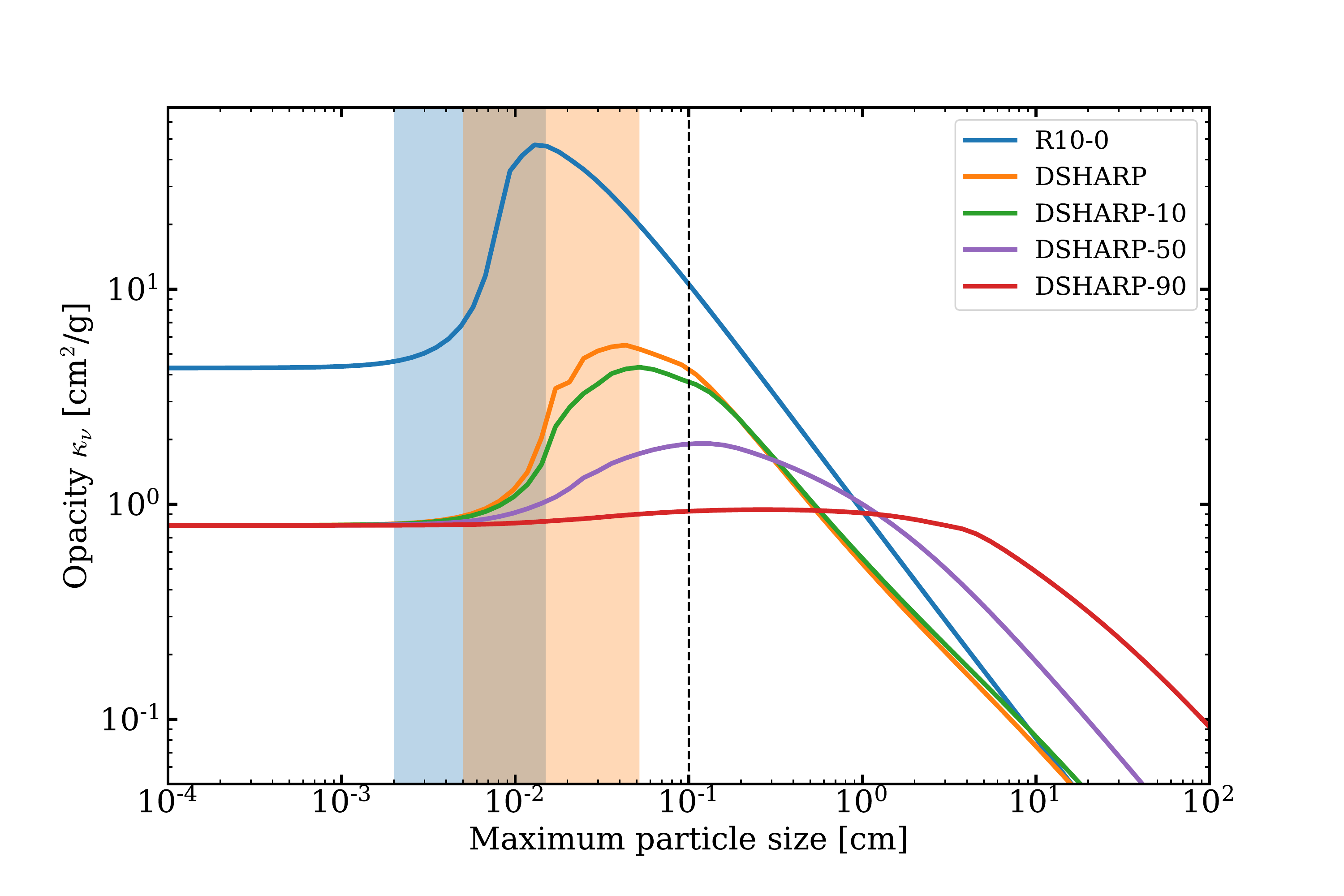}
    \caption[]{Comparison between the opacity models that we used at 850 $\mathrm{\mu m}$ as a function of the maximum particle size and for a power-law size distribution with an exponent of $\rm{q=2.5}$. Shown is  the \textit{opacity cliff} (see text for details) at a wavelength of $\SI{850}{\mu m}$ (blue and orange shaded regions). The blue line refers to the opacity model from \citet{Ricci_2010} with compact grains (labeled R10-0) and the orange line  to that of  \citet{Birnstiel_DSHARP} with compact grains (labeled  DSHARP). The green, purple, and red lines refer to $\mathrm{10\%}$, $\mathrm{50\%,}$ and $\mathrm{90\%}$ porous grains in the DSHARP model. The R10-0 and DSHARP opacity values  differ by a factor of $\mathrm{\sim 8.5 }$ at the position of the opacity cliff. As the porosity of the DSHARP model increases, the opacity cliff starts to flatten out, until it completely disappears for very porous grains ($\mathrm{90\%}$). The location of the cliff shifts to larger particle sizes as it diminishes in porosity. The black dashed line shows the particle size at $\mathrm{\SI{1}{mm}}$. The value for the R10-0 model at this size corresponds to $\mathrm{\kappa_{\nu}^{R10-0}=\SI{9.8}{cm^{2}/g}}$, while for the DSHARP to $\mathrm{\kappa_{\nu}^{DSHARP}=\SI{4}{cm^{2}/g}}$.}
    \label{fig:Opacities}
\end{figure}

We also examined the effect of different opacity models and different grain porosities. As a base model we used the composition from the \citet{Ricci_2010} opacities (hereafter denoted R10-0), but for compact grains (i.e., without porosity) as in the model of \citet{rosotti2019millimetre}. Furthermore, we used the DSHARP opacity model \citep{Birnstiel_DSHARP}  and we altered the grain porosity to $\mathrm{10\%}$ (little porous, DSHARP-10), $\mathrm{50\%}$ (semi-porous, DSHARP-50), and $\mathrm{90\%}$ (very porous, DSHARP-90). The particle bulk densities for the different porous grains are $\rm{\rho_{s}}=\SI{1.508}{g/cm^3}$, $\rm{\rho_{s}}=\SI{0.838}{g/cm^3}$, and $\rm{\rho_{s}}=\SI{0.168}{g/cm^3}$, respectively. An important feature of the opacity models that we used is   called the  \textit{opacity cliff}, which  refers to the sharp drop in the opacity at $\SI{850}{\mu m}$, at a maximum particle size around $\SI{0.1}{mm}$, as defined in \citet{rosotti2019millimetre} (see \autoref{fig:Opacities}). In all the figures that are shown in this paper the R10-0 opacity model from \citet{Ricci_2010}  is used, unless it is explicitly stated otherwise.

\subsection{Matching simulations}
\label{sub:survival_frequency}

The behavior of the simulations on the size-luminosity diagram ($\mathrm{L_{mm}-r_{eff}}$ plane, hereafter \textit{SL Diagram}) depends on the time evolution of disks. According to \citet{Andrews2018a}, the linear regression of the joint data between \citet{tripathi2017millimeter} and \citet{Andrews2018a} gives a relation between the disk size and the $\SI{340}{GHz}$ luminosity. The effective radius $\mathrm{r_{eff}}$ and the luminosity $\mathrm{L_{mm}}$ are correlated as
\begin{equation}
\mathrm{log} \mathrm{r_{eff}} = (2.10^{+0.06}_{-0.03}) + (0.49^{+0.05}_{-0.03}) \mathrm{log} \mathrm{L_{mm}}\,,
\label{eq:size-lum}
\end{equation}
with a Gaussian scatter perpendicular to that scaling with a standard deviation (1$\mathrm{\sigma}$) of $\mathrm{0.20^{+0.02}_{-0.01}}$ $\rm{dex}$ (where $\mathrm{r_{eff}}$ is in $\rm{au}$ and $\mathrm{L_{mm}}$ is in $\rm{Jy}$ at a distance of $\SI{140}{pc}$).

In \autoref{fig:param_effects} (top left) we show an evolution track. It is the path that the simulations follow on the luminosity-radius diagram in a chosen  time span. They move from the top right  (higher luminosity and radius)  to the bottom left as the disk evolves. We  plot the evolution track from $\SI{300}{kyr}$ to $\SI{3}{Myr}$. The thought behind this is that at roughly $\SI{300}{kyr}$ our disks reach a quasi-steady state. 
Specifically,  the drift speed in the drift limit is $\mathrm{V_{r} = \epsilon V_{K}}$, with $\mathrm{\epsilon}$ being the dust-to-gas ratio and $\mathrm{V_{k}}$ the Keplerian velocity. So after at least one order of magnitude is lost in the dust, the evolutionary time becomes too long. We refer the reader to \autoref{sec:discussion}, where we show the evolution of disk dust-to-gas ratio as a function of time for different cases. Longer evolutionary times than  we are exploring here do not alter our results significantly, and are not included to simplify the discussion. They will be included as a topic of future research since at later stages disks are more strongly affected by dispersal. Moreover, our chosen time span covers the observed disks from the \citet{Andrews2018a} and \citet{tripathi2017millimeter} joint sample.

In order to filter our simulations, we divided them in categories. A simulation that  lies within $\mathrm{1\sigma}$ (blue shaded region in \autoref{fig:param_effects}, top left, of the SLR) (\autoref{eq:size-lum}) at all times in our chosen time span is considered matching (see \autoref{fig:param_effects}, green evolution track). On the other hand, if at any time a simulation does not lie within the area, it is considered  \textit{discrepant}. The \textit{discrepant} simulations can be further divided into two subcategories. One that is above the SLR (see \autoref{fig:param_effects}, purple and
 yellow tracks, top left) and one below (red track). With this classification we can identify the main parameters that drive a simulation to be located in a certain spot on the SL Diagram. It is worth mentioning that a fraction ($\mathrm{\sim32\%}$) of the observational data points lie outside the 1$\mathrm{\sigma}$ region by definition. This highlights that the definition of the matching simulations is conservative with respect to the observational data. 

Later on we define the term \textit{matching fraction} as the percentage of the matching simulations against the total number of simulations performed with a certain initial condition (see \autoref{sub:evolution_tracks}).

%--------------------------------------------------------------------
\section{Results}
\label{sec:results}
In this section we present the main results of this analysis. In \autoref{sub:evolution_tracks} we explain the effect of every parameter on the path of the disk in the SL diagram. In \autoref{sub:heatmaps} we present the general properties that disks should have to follow the SLR, and we derive a theoretical SLR for disk with substructures in \autoref{subsub:scaling_relation}. In \autoref{app:corner} we present an additional analysis for the results discussed below.

\subsection{Evolution tracks}
\label{sub:evolution_tracks}

\begin{figure*}
    \centering
    \includegraphics[width=0.49\linewidth]{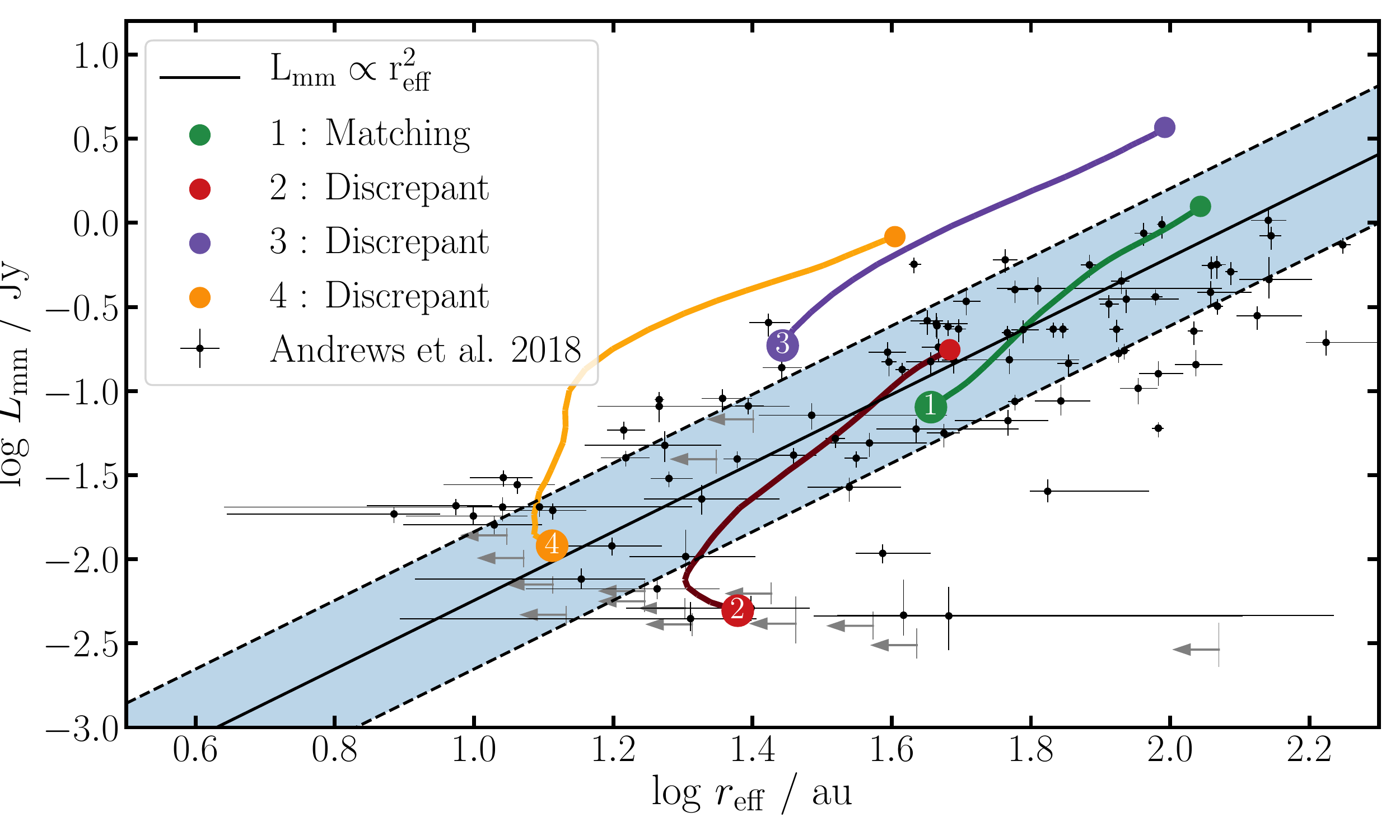}
    \includegraphics[width=0.49\linewidth]{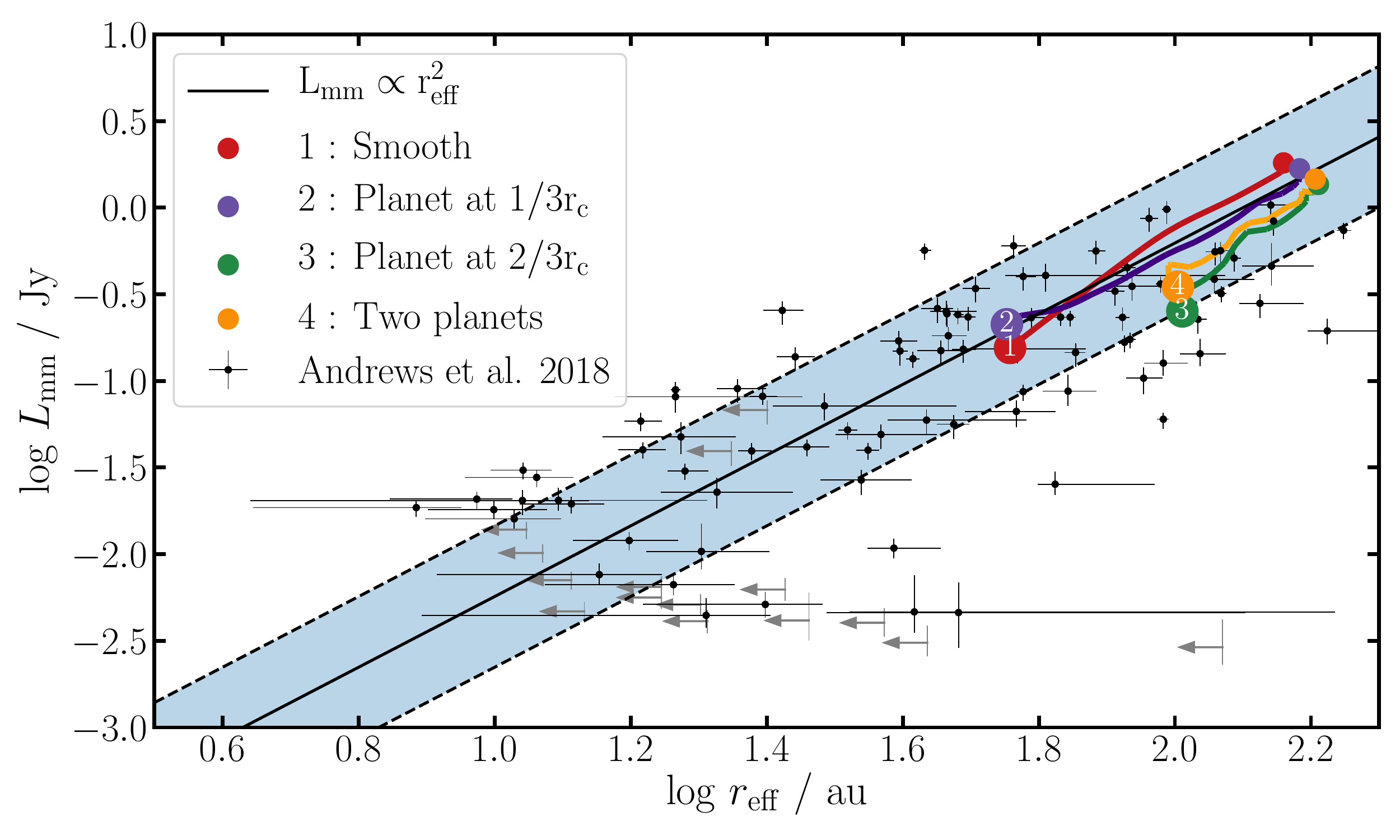}\\
    \includegraphics[width=0.49\linewidth]{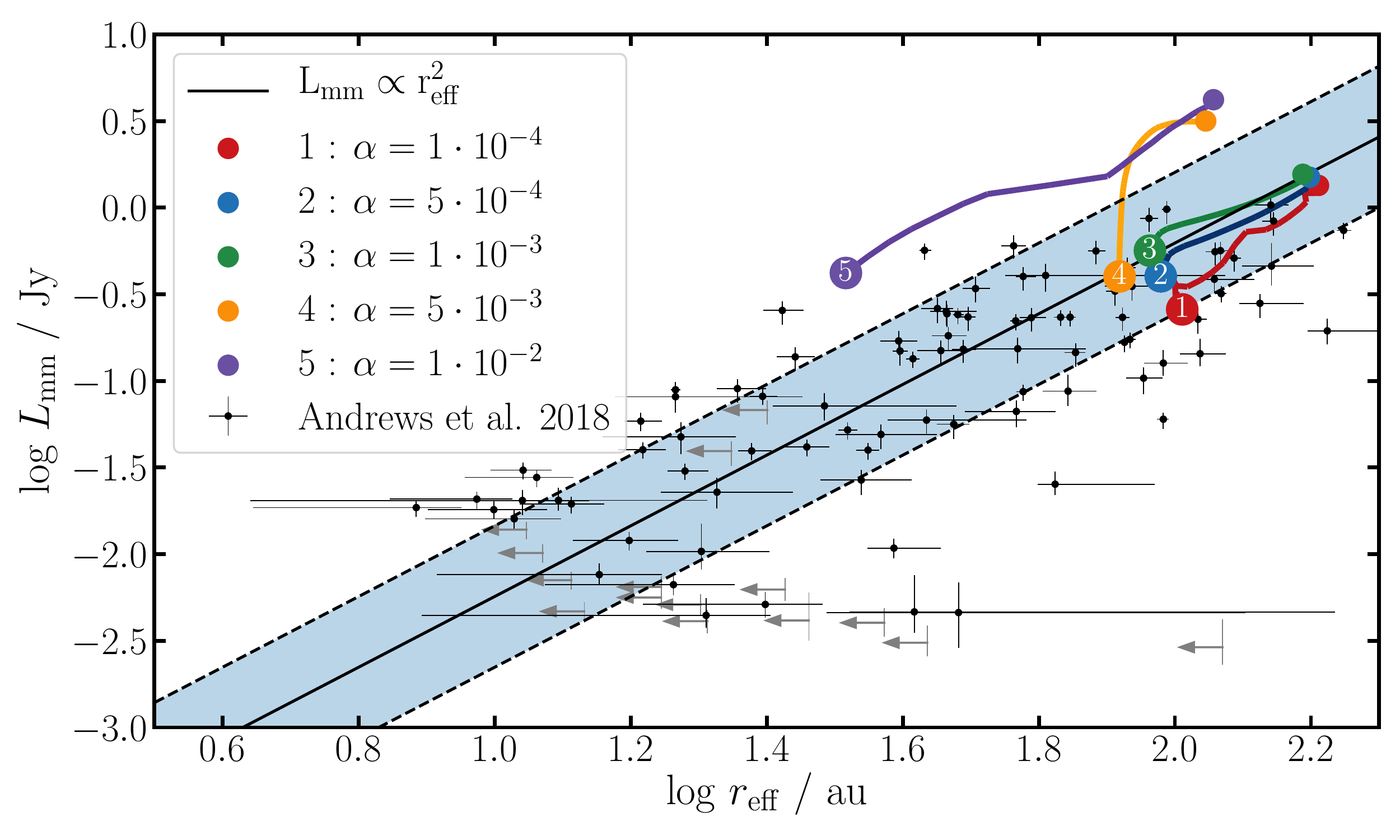}
    \includegraphics[width=0.49\linewidth]{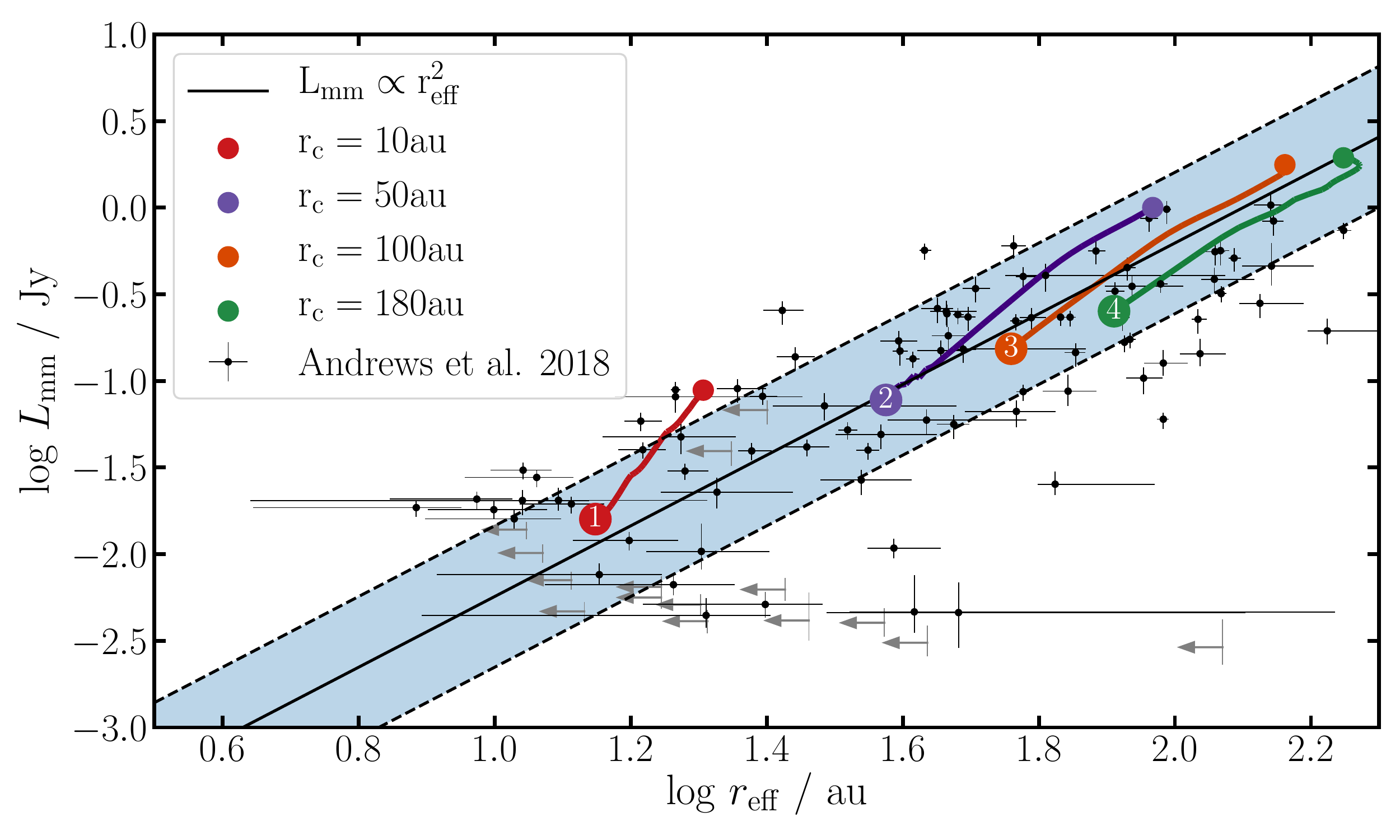}
    \caption[]{Evolution tracks and explanation of matching and discrepant simulations. Examples of a disk with the same initial conditions varying only one parameter at a time. \textbf{Top left}:  SLR according to \citet{Andrews2018a} and examples of evolution tracks. The black points correspond to the observational data, and the black dashed line is  \autoref{eq:size-lum}, which shows the relation between the luminosity and the effective radius. Finally, the blue shaded region is the area within $1\sigma$ of the SLR where the simulations are  considered to be \textit{matching}. The green track  labeled  number 1 is considered  a \textit{matching} simulation since it starts and ends inside the SLR. The other tracks are considered  \textit{discrepant}. The beginning of the track is where the empty bullet is (top right) and the end is where the number is found (bottom left). \textbf{Top right}: Varying only the presence and the position of a Jupiter-mass planet. The red line (number 1) is for a  smooth disk and the green line (number 3)   for a disk with a Jupiter-mass planet at 2/3 of the \rc;   the purple line (number 2) corresponds to a planet at 1/3 of the \rc and the yellow line (number 4) to a simulation with two planets at the positions mentioned above. \textbf{Bottom left}: Varying only the turbulence parameter $\mathrm{\alpha}$. Higher $\mathrm{\alpha}$ values lead to higher luminosity. \textbf{Bottom right}: Varying only the characteristic radius \rc. Higher \rc values lead to larger and more luminous disks.}
    \label{fig:param_effects}
\end{figure*}

In \autoref{sub:survival_frequency} we explain what an evolution track is, while we show some examples in \autoref{fig:param_effects} (top left). Every track is affected by the initial conditions of the parameters chosen, and by the presence or not of a planet. In the following sections we explore the effect of the most important parameters of our grid model and we show how every parameter affects the evolution track on the SL Diagram in \autoref{fig:param_effects}, \autoref{fig:mass_acc_1e-3}, and \autoref{fig:parameter_tracks}.

Since the grid consists of 100.000 simulations, single evolution tracks do not show the preferred initial conditions that allow a simulation to stay in the SLR, but only a representative case. In order to identify trends between the initial conditions and the matching fraction of every disk, we constructed histograms where on the y-axis we have the matching fraction (i.e., the percentage of simulations that stay on the SLR for the chosen time span) and on the x-axis we have the value of each parameter in our grid model. Different colors represent different simulation grids, with or without planets and with varying the planetary mass or positions (e.g., \autoref{fig:Histograms}). In black we show the simulations where we used a smooth surface density profile, as in \citet{rosotti2019millimetre}. In green we show the case where the planet/star mass ratio is $\mathrm{q=1\cdot10^{-3}}$ at a location of 1/3 of the \rc (inner planet), in red the same planet at a distance of 2/3 of the \rc (outer planet), and in blue two planets of $\mathrm{q=1\cdot10^{-3}}$ at 1/3\rc and 2/3\rc. The white hatched bars show the same cases, but using a different opacity model DSHARP \citep{Birnstiel_DSHARP}. 

\subsubsection{Effect of planetary parameters}
\label{subsusb:planet_tracks}

In a smooth disk, the evolution track evolves toward smaller radii and lower luminosity as the dust drifts inward, so the emission (and size) decreases. Moreover, the opacity cliff moves farther in as the radius in the disk where the maximum particle size $\mathrm{a_{max}}$ is (i.e., the value for the peak opacity) decreases, due to radial drift and grain growth, as in \citet{rosotti2019time_evolution}.

In contrast, in the case where a planet is present the pressure bump that is formed stops the dust from drifting toward the host star,  delaying the evolution of the disk on the SL Diagram, thus keeping the tracks on the SLR for longer times. With this in mind, we expect a less extended evolution track when we include planets, since planets are included early in the disk evolution. In \autoref{fig:param_effects} (top right), we show an example of the evolution tracks of a disk with the same initial conditions, varying only the presence and the position of a planet. The red line represents the evolution track of a disk with a smooth surface density profile. If we include a planet with planet/star mass ratio $\mathrm{q=10^{-3}}$, Jupiter mass in this case, at a location that is close to the characteristic radius of the disk, in this case at 2/3 of the \rc, we see from the green line (number 3) that both the effective radius and the luminosity increase relative to the planet-less case, and the evolution track is shorter at the same time span as in all planet cases. This is clear since the pressure bump is trapping particles and the dust mass is retained. Therefore, the luminosity does not decrease as quickly. At the same time, the fixed position of the pressure bump causes the effective radius to remain the same. Together this means that the track in the SLR comes to a halt.

On the other hand, if we place the planet close to the star, in this case at 1/3 of the \rc, we observe a much longer track similar to the one with the smooth profile (purple line). The size and the luminosity of the disk change only slightly. The smaller radius compared to the case where we have an outer planet is explained because the dust now stops at the inner pressure bump and the luminosity is roughly the same for the two cases. This could lead to the conclusion that a planet close to the star will not affect dramatically its position on the SL Diagram, but this is not true for all cases (see \autoref{subsusb:rc_tracks}). Instead, the evolution track here is similar to the smooth one because the disk is too large and massive and the inner planet cannot affect the evolution track too much.

As a last point, we included two planets, one at the 2/3 of the \rc and one at 1/3. We observe a similar evolution track as in the case where we have only one planet close to the outer radius, with the disk being slightly more luminous. This is also explained by the fact that the dust from the outer disk stops at the outer pressure bump, while the dust that exists inside the outer planet stops at the pressure bump of the inner planet. Since most of the dust mass initially resides outside the outer planet, the dust that is trapped between the two planets contributes only partially to the total luminosity. When two or more planets are present the location of the outermost planet is more dominant in the evolution track. 

\subsubsection{Effect of the turbulence $\mathrm{\alpha}$-parameter}
\label{subsusb:alpha_tracks}

The effect of the turbulence parameter $\mathrm{\alpha}$ on the evolution track is straightforward: higher $\mathrm{\alpha}$ leads to higher luminosity. In \autoref{fig:param_effects} (bottom left) we show the evolution tracks of a disk with the same initial conditions, varying only the $\mathrm{\alpha}$-parameter. In this case, we choose a disk where we have also inserted a Jupiter mass planet at the 2/3 of the characteristic radius (\rc) since the effect is more prominent on these disks.

\begin{figure}
    \centering
    \includegraphics[width=\linewidth]{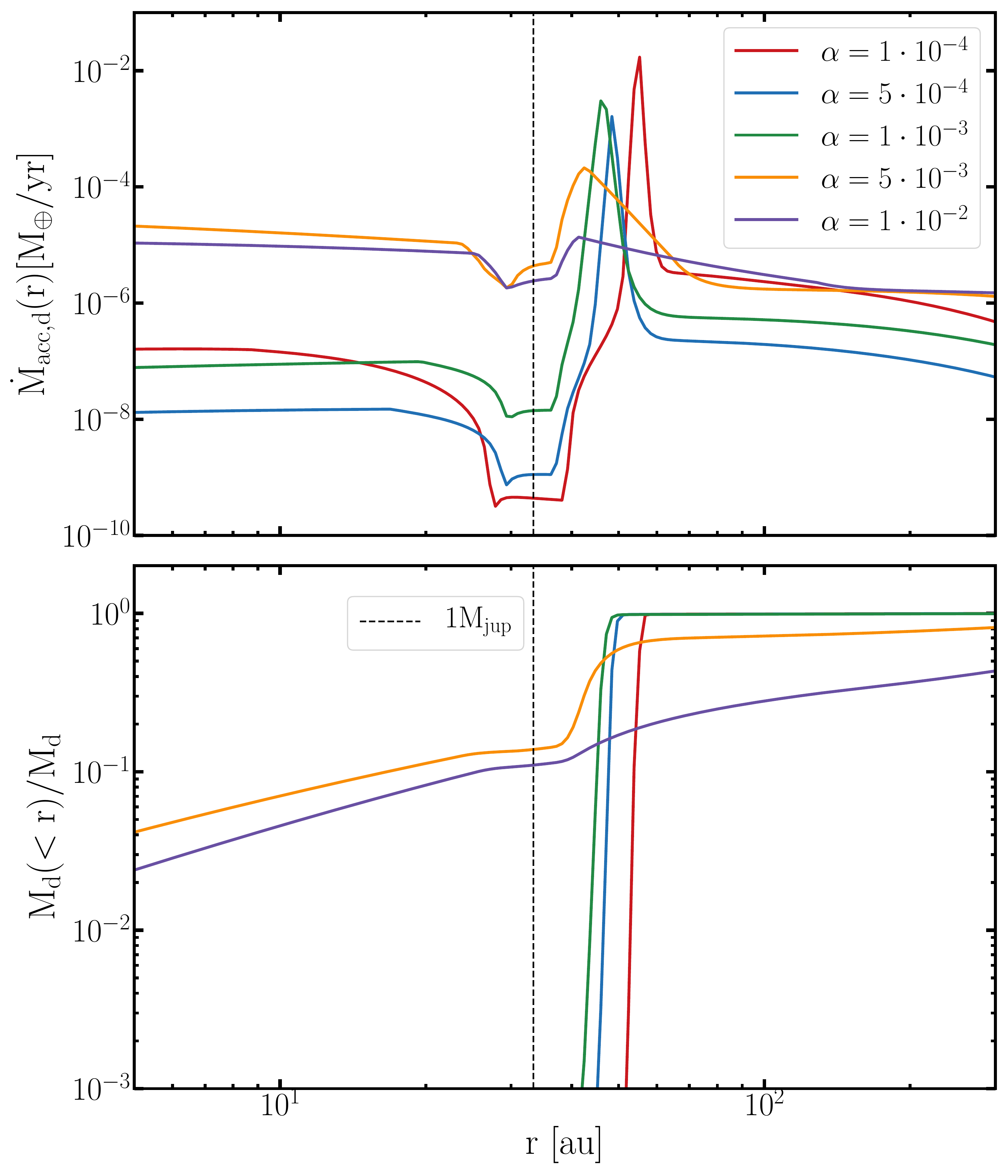}
    \caption[]{\textbf{Top panel}: Local flow rate of the dust mass in $\mathrm{M_\oplus/year}$ as a function of radius for a disk with a Jupiter mass planet at $\SI{31}{au}$. For the low $\mathrm{\alpha}$-values $\mathrm{1\cdot10^{-4}}$ (red line), $\mathrm{5\cdot10^{-4}}$ (blue) and $\mathrm{1\cdot10^{-3}}$ (green), we observe that the bump outside the planet location is large enough to hold the dust from drifting toward the star, while for larger $\mathrm{\alpha}$-values $\mathrm{5\cdot10^{-3}}$ (yellow), $\mathrm{10^{-2}}$ (purple), there is only a weak accumulation outside the planetary gap.

    \textbf{Bottom panel}: Cumulative dust mass contained within a radius $r$, as function of radius. For $\mathrm{\alpha}$-values $\mathrm{1\cdot10^{-4}, 5\cdot10^{-4}, 1\cdot10^{-3}}$, roughly all the mass of the disk is inside the bump that is created from the planet. For larger $\mathrm{\alpha}$-values $\mathrm{5\cdot10^{-3}, 1\cdot10^{-2}}$, the bump is minimal and not able to hold the dust from drifting.}
    \label{fig:mass_acc_1e-3}
\end{figure}

To understand the trends in \autoref{fig:param_effects} (bottom left), it is instructive to consider \autoref{fig:mass_acc_1e-3}, where we show an example of how different $\mathrm{\alpha}$-values affect the efficiency of trapping. In the top panel we show the dust mass local flow rate $\rm{\dot M_{acc,d}(r)}$ $[\mathrm{M_\oplus/yr}]$ as a function of radius. For low $\mathrm{\alpha}$-values  $\mathrm{1\cdot10^{-4}}$ (red line), $\mathrm{5\cdot10^{-4}}$ (blue), and $\mathrm{1\cdot10^{-3}}$ (green), the mass   flows toward the bump and the local flow rate for $\mathrm{r < r_{p}}$ is low ($\mathrm{\leq10^{-8} M_\oplus/yr}$), meaning that the trapping is efficient enough to stop the dust from drifting toward the star. On the other hand, for the high $\mathrm{\alpha}$-values $\mathrm{5\cdot10^{-3}}$ (yellow) and $\mathrm{1\cdot10^{-2}}$ (purple), the dust mass local flow rate stays almost constant ($\mathrm{\leq 10^{-5} M_\oplus/yr}$) throughout the whole disk, meaning that  the bump does not trap the particles, just locally slows them down. In the bottom panel we show the cumulative mass of the disk integrated from inside out as a function of radius. For the low $\rm{\alpha}$-values, most of the mass is located in the bump, while for $\rm{\alpha}=\rm{5\cdot10^{-3}}$ and $\rm{\alpha}=\rm{1\cdot10^{-2}}$ it increases with the radius, meaning that the bump allows more  grains to escape and therefore it does not contain a significant fraction of the disk mass.  

Returning to \autoref{fig:param_effects} (bottom left), disks with low to medium $\mathrm{\alpha}$-values ($10^{-4}$, $5\cdot10^{-4}$, $10^{-3}$) are less luminous than disks with higher $\mathrm{\alpha}$-values. This is so because the ring that is formed due to the pressure bump is becoming too narrow and optically thick. The total flux emitted by an optically thick ring of a given temperature is just a function of the emitting area. Lower $\mathrm{\alpha}$ will lead to a narrower ring \citep{Dullemond2018DSHARP}, and thus less emitting area and therefore lower luminosity, independently of the amount of mass in the ring.
On the other hand, high $\mathrm{\alpha}$-values work against trapping in various ways, leading to more luminous disks. A higher $\mathrm{\alpha}$-value decreases the particle size in the fragmentation limit, which are less efficiently trapped by radial drift (e.g., \citealt{Zhu2012DustFiltration}). It increases the diffusivity, which allows more grains to escape the bump and it increases the viscosity in the same way, so more dust sizes are traveling with the accreting gas. Furthermore it smears out the pressure peak, causing less efficient trapping by radial drift \citep{Pinilla2012}. Moreover, with high $\alpha$-values a higher dust-to-gas ratio is retained because grain growth is impeded by fragmentation, hence radial drift is much slower. 

If we consider the two cases where the $\mathrm{\alpha}$-value is high ($5\cdot10^{-3}$, $10^{-2}$), the planet cannot efficiently trap the dust and the disk evolves farther along the SLR to lower luminosities. A disk that contains a planet with $\mathrm{\alpha=10^{-2}}$ behaves the same as a smooth disk without a planet on the SL Diagram, implying that in this case    a very massive planet (several Jupiter masses) would be needed to significantly affect disk evolution. This results in the dust grains  becoming smaller since the gas turbulent velocity is increasing and the collisions between them are more destructive. Consequently, the gap is becoming shallower while there is also more diffusion.

In \autoref{fig:Histograms} (top left), we show the dependence of the $\mathrm{\alpha}$-viscosity parameter on the matching fraction. As expected, all simulations tend to favor low values of the turbulence parameter $10^{-4}\leq \mathrm{\alpha} \leq10^{-3}$. Smooth simulations show a clear tendency toward low $\rm{\alpha}$-values because when they are drift dominated they remain in the SLR. On the other hand substructured disks show a preference toward $5\cdot10^{-4}\leq \mathrm{\alpha} \leq10^{-3}$. The dust trapping is efficient enough and allows the disks to retain their mass, but while moving to higher $\mathrm{\alpha}$-values $\mathrm{\alpha > 2.5\cdot 10^{-3}}$, the trapping stops being efficient and  leads to higher chances that the evolution track will leave the SLR in the selected time span. If $\rm{\alpha}<5\cdot10^{-4}$ the dust rings become narrow and the luminosity is not large enough to place them in the SLR, and therefore the matching fraction decreases.

\subsubsection{Effect of the characteristic radius  \rc}
\label{subsusb:rc_tracks}
In \autoref{fig:param_effects} (bottom right), we plot the evolution tracks of a smooth disk with the same initial conditions while varying only the characteristic radius ($\rm{\rc=\SIlist{10;50;100}{au}}$, and $\SI{180}{au}$). The same trend applies to disks where a planet is included, but for a smooth disk the evolution tracks are longer and the effect is more easily visible. The effect of the characteristic radius on the evolution tracks is straightforward. The larger the \rc the more the evolution track moves toward the top right of the plot, meaning that for a larger disk we expect higher luminosity. In more detail, increasing the characteristic radius (going from the red to green line) explains why the effective radius increases, while the luminosity increases because the total disk mass remains fixed on all these simulations. This result is consistent with the SLR from \citet{Andrews2018a}.

Taking a look at \autoref{fig:Histograms} (middle left), we do not observe a continuous pattern as in the other histograms. Smooth disks do not seem to depend on the characteristic radius as disks of all sizes can reproduce the SLR. On the other hand, substructured disks with small \rc ($\SI{10}{au}$) are mostly above the correlation because they have high luminosity relative to their size (as explained by our size-luminosity estimate in \autoref{disc:opacity}) and they are unable to enter the correlation in time. 
For large radii ($>\SI{150}{au}$) the disks can become too large but with low luminosity, and can end up below the correlation to the very right part of the SL diagram (see \autoref{fig:heatmaps_ricci}).

Therefore, we observe a peak toward a specific characteristic radius, around $\mathrm{80-\SI{130}{au}}$ when a planet is at a location of 1/3 of the \rc and around $\mathrm{30-\SI{80}{au}}$ for the case where a planet is at a location of 2/3 of the \rc. The inner planet constrains the disk to a small size but with relatively high luminosity, therefore placing it above the SLR before $\mathrm{\SI{300}{kyr}}$, while the opposite effect occurs when an outer planet exists. When two planets are included we observe a mixed situation of the single cases. The reason is that the two pressure bumps compete with each other and each  contributes in one of the ways described above.

\subsubsection{Effect of the disk mass - $\rm{M_{d}}$}
\label{subsusb:disk_mass_tracks}

\begin{figure*}
    \centering
    \includegraphics[width=0.49\linewidth]{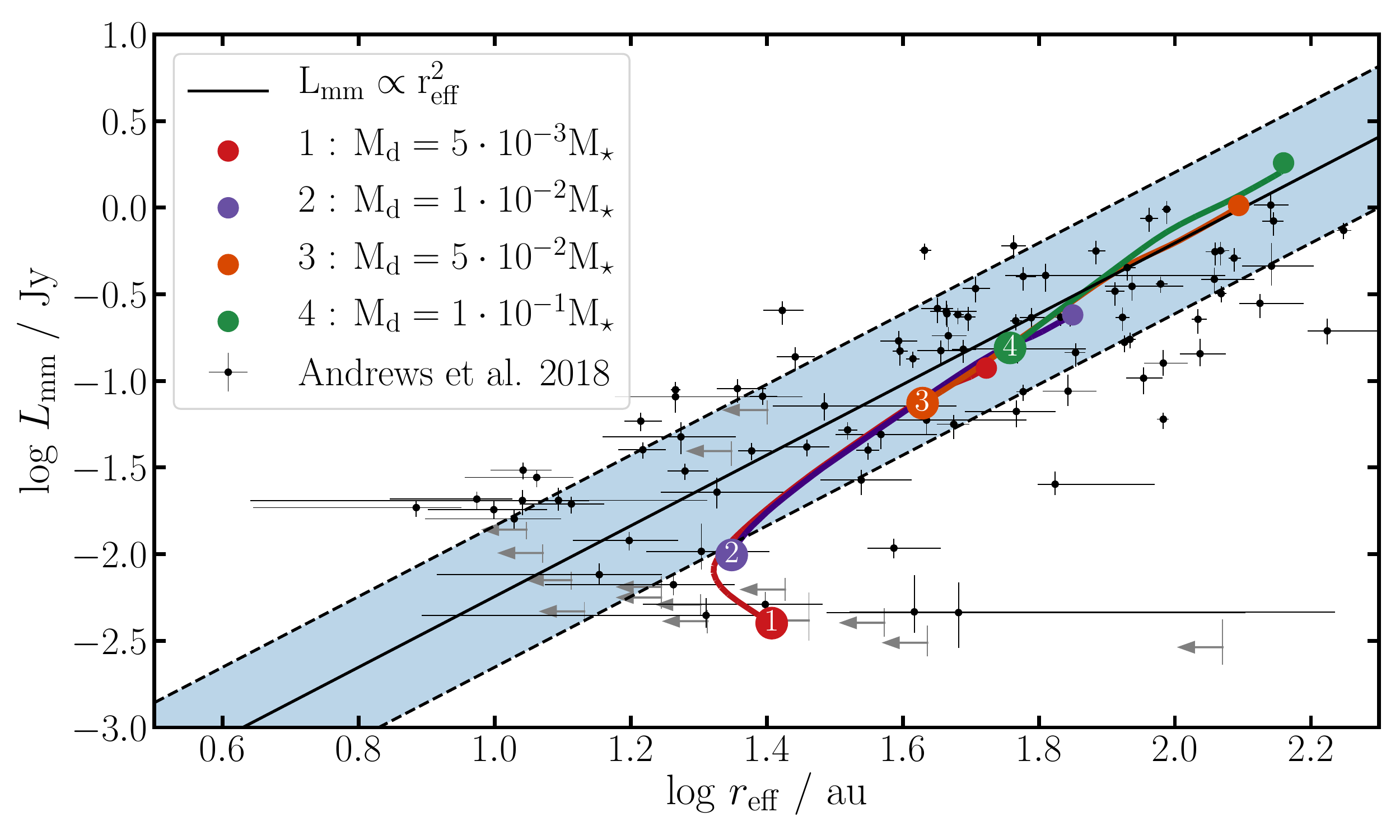}
    \includegraphics[width=0.49\linewidth]{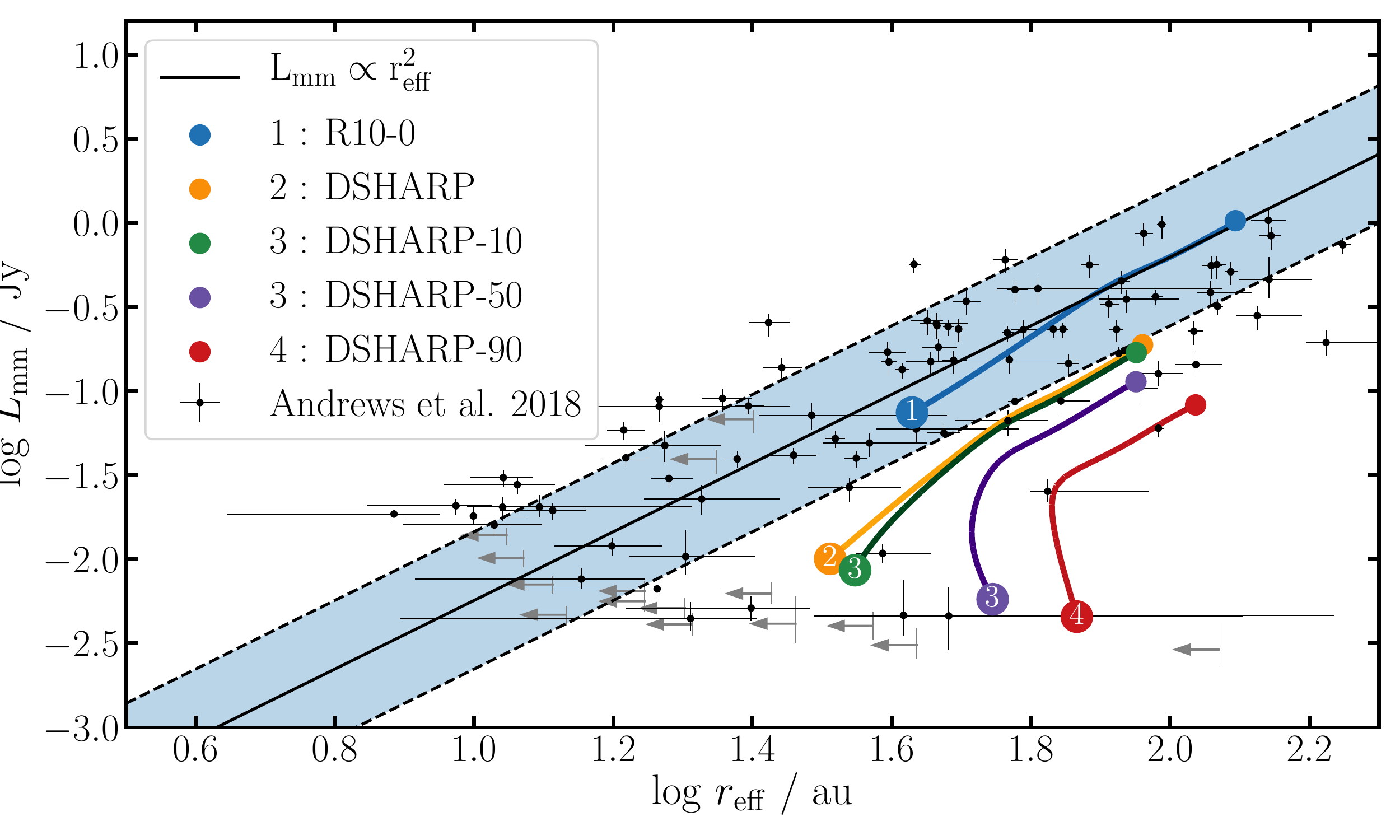}\\
    \includegraphics[width=0.49\linewidth]{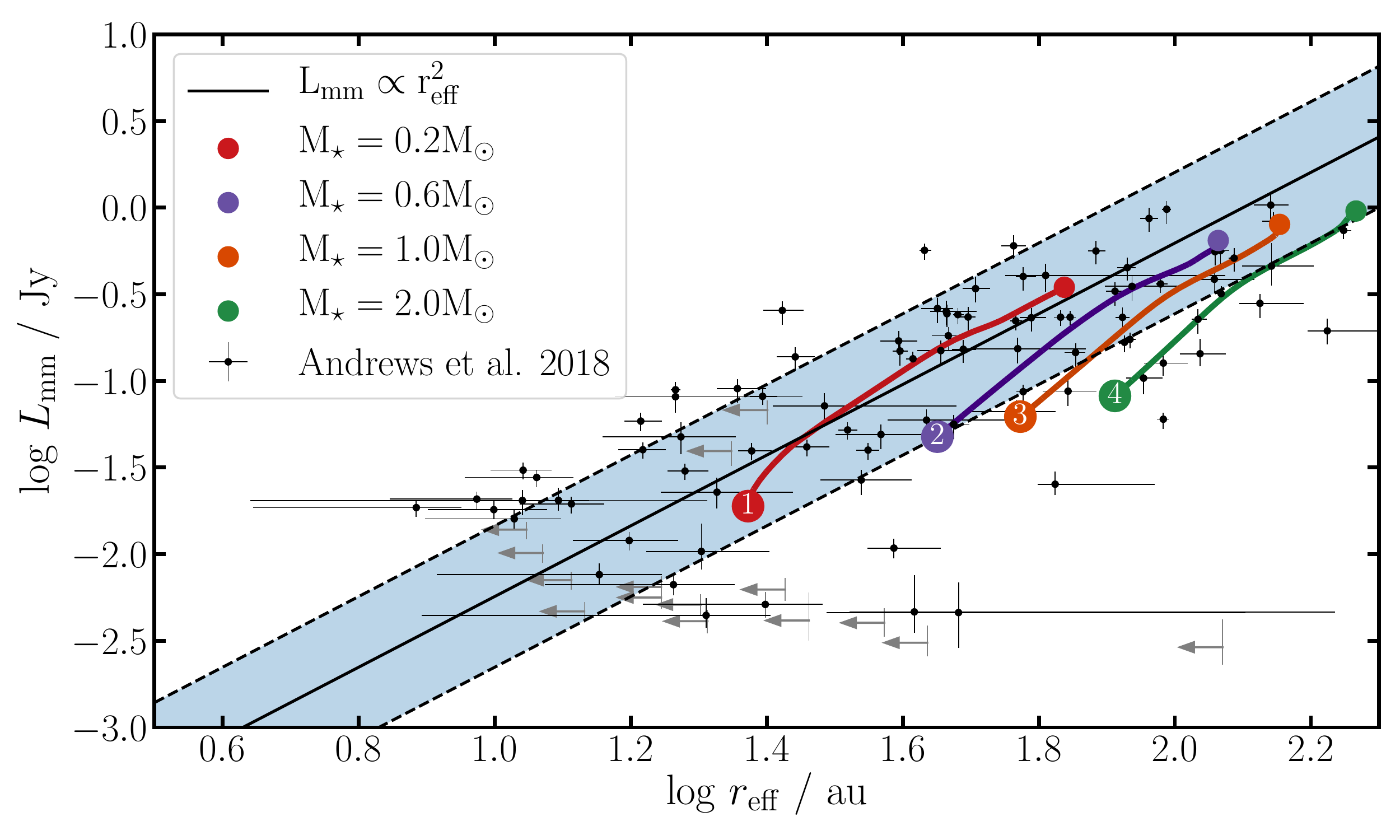}
    \includegraphics[width=0.49\linewidth]{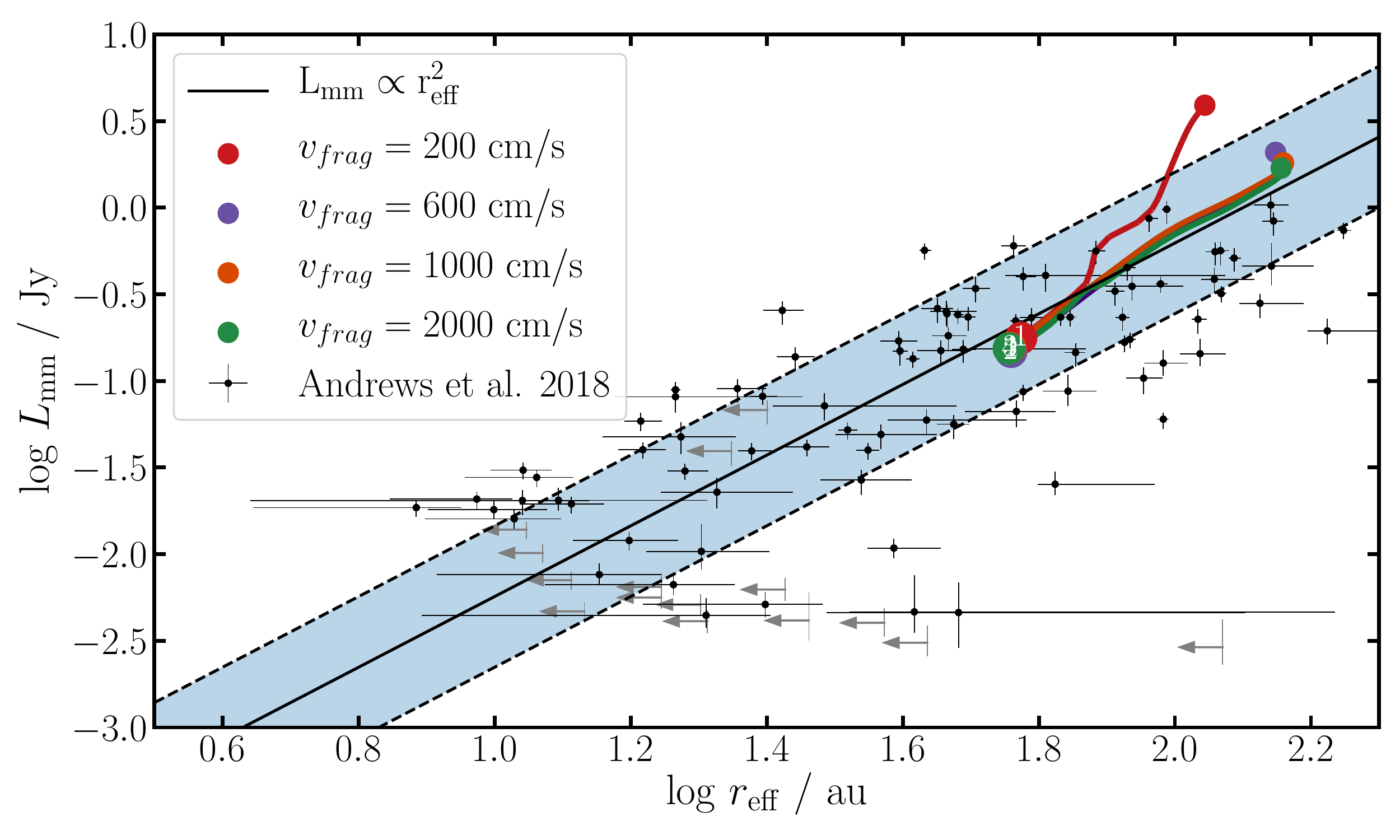}
    \caption[]{evolution tracks with the same initial conditions varying only one parameter at a time. \textbf{Top left}: Varying the disk mass of a smooth disk. \textbf{Top right}: Varying the different opacity models and the porosity of the DSHARP model of a smooth disk. \textbf{Bottom left}: Varying the stellar mass of a disk that contains a planet. \textbf{Bottom right}: Varying the fragmentation velocity of a smooth disk.}
    \label{fig:parameter_tracks}
\end{figure*}

In \autoref{fig:parameter_tracks} (top left) we plot the evolution tracks of a smooth disk with the same initial conditions varying only the disk mass ($\mathrm{M_{d}}$). We choose a smooth disk to show the effect more clearly, but the same principle applies to most disks.

The disk mass contributes to both luminosity and radius. Higher disk masses lead to higher luminosities, both at the beginning and at the end of the track. Since for a fixed \rc, disks with higher $\mathrm{M_{d}}$ have higher $\mathrm{L_{mm}}$, it is only logical that more material will lead to higher luminosity and vice versa. By the end of the evolution tracks, the less massive disks have left the SLR. The dramatic curvature of the SLR for the lowest $\mathrm{M_{d}}$ case ($\mathrm{M_{d}=5\cdot 10^{-3}M_{\star}}$) occurs because all grain sizes become smaller than the opacity cliff. If we choose a disk that contains a planet, the massive disks ($\mathrm{M_{d}\geq5\cdot 10^{-2}} M_{\star}$) will still evolve toward lower radii and luminosities on the SLR, but the less massive ones will have shorter tracks. The pressure bump will trap all the material outside of the planet position, so the emission and the effective radius will both remain almost constant. The only case where the track of a low mass disk can be long is if the planet mass is small and the pressure bump is not large enough to retain dust.

In \autoref{fig:Histograms} (top right) we plot the matching fraction of the disk mass values. The tendency here is that higher disk mass leads to more simulation inside the SLR. The reason for this is that high initial disk mass places the disks above the SLR until they reach a stable state. While the dust is drifting toward the host star the luminosity decreases, allowing them in our chosen time span to reach the SLR and stay there for the remaining time. Since most of the dust is in the trap at this point, the remaining evolution time is set by the trap life time.

The difference is noticeable between the smooth and the planet(s) cases. As we see from the yellow bars, a disk with a smooth surface density profile must be initially massive ($\rm{M_{d}\geq\SI{0.025}{M_{\star}}}$) to remain in the SLR.
The probability of a smooth simulation to match is even greater than the cases where a planet is included.

Some of these results though are an effect of the opacity model used and the chosen time span, as we  discuss in \autoref{subsusb:porosity_tracks}.

\subsubsection{Effect of different opacity and grain porosity}
\label{subsusb:porosity_tracks}
In \autoref{fig:parameter_tracks} (top right) we represent the behavior of several similar tracks varying only the opacity model. The simulations with the \citet{Ricci_2010} opacity model (R10-0, blue line) produce more luminous and larger disks than those with the DSHARP model (orange line)  due to the higher value of the opacity. The R10-0 opacity is 8.5 times higher than the DSHARP at the peak of the opacity cliff. If we use slightly porous grains (DSHARP-10, green line) by altering the DSHARP opacity, we observe that the effect is insignificant as the shape of the opacity is roughly the same. On the contrary, for semi-porous and very porous grains (DSHARP-50 and DSHARP-90, purple and red line, respectively) the opacity cliff starts to flatten out \citep{Kataoka2014}, leading to a disk with low luminosity and no significant change in disk size. 

\begin{figure*}
    \centering
    \includegraphics[width=0.49\linewidth]{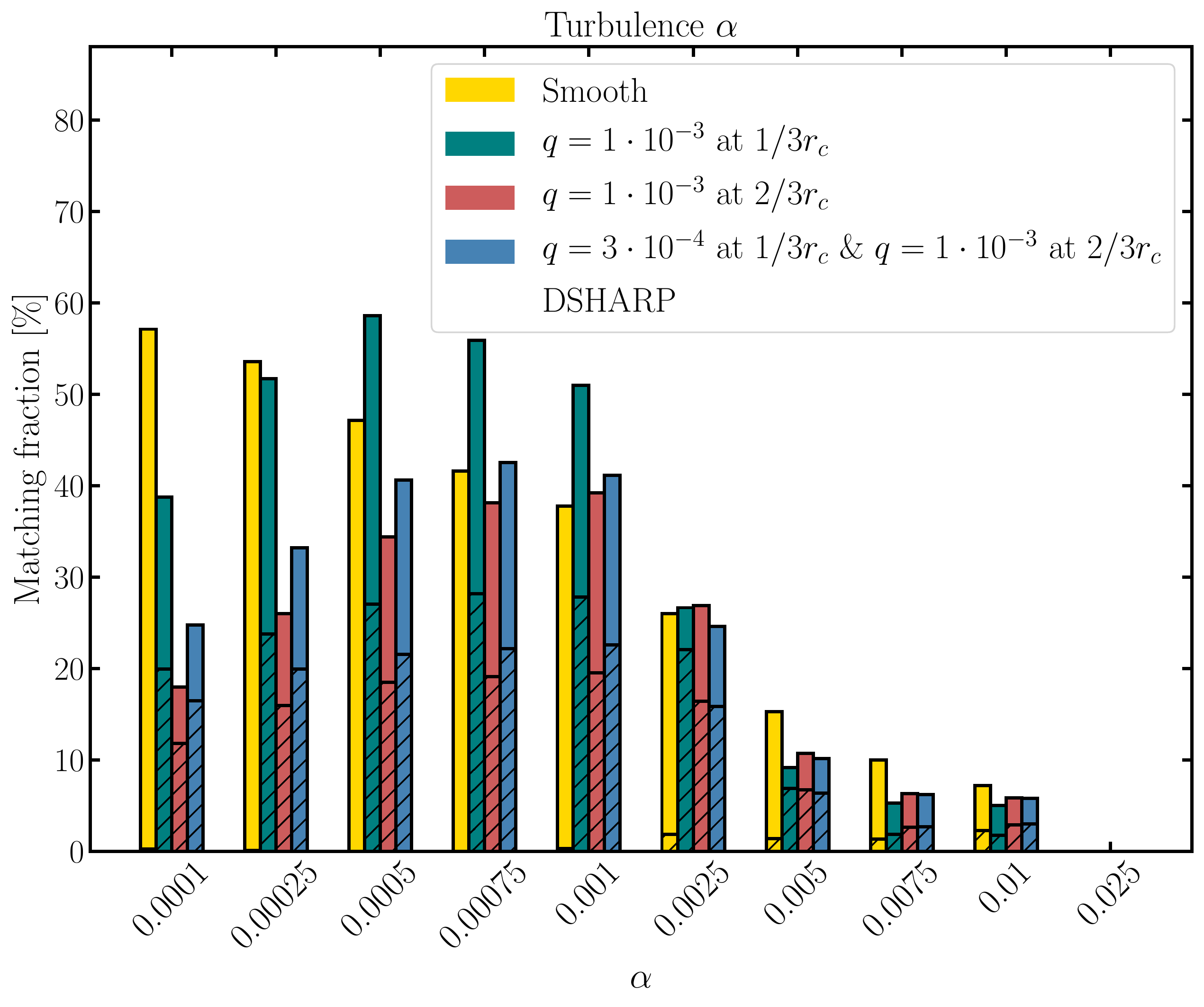}
    \includegraphics[width=0.49\linewidth]{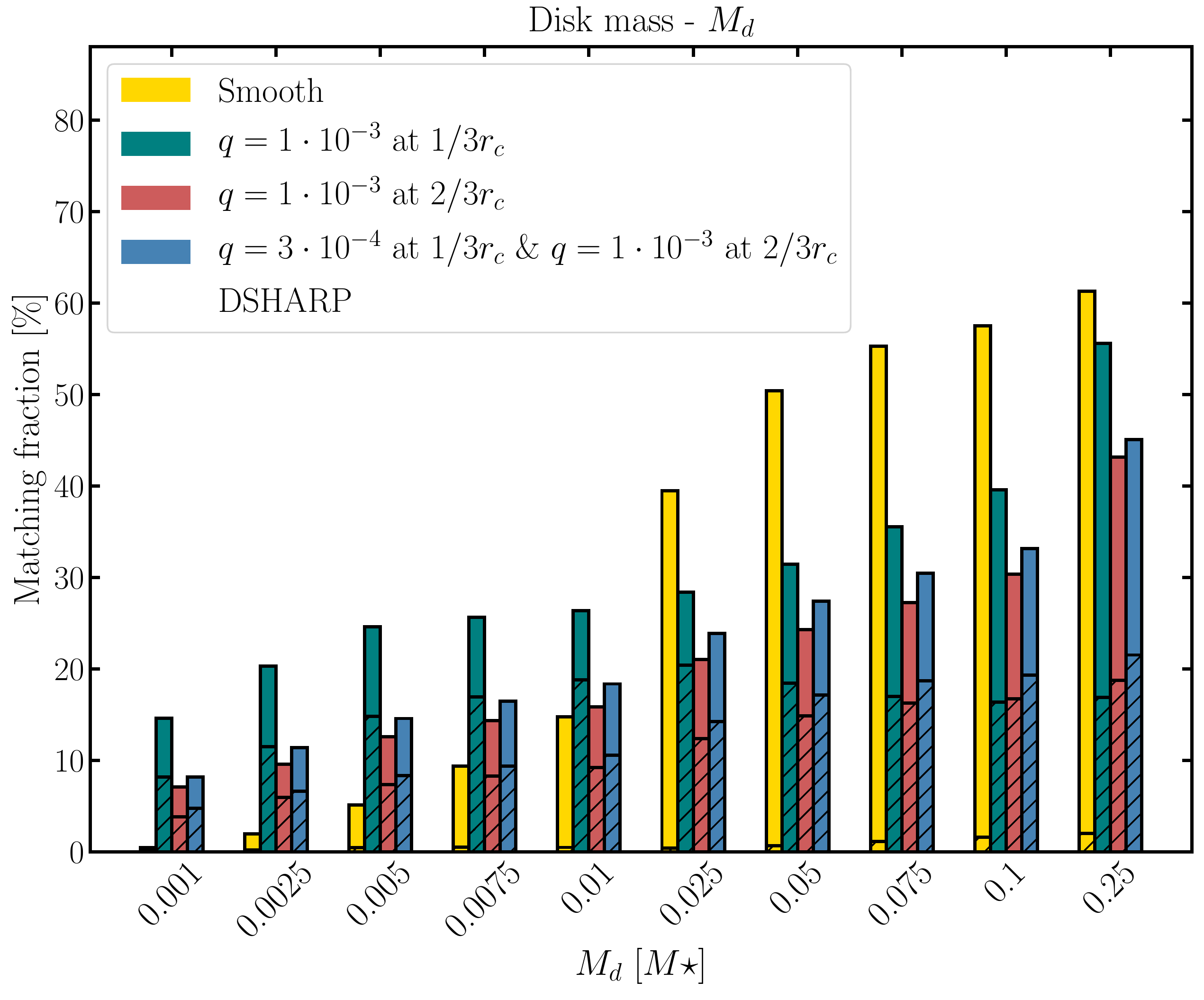}\\
    \includegraphics[width=0.49\linewidth]{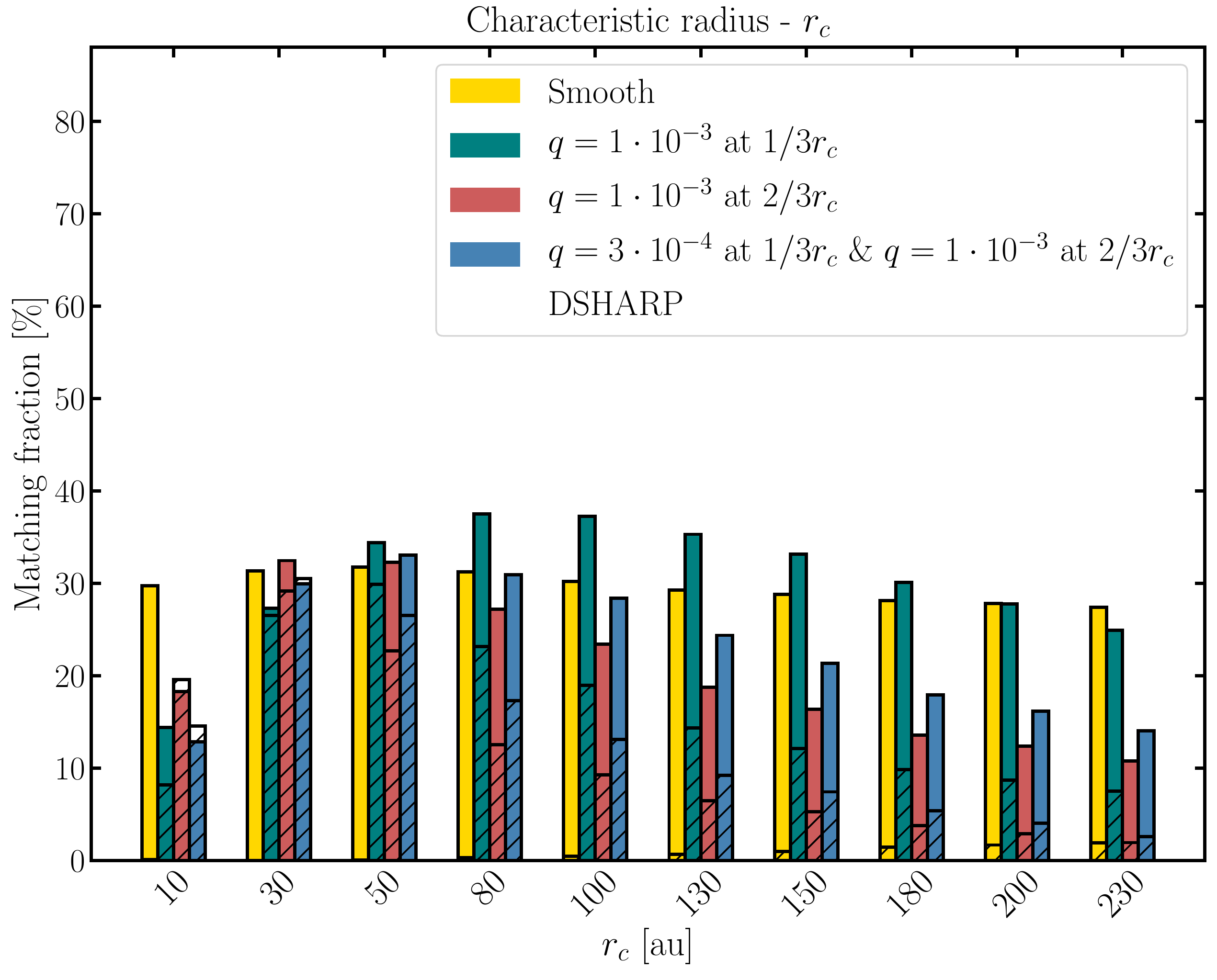}
    \includegraphics[width=0.49\linewidth]{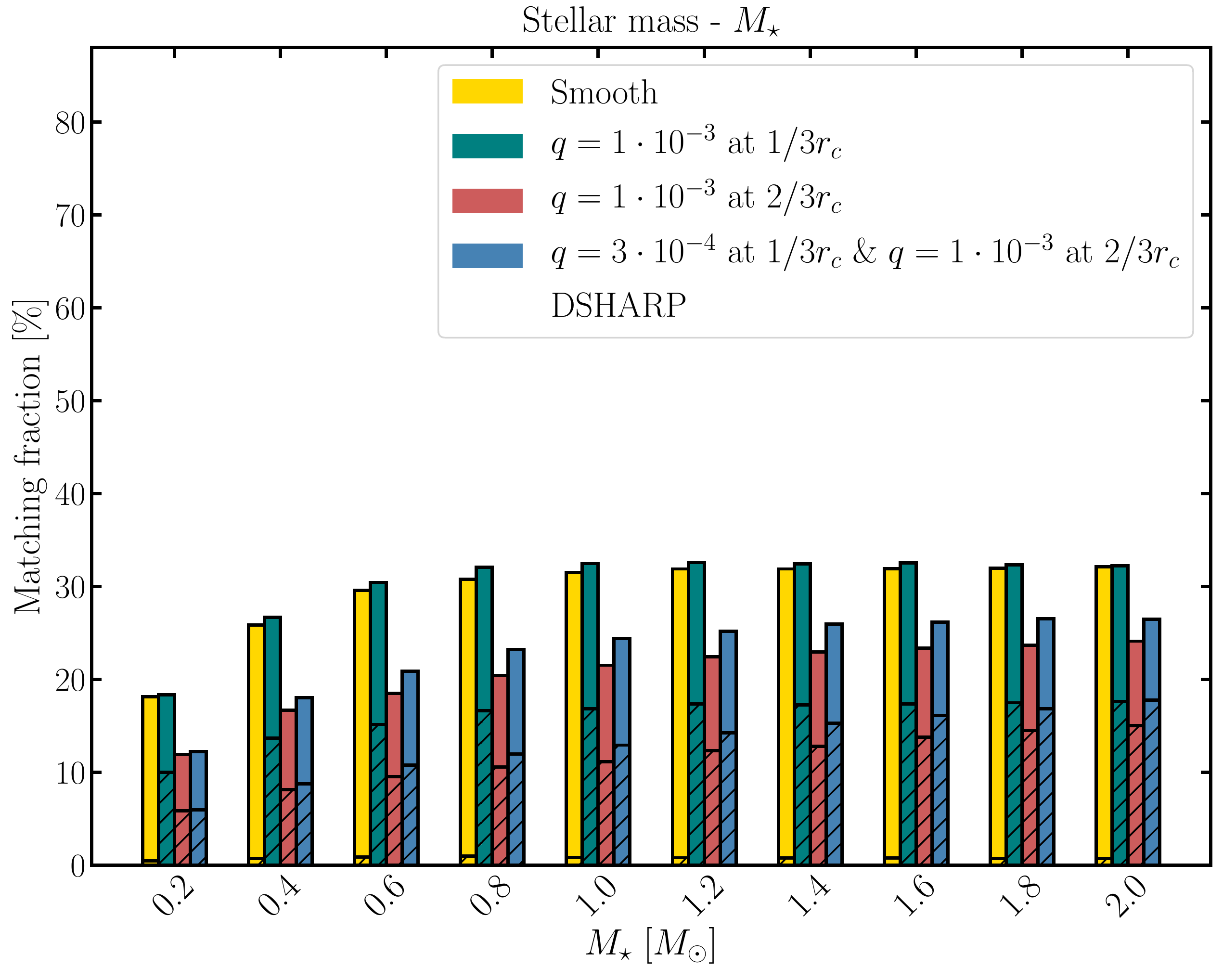}\\
    \includegraphics[width=0.49\linewidth, left]{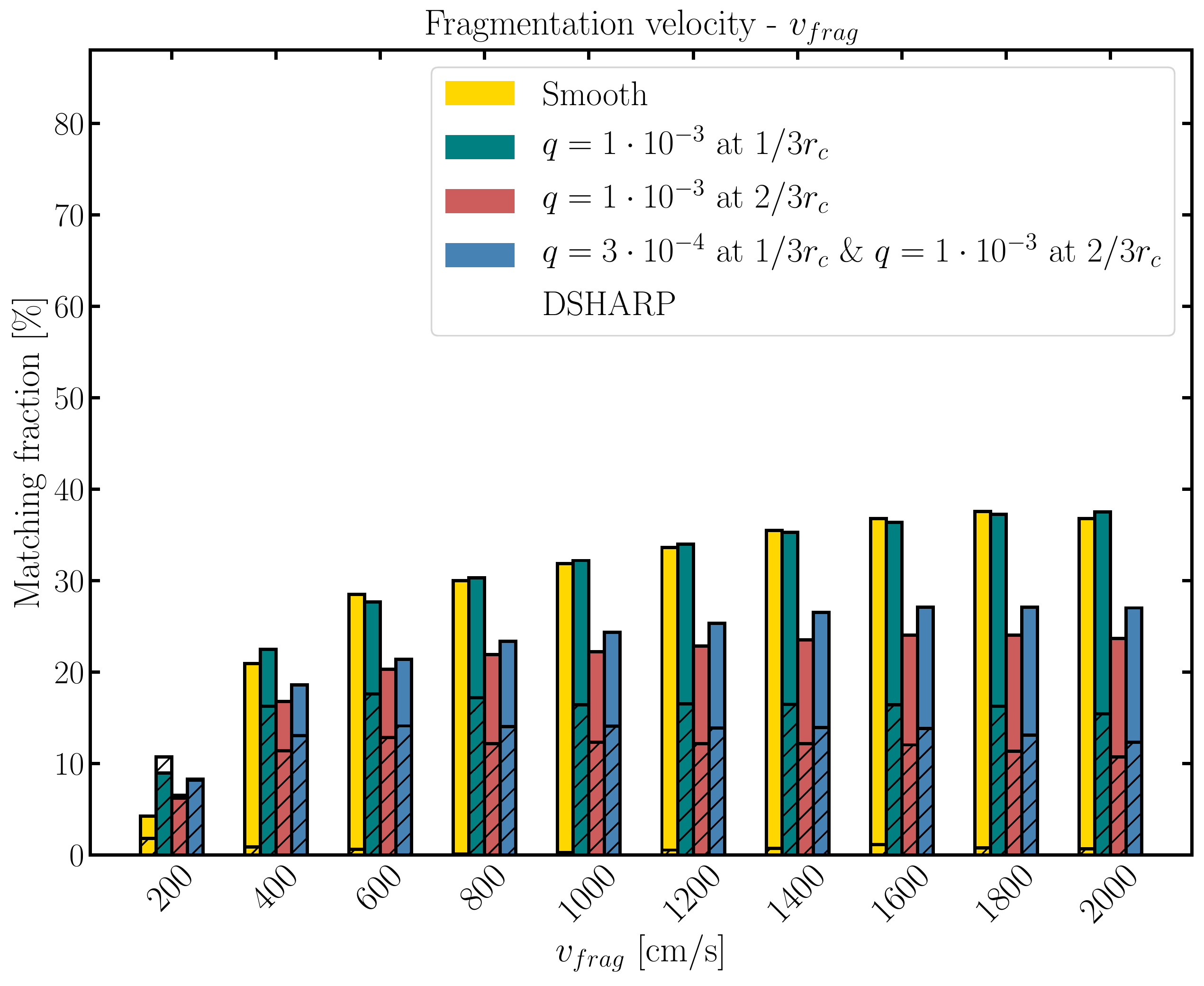}
    \caption[]{Histograms of the matching fraction for disk mass, characteristic radius, stellar mass, and fragmentation velocity. The matching fraction shows the percentage of the simulations that remained on the SLR for the chosen time span ($\rm{\SI{300}{kyr} - \SI{3}{Myr}}$). \textbf{Top left}: Dependence on the $\alpha$-value. There is a preference to low $\alpha$-values $\rm{(10^{-4}\leq \alpha \leq 10^{-3})}$. \textbf{Top right}: Dependence on the disk mass. There is a preference to high disk masses $\mathrm{(0.025 \leq \frac{M_{d}}{M_{\star}} \leq 0.25)}$. \textbf{Middle left}: Dependence on the characteristic radius. Smooth disks do not depend on the \rc. \textbf{Middle right}: Dependence on the stellar mass. There is a slight preference to higher values. \textbf{Bottom left}: Dependence on the fragmentation velocity. There is a slight preference to higher values.}
    \label{fig:Histograms}
\end{figure*}

In all the histogram figures (\autoref{fig:Histograms}) the same trend stands for either the opacity from R10-0  (solid color bars) or  DSHARP \citep{Birnstiel_DSHARP} (hatched bars). The difference is that more simulations match when the R10-0 opacity model is used as opposed to the DSHARP model. Disks with the DSHARP opacity are generally less bright because they do not become optically thick in the rings and end up below the SLR. Therefore, it would need more dust (i.e., stronger traps) for them to be luminous enough. Especially for the smooth case (yellow bars), there are only a few simulations that match, hence the hatched bars are barely visible. For smooth disks the total matching fraction is $\rm{29.6\%}$ with the R10-0 opacity, while it is $\rm{0.8\%}$ with the DSHARP value. For disks with an inner planet the matching fractions are $\rm{30.2\%}$ and $\rm{15.9\%,}$ respectively (see  \autoref{disc:opacity}, where we explore the overall impact of porous grains for the entire grid of models).

\subsubsection{Effect of the stellar mass - $\mathrm{M_\star}$}
\label{subsub:Mstar_tracks}
In our models the stellar mass is assumed to be directly correlated with the disk mass because we varied the stellar mass, but we always kept the disk-to-star mass ratio constant. Therefore, a higher star mass implies a higher total mass of the dust, leading to higher continuum luminosities.

In \autoref{fig:parameter_tracks} (bottom left) we plot the evolution tracks of a disk that contains a planet with the same initial conditions, varying only the stellar mass for $\SI{0.2}{M_{\odot}}$, $\SI{0.6}{M_{\odot}}$, $\SI{1.0}{M_{\odot}}$, and $\SI{2.0}{M_{\odot}}$. As expected, the highest value of $\mathrm{M_\star=2.0M_{\odot}}$ leads to the largest and most luminous disk (green line), while the opposite is true for $\mathrm{M_\star=0.2M_{\odot}}$ (red line). There is similar behavior for smooth disks, but in that case the radius of each disk is  much smaller due to radial drift.

Furthermore, the stellar mass is the least important parameter when defining whether a simulation matches. The histogram in \autoref{fig:Histograms} (middle right) confirms this. Even though the trend shows that using higher stellar mass has a greater matching fraction, this occurs because the stellar mass scales to the disk mass, and higher disk mass leads to more matching simulations (see \autoref{subsusb:disk_mass_tracks}).

This scaling implies that the luminosity ($\mathrm{L_{mm}}$) scales with the stellar mass ($\mathrm{M_{\star}}$). Our models follow a relation that is not as steep as the observed $\mathrm{L_{mm}\propto M_{\star}^{1.5}}$ in \citet{Andrews2018a}, because there is not a correlation between disk size and stellar mass in our simulations (see  \autoref{disc:Lmm-Mstar} and \autoref{fig:L_Mstar} where we explore further this relation).

\subsubsection{Effect of the fragmentation velocity - $\mathrm{v_{frag}}$}
\label{subsub:vfrag_tracks}

In \autoref{fig:parameter_tracks} (bottom right) we plot the evolution tracks, of a smooth disk with the same initial conditions, varying only the fragmentation velocity for values of $\SI{200}{cm/s}$, $\SI{600}{cm/s}$, $\SI{1000}{cm/s}$, and $\SI{2000}{cm/s}$. We observe that for medium and high values of $\mathrm{v_{frag}}$ (in this case for $\mathrm{v_{frag}} \geq \SI{600}{cm/s}$), the evolution tracks overlap. Since most of our simulations are drift-limited we expect that no effect from the fragmentation velocity will take place for these values since particles do not grow big enough to drift, so more mass remains at large radii to produce more emission. Therefore, this effect only arises when the fragmentation velocity  becomes too low, which leads to a higher luminosity in the first snapshots.

Moreover, if a disk is fragmentation-limited then it is so mostly in the inner part. Therefore, considering that the main bulk of the disk is in the outer part, the emission that defines the luminosity will still depend on the drift-limited regime hence leading to the overlapping tracks. The effect of a planet in these tracks is minimal and we expect a similar behavior to the case shown here. 

This can be validated in \autoref{fig:Histograms} (bottom left), where we plot the matching fraction compared to the fragmentation velocity and the tendency is to higher fragmentation velocities for all cases. More specifically, if $\mathrm{v_{frag}\geq 600cm/s}$, there is a large number of matching simulations and it only gets larger with increasing fragmentation velocity. In this range the simulations are mostly drift-limited. For low values of $\mathrm{v_{frag}}$, most of the simulations are fragmentation-limited and they lose luminosity relatively quickly, moving them out of the SLR. Low values of $\mathrm{v_{frag}}$ lead to smaller particles and less efficient trapping. Therefore, those disks lose their solids too quickly. This is analyzed in more detail in \autoref{app:corner}.

For reference, in the  recent lab experiment of  \citet{Wurm2018SSRv..214...52B}, $\SI{100}{cm/s} \leq \mathrm{v_{frag}\leq \SI{1000}{cm/s}}$ is considered  a value consistent with lab work and $\mathrm{v_{frag}>\SI{1000}{cm/s}}$ a high fragmentation velocity.

\subsection{Heat maps}
\label{sub:heatmaps}
Single evolution tracks give us an idea of how a single simulation evolves on the SLR, but with a large sample of simulations like the one we have it is not easy to extract global results. Since there are too many tracks to plot all of them in a single diagram, we treat the position of every simulation at every snapshot as an independent sample, and we plot them on a heat map. In these figures we plot the position of every simulation for a specific case (smooth and planet in different locations), for three different snapshots ($\mathrm{\SI{300}{kyr}}$, $\mathrm{\SI{1}{Myr}}$, $\mathrm{\SI{3}{Myr}}$).

\begin{figure*}
    \centering
    \includegraphics[width=\textwidth]{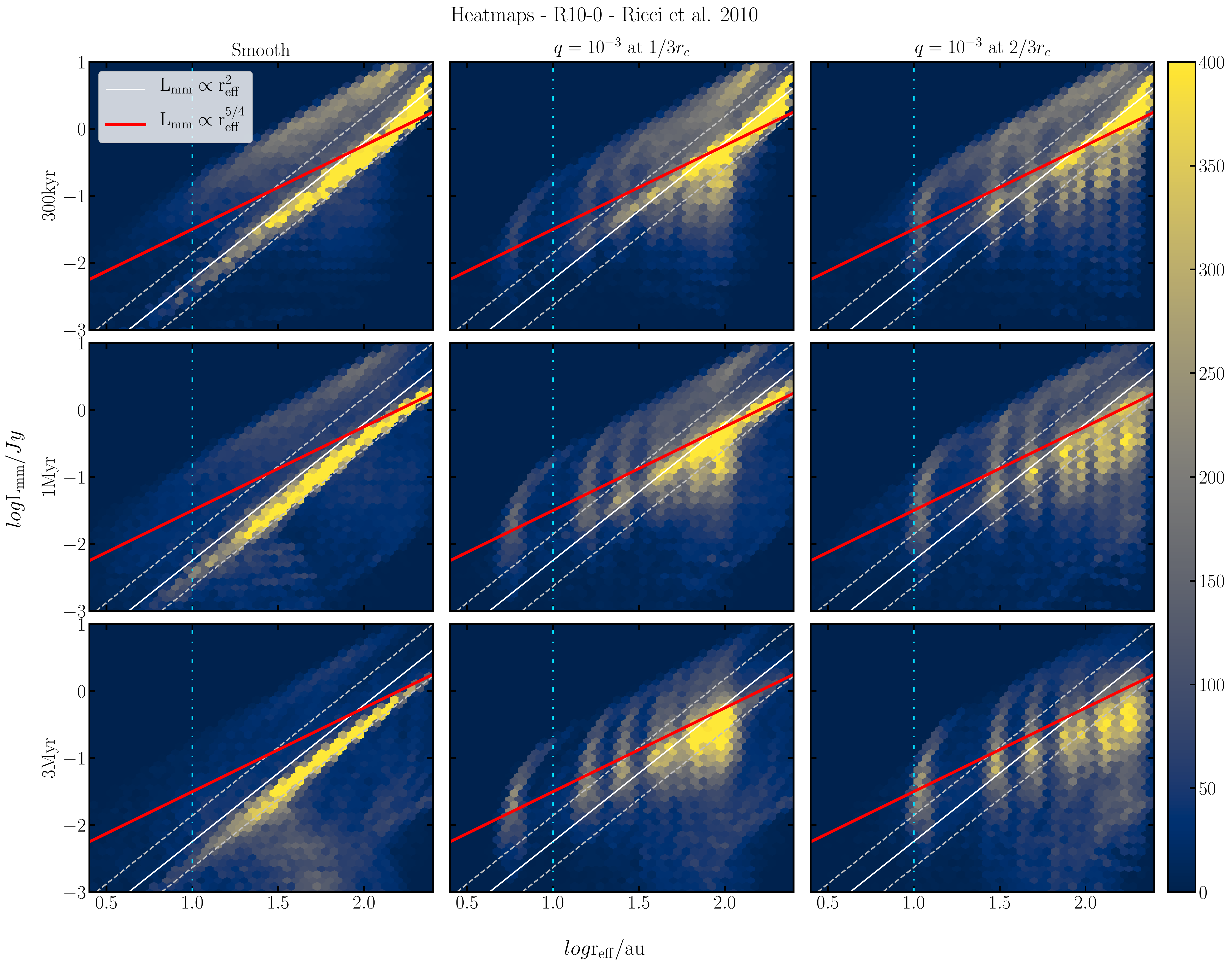}
    \caption[]{Heat maps of simulations with the R10-0 opacities. From left to right, the three  columns represent the smooth case, a  planet at $\mathrm{1/3r_{c}}$, and a planet at $\mathrm{2/3r_{c}}$. From top to bottom, the rows represent  three different snapshots at $\mathrm{300kyr}$, $\mathrm{1Myr,}$ and $\mathrm{3Myr}$. The white solid line is the SLR from \citet{Andrews2018a} and the red solid line our fit for the cases where a planet is included. The color bar shows the number of simulations in a single cell. The blue dash-dotted line shows the minimum limit ($\mathrm{r_{eff} \sim \SI{10}{au}}$) where observational results are available.}
    \label{fig:heatmaps_ricci}
\end{figure*}

In \autoref{fig:heatmaps_ricci} we plot three different cases. In    Col. 1 we show the smooth case, in the Col.
2   the case where we use a planet/star mass ratio $\mathrm{q=10^{-3}}$ at 1/3 of the \rc, and in the last column $\mathrm{q=10^{-3}}$ at 2/3 of the \rc. The snapshots are at three times as shown, as are the SLR and the standard deviation. The red line is our prediction for the cases where we include a planet (see \autoref{subsub:scaling_relation}). Instead of following the relation from \citet{Andrews2018a}, they seem to follow a relation of $\mathrm{L_{mm}\propto r_{eff}^{5/4}}$. In \autoref{subsub:width} and \autoref{subsub:scaling_relation} we perform a more detailed analysis on the this topic. 

%From top to bottom the rows  refer to the snapshots, \SI{300}{kyr}, \SI{1}{Myr,} and \SI{3}{Myr} . The solid white line refer to the SLR from \citet{Andrews2018a} and the dashed white lines show the $\mathrm{1\sigma}$ deviation (same as the blue shaded region in the single evolution tracks). The color bar to the right shows the number of simulations in a single cell. The red line is our prediction for the cases where we include a planet (see \autoref{subsub:scaling_relation}). Instead of following the relation from \citet{Andrews2018a}, they seem to follow a relation of $\mathrm{L_{mm}\propto r_{eff}^{5/4}}$. In \autoref{subsub:width} and \autoref{subsub:scaling_relation} we perform a more detailed analysis on the this topic. 

We observe that   most  of the disks start inside and above the correlation (first row at $\mathrm{\SI{300}{kyr}}$). In the smooth case the disks lose a great deal  of their luminosity relatively quickly and also shrink in size, making them move to the lower radii due to radial drift. We end up with a large number \textbf{($\rm{29.6\%}$)} of simulations occupying and following the SLR. As we explain in \autoref{subsub:scaling_relation}, the slope is expected from \citet{rosotti2019millimetre}, but the normalization depends on the choice of opacities. 

On the other hand, when we include a planet and the time increases the disks do not decrease in size, but mainly in luminosity. This is due to the formation of the pressure bump that keeps the dust from drifting farther in, thus keeping the same effective radius. The consequence is that if the disks leave the SLR, it is due to luminosity decrease since they move vertically in the diagram. The clustering in $\mathrm{r_{eff}}$ that is formed in the plots with a planet are an artifact of our parameter grid. A randomly chosen value of \rc and planet position would result in a continuous (non-clustered) distribution.
From the comparison of the two cases where a planet is included, there are more matching simulations when a planet is in the inner part of the disk \textbf{($\rm{30.2\%}$)}. Having a planet in the outer part leads the disks to the right (large radii) and bottom part of the diagram and consequently leaves them outside of the relation ($\rm{20.6\%}$ of the disks match).

\begin{figure*}
    \centering
    \includegraphics[width=\textwidth]{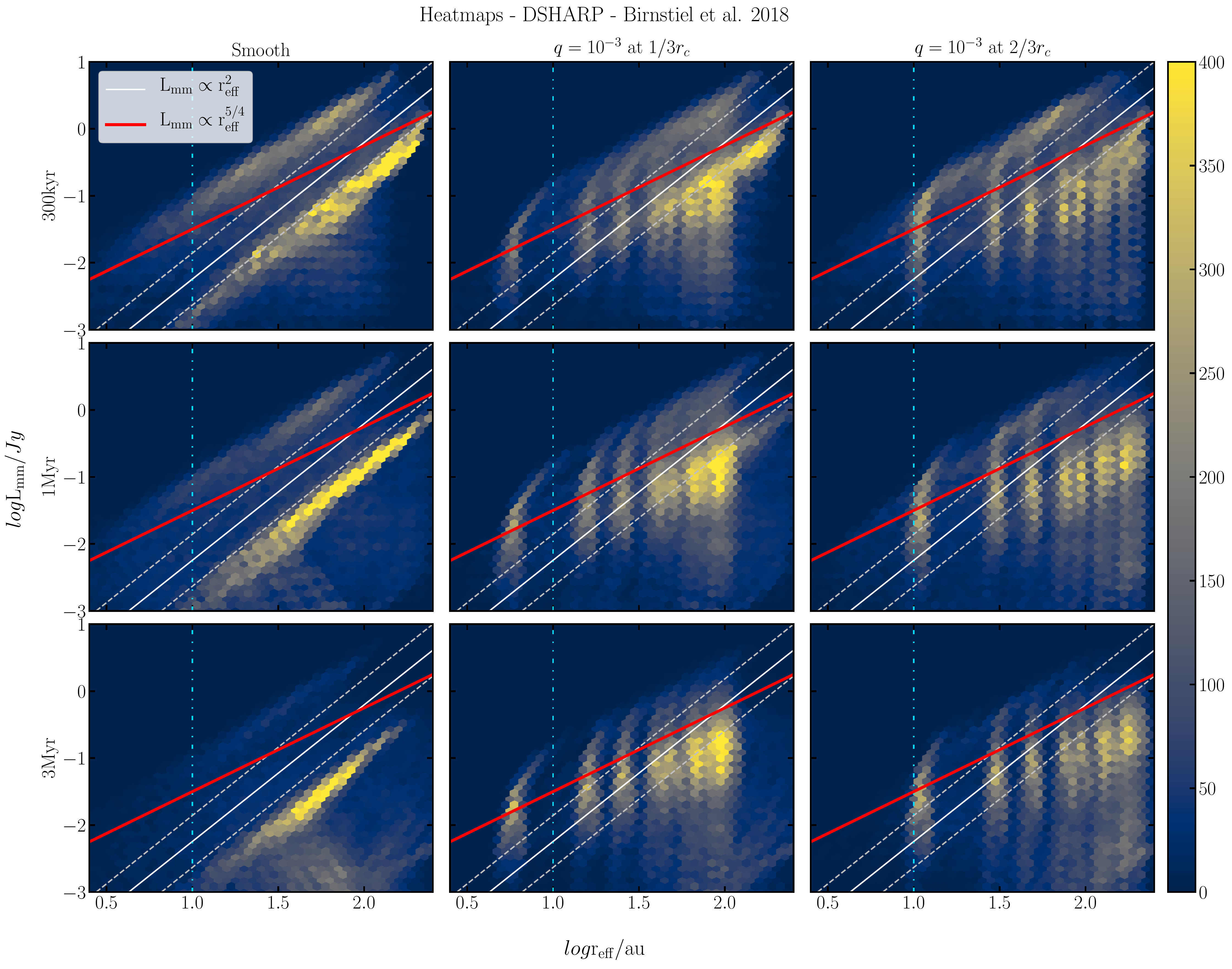}
    \caption[]{Heat maps of simulations with the \citet{Birnstiel_DSHARP} opacities. From left to
right, the three different columns represent the smooth case, a planet at $\mathrm{1/3r_{c}}$,  and a planet at $\mathrm{2/3r_{c}}$. From top to bottom, the rows represent three different snapshots at $\mathrm{300kyr}$, $\mathrm{1Myr}$, and $\mathrm{3Myr}$. The white solid line is the SLR from \citet{Andrews2018a} and the red solid line our fit for the cases where a planet is included. The color bar shows the number of simulations in a single cell. The blue dash-dotted line shows the minimum limit ($\mathrm{r_{eff} \sim \SI{10}{au}}$) where observational results are available.}
    \label{fig:heatmaps_dsharp}
\end{figure*}

The difference becomes striking when we use the dust opacities from DSHARP in \autoref{fig:heatmaps_dsharp}. Since the DSHARP opacity is lower than the R10-0 (see \autoref{fig:Opacities}), many of the smooth disks start below the SLR. This leads the majority of them outside of the relation by the last snapshot ($\rm{\SI{3}{Myr}}$) and only ($\rm{0.8\%}$) of the disks match. The same stands for the case where a planet is included. The disks have lower luminosity in the first snapshot, but the pressure bump formed is big enough to keep them in the relation for the remaining time ($>\rm{11.1\%}$) of the disks match depending on the model.

\begin{figure*}
    \centering
    \includegraphics[width=\textwidth]{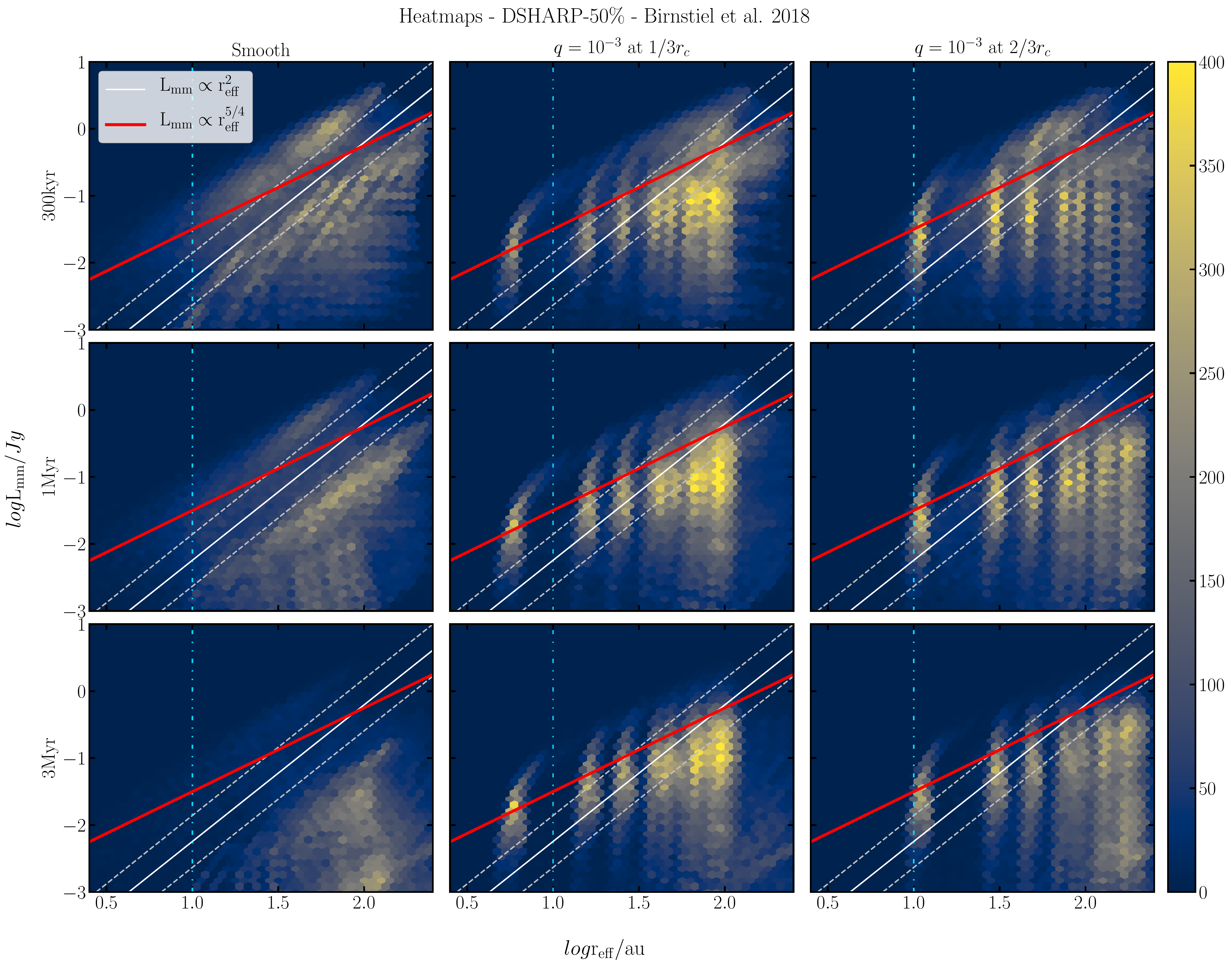}
    \caption[]{Heat maps of simulations with the \citet{Birnstiel_DSHARP} D-50 opacities with 50\% porostiy. From left to right, the three  columns represent the smooth case, a planet at $\mathrm{1/3r_{c}}$, and a planet at $\mathrm{2/3r_{c}}$. From top to bottom, the rows represent three different snapshots at $\mathrm{300kyr}$, $\mathrm{1Myr}$, and $\mathrm{3Myr}$. The white solid line is the SLR from \citet{Andrews2018a} and the red solid line our fit for the cases where a planet is included. The color bar shows the number of simulations in a single cell. The blue dash-dotted line shows the minimum limit ($\mathrm{r_{eff} \sim \SI{10}{au}}$) where observational results are available.}
    \label{fig:heatmaps_d50}
\end{figure*}

A similar behavior for the cases where we include strong substructures is seen for the opacity model with semi-porous grains, DSHARP-50, in \autoref{fig:heatmaps_d50}. Even though there are fewer matching simulations in total, if the substructures are large enough, they are able to keep a significant number of simulations in the clusters on the SLR \textbf{($>\rm{10.7\%}$)}. The same argument cannot be made for simulations with the smooth surface density profile. With semi-porous grains there is only a small fraction of matching simulations \textbf{($\rm{0.7\%}$)}. The absence of a strong opacity cliff in the opacity profile leads to low luminosities and consequently all the simulations below the SLR. 
An almost identical point is true looking at the heat map (\autoref{fig:heatmaps_d90}) for the case  where very porous grains are used (DSHARP-90). The complete absence of the opacity cliff does not allow a considerable fraction of smooth disks to enter the SLR \textbf{($\rm{1.3\%}$),} while a similar fraction of substructured disks match, as in the DSHARP-50 case \textbf{($>\rm{10.2\%}$)}. This heat map is included in \autoref{app:heatmap}.

From these heat maps we can extract three important results: 
\begin{itemize}
    \item Disks with strong traps (i.e., massive planets) follow a different SLR than smooth disks, while smooth disks are more consistent with the SLR in terms of the shape of the relation.
    
    \item Whether a smooth disk matches or not depends heavily on the opacity model. \citet{Birnstiel_DSHARP} DSHARP opacities produce significantly fewer simulations in the SLR than the \citet{Ricci_2010} R10-0 model and only a fraction of the simulations  match the  semi-porous and very porous grains and the model DSHARP-50 and DSHARP-90. Therefore, the porosity should be lower than $\mathrm{50}\%$ when the \citet{Birnstiel_DSHARP} opacities are used. However, the distribution of simulations is significantly tighter than the observed correlation for the smooth disks with the R10-0 opacity. As  discussed in \autoref{sec:discussion}, the observed correlation can be a mixture of smooth and substructured disks that adds scatter to the simulated SLR.
    
    \item A bright disk (top right on the SL diagram) is more likely to remain in the SLR if there is a pressure bump formed in the first $\mathrm{1Myr}$, regardless of the opacity model.
\end{itemize}

\subsubsection{Width of the pressure maxima}
\label{subsub:width}
To understand the overall shape of the heat map for the case with massive planets (i.e., the red lines in \autoref{fig:heatmaps_ricci} and \autoref{fig:heatmaps_dsharp}), we  derive a theoretical estimate in the following. This estimate depends on the width and position of the pressure maximum formed outside the position of the gap-opening planets. We therefore first derive a relation of the gas width versus radius $r$, planet/star mass ratio $q$, and scale height $h$, using hydrodynamical simulations of planet-disk interaction with the FARGO-3D code \citep{FARG03D_2015ascl.soft09006B}. In \autoref{subsub:scaling_relation} we estimate the SLR based on these empirically determined widths. For a complete derivation of the two sections, we refer the reader to \autoref{app:width_slr_derivation} and \autoref{app:scaling_relation}.

In addition to our two-pop-py models, we performed 24 hydrodynamical simulations with FARGO-3D \citep{FARG03D_2015ascl.soft09006B} for different planet/star mass ratios, planet locations, and $\mathrm{\alpha}$-values (see \autoref{tabel:gaps_param}). We used these simulations to calculate the width of the outer pressure bump caused by the planet in the gas. The surface density maximum is locally well fitted by a Gaussian, which allows us to measure the width (i.e., the standard deviation) using the curvature at the maximum. By measuring all the widths of our hydrodynamical simulations we fit as a multiple power law to search how they scale with radius, scale height, and the $\mathrm{\alpha}$-parameter,
\begin{equation}
    \sigma_{g} = C \cdot h^{p} \cdot q^{k} \cdot \alpha^{l}\,     
,\end{equation}
where $\mathrm{C}$ is a constant, $h$ is the scale height, $q$ the mass ratio, and $\mathrm{\alpha}$ the turbulence parameter. We find that the width in the measured range scales approximately as
\begin{equation}
\label{eq:power_law}
    \sigma_{g} \propto h^{0.81} \cdot q^{0.14} \cdot \alpha^{0.05}\,
.\end{equation}

\subsubsection{Size-luminosity relation of disks with companions}
\label{subsub:scaling_relation}

In \autoref{fig:heatmaps_ricci} we show a red line that scales differently from the SLR when we include planets. We predict that this line fits with a correlation $\mathrm{L_{mm}\propto r_{eff}^{5/4}}$. If we assume that all the luminosity of a disk comes from rings that are approximately optically thick, we can approximate as 
\begin{equation}
    L \simeq A \cdot B_{\nu}\,
,\end{equation}
where $\mathrm{A}$ is the area and $\mathrm{B_{\nu}}$ is the Planck function. If we assume the Rayleigh--Jeans approximation to approximate the Planck function with the temperature, the equation becomes
\begin{equation}
    L \simeq A \cdot T\,
.\end{equation}
We make the assumption that the area of the pressure bump scales as $\mathrm{A} \propto r \cdot \sigma_{d}$, where $\mathrm{r}$ is the radius and $\mathrm{\sigma_{d}}$ is the width of the pressure bump in the dust and there is linear scaling of $\mathrm{\sigma_{d}}$ with $\mathrm{h}$. The width of the dust ring depends on the width of the gas,
\begin{equation}
    \sigma_{d} \propto \sigma_{g} \cdot \sqrt{\frac{\alpha}{St}}\,
,\end{equation}
as in \citep{Dullemond2018DSHARP}, where $\mathrm{St}$ is the Stokes number, while in the ring there is also effective dust trapping and diffusion. Using the relation that previously measures the gas width in \autoref{subsub:width}, we find that the luminosity ($\mathrm{L_{mm}}$) scales with the radius as
\begin{equation} 
\label{eq:luminosity_hydro}
L_{mm} \propto r_{eff}^{5/4}\,
\end{equation}
which is the relation that we  plot with the red line in \autoref{fig:heatmaps_ricci}, \autoref{fig:heatmaps_dsharp}, \autoref{fig:heatmaps_d50}, and \autoref{fig:heatmaps_d90}. We find that this theoretical estimate nicely explains the size-luminosity scaling seen when strong substructure is present. However, toward larger radii, this relation slightly  overpredicts the luminosity. For example, we could get a shallower slope looking at the heat map because toward large radii our fitting line is above the main bulk of the simulations. For a complete derivation of the two sections, we refer the reader to \autoref{app:width_slr_derivation} and \autoref{app:scaling_relation}.

\section{Discussion}
\label{sec:discussion}

To summarize the results discussed above, we  explored the observed trend among the (sub-)mm disk continuum luminosity ($\mathrm{L_{mm}}$) and the $\mathrm{68\%}$ effective radius ($\mathrm{r_{eff}}$) of protoplanetary disks. Following the size-luminosity relation (SLR) obtained from \citet{tripathi2017millimeter} and \citet{Andrews2018a}, $\mathrm{L_{mm}\propto r_{eff}^{2}}$, we  showed which initial conditions are favorable for a disk to remain on the SLR for a time span of \SI{300}{kyr} - \SI{3}{Myr}. We explored the effect of every parameter on the disk evolution tracks on the SL Diagram, we got a visual representation of how the disk population moves on the same diagram, and we found relations between the parameters (\autoref{app:corner}). We present a different correlation for disks that are dominated by strong substructures compared to the disks that have a monotonically decreasing surface density profile. Moreover, we investigated the effect of different opacity models with compact or porous grains and we conclude that it is a major factor in reproducing the observational results. In the following sections we briefly recap these results and we discuss in detail some of the implications.

\subsection{Dominant parameters}
\label{disc:dominant_params}
In summary, our results imply that the most dominant parameters for the evolution of disks are the viscosity parameter $\mathrm{\alpha}$, the initial disk mass $\mathrm{M_{d}}$, the location of a giant planet if present, and the opacity model that is used to derive the continuum intensity. The disks that match the SLR are characterized by low turbulence ($\mathrm{\alpha} \leq 10^{-3}$) and high disk mass ($\mathrm{M_{d}} \geq 2.5\cdot10^{-2} M_{\star}$), and they are affected strongly by the existence of a strong trap (in this study caused by a giant planet). 

Turbulence-$\mathrm{\alpha}$ values greater than $\rm{10^{-3}}$ lead to smaller grains due to fragmentation, and consequently to less luminous disks that do not enter the SLR. Moreover, particles are diffused more efficiently and, due to their size, are less efficiently trapped. Finally, the dust trap is not as pronounced if the alpha viscosity is higher. All of this  acts in concert to make dust trapping ineffective, and causes the disks to behave as if they were smooth (see \autoref{fig:mass_acc_1e-3} in \autoref{sec:results}). If the fragmentation velocity is high enough, there can be simulations that stay on the SLR, but we consider these disks to have unrealistic initial conditions according to the known literature.

High initial disk mass locates the disks initially either inside or above the SLR (i.e., too bright for the given size) until they reach a quasi-steady state. This allows them to migrate to lower luminosities while they evolve up to $\SI{3}{Myr}$ and still remain in the SLR. This effect is aided by the right choice of opacity model and grain porosity. Compact grains shift the position of disks to higher luminosity (see \autoref{disc:opacity}). On the other hand, most of the less massive and smaller disks end up below the correlation at lower luminosities, characterizing them as discrepant.

Planets can alter the evolution path of the disk on the SL Diagram significantly. An effectively trapping planet causes the disk to quickly settle to a quasi-steady state on the SL Diagram, thereby leading to a shorter track and thus delaying the evolution of disks toward lower luminosity and radius. Disks with a massive planet in the inner part of the disk ($\mathrm{1/3 \rc}$) have more extended evolutionary tracks that are overall less luminous. In contrast, disks with a planet in the outer part ($\mathrm{2/3 \rc}$) have shorter and more luminous disks if all the other parameters remain the same. This can be explained by the planet trapping a large part of the disk solids at large radii. When two planets are included then both of them contribute to the luminosity while the outer one defines the effective radius of the disk. Overall, the presence of planets increases the fraction of matching simulations on the SLR, but this result is also a function of the opacity (see \autoref{disc:opacity}).

\subsection{Position along the SLR}

\begin{figure*}
    \centering
    \includegraphics[width=\textwidth]{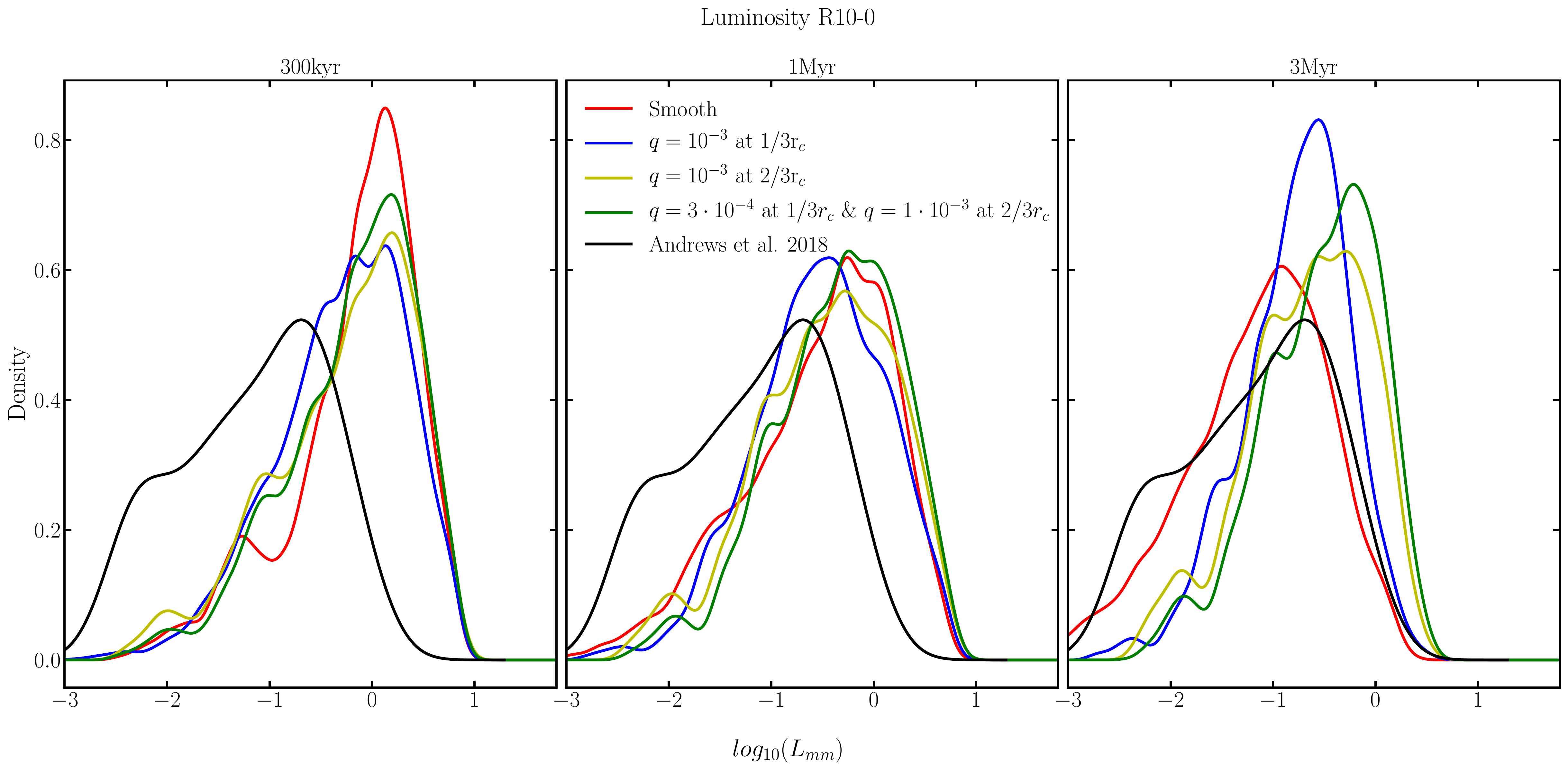}
    \caption[]{Kernel density distribution of the luminosity for all matching simulations using the \citet{Ricci_2010} R10-0 opacity model, for four different cases and three different snapshots, from $\mathrm{300kyr-3Myr}$. The black line refers to the disks from the \citet{Andrews2018a} sample. Disks with planets have higher luminosity, while smooth disks have low luminosity at $\mathrm{3Myr}$. When two planets are included the luminosity is higher than with a single planet.} 
    \label{fig:lum_R10}
\end{figure*}

The position of a disk along the SLR is determined mainly by the disk mass $\mathrm{M_{d}}$ and the disk size \rc, as  can be seen in \autoref{fig:param_effects} and \autoref{fig:parameter_tracks}. More massive and large disks are located in the top right part of the SL diagram, while small and less massive are in  the middle and left parts. In \autoref{fig:lum_R10} we show the kernel density estimate (kde) of the luminosity for all matching simulations for four different cases and three different snapshots, and also  plot the observational kde from \citet{Andrews2018a}. The brightest disks that stay on the SLR are those that contain planets that are located at the outer part of the disk (2/3 of the \rc, yellow and green lines). A planet in the outer region leads to larger,  more luminous disks as explained in \autoref{subsec:planets} and \autoref{disc:dominant_params}. Massive planets at 1 and \SI{3}{Myr} reproduce the peak of the observed brightness distribution, but overall produce too many bright and too few faint disks. The peak of the luminosity distribution for smooth disks is generally at much lower luminosities. Given these results, it is conceivable that the observed sample consists of two distinct categories of disks: a brighter and larger category due to massive outer planets that trap the dust while planets in the second category are not massive enough to trap the dust effectively and these disks evolve similarly to a smooth disk.

\subsection{Opacity model preferences}
\label{disc:opacity}
We used different opacity models for our study. As a base model, we made use of the composition of \citet{Ricci_2010} opacities (R10-0), but for compact grains (i.e., no porosity) as in \citet{rosotti2019millimetre}. Moreover we used the non-porous \citet{Birnstiel_DSHARP} opacities (DSHARP) and varied the porosity between $10\%$ DSHARP-10, $50\%$ DSHARP-50, and $90\%$ DSHARP-90. 

Therefore, we find that independent of the model used, relatively compact grains ($\mathrm{<50\%}$) are preferred instead of the highly porous grains. When compact grains are included the initial position of the disks on the SL Diagram is shifted toward higher luminosity giving it more time to evolve in the SLR on our chosen time span. Disks with the DSHARP opacity are generally less bright and end up below the SLR. We recall that the opacity at our wavelength is a factor of $\mathrm{\sim 8.5}$ higher in the R10-0 case compared to DSHARP at the opacity cliff location ($\mathrm{\sim 0.1-1mm}$), with the difference mainly stemming from the choice of carbonaceous material (\citealp{Zubko1996MNRAS.282.1321Z} versus \citealp{Henning1996}, see the comparison in \citealp{Birnstiel_DSHARP}). However, this point holds only for smooth disks and disks with weak substructures (where a disk behaves as smooth). If this is the case then only compact grains can explain the SLR; instead,  when substructures are strong then any of the opacity models and the porosities tested in this work can explain the substructured SLR. The latter applies because most of the substructures become optically thick.

It  can be argued that alternative compositions that also exhibit a strong opacity cliff and a high opacity would be equally suitable.

\subsection{Types of disks on the SLR}

We show that when strong traps (i.e., massive planets) are included, then disks follow a different SLR than the smooth ones. Based on measurements of the width of the pressure maxima formed by the planet in hydrodynamical simulations (see \autoref{app:width_slr_derivation}), we derived a theoretical prediction for disks with substructures (\autoref{subsub:scaling_relation}). Smooth disks with compact or slightly porous grains seem to follow the \citet{Andrews2018a} relation $\mathrm{L_{mm} \propto r_{eff}^{2}}$, while disks with massive planets follow a relation $\mathrm{L_{mm} \propto r_{eff}^{5/4}}$.

This result does not imply that the observed disks from the \citet{tripathi2017millimeter} sample are all free of substructure, but  they might not show strong enough and optically thick substructures, such  as  AS209 or HD163296 \citep[see][]{Huang2018}. \autoref{fig:heatmaps_ricci} shows how smooth disks follow the SLR, while disks with strong substructures follow a different relation but intersect the SLR at the bright end (top right part of the SLR). In contrast, the less luminous disks follow the SLR if they are smooth, but disks of the same effective size with substructure are too luminous (cf. bottom left part of the SLR).

In \citet[][Fig. 6]{Hendler2020ApJ...895..126H}, disks seem to follow a universal relation (close to the SLR) in all star-forming regions (Ophiuchus, Tau/Aur, Lupus, Chal) except for USco, the oldest region ($\mathrm{\sim 10Myr}$). The observed SLR therefore might flatten with age of the region. We could examine these results since we evolve our simulations to $\rm{10Myr}$, but our models do not include photo-evaporation and that would lead to uncertainties in the results. For example, at the age of USco the detectable disk fraction is $\rm{<20\%}$, while in our models it would be $\rm{100\%}$.

This raises a question. If a planet is not massive at early times, but around \SI{1}{Myr} has a planet/star mass ratio of $\mathrm{q=10^{-3}}$, will the disk follow the observed SLR or the SLR with strong substructures? According to the analysis of the evolution tracks in \autoref{sub:evolution_tracks}, most of the small and less luminous disks that do not initially have a giant planet drift toward lower radii and luminosities and even below the SLR. Therefore, strong substructures need to form in the first $\sim$\SI{1}{Myr} for the disk to follow the $\mathrm{L_{mm} \propto r_{eff}^{5/4}}$ relation. This result might imply that in most of the star-forming regions, strong substructures might not have formed early enough for small disks, or that the substructure is weak. On the other hand, bright and large disks can very clearly show strong substructures and follow the SLR at the same time. This is indeed the case for the DSHARP sample \citep{Andrews_DSHARP_2018ApJ...869L..41A}, which is biased toward bright disks and shows significant substructures in every source. The latter can be confirmed from \autoref{fig:lum_R10}   in the previous section. The brightest disks that stay on the SLR are those that contain planets that are located in the outer part of the disk (yellow and green lines).

The SLR can be explained if there is a mixture of both smooth and strong substructured disks. Smooths disks always follow the SLR, as shown  in \autoref{fig:heatmaps_ricci}, while the bright substructured disks populate the upper right part of the SLR (Figures \ref{fig:heatmaps_ricci}, \ref{fig:heatmaps_dsharp}, and \ref{fig:heatmaps_d90}). Disks with substructures that have large \rc and low disk mass $\mathrm{M_{d}}$ populate the lower right part of the plot under the SLR. These disks are not favored by \citet{Andrews2018a}, who find a tentative positive correlation between the mass of the star (or the disk) and the size of the disk. If massive small disks are excluded from the plot then the SLR could be reproduced by both substructured disks that occupy the upper right part and smooth (or weakly substructured) disks that occupy the lower left part of the SLR. Our results seem to be in agreement with the observational classification of \citet{VanDerMarel2021AJ....162...28V} who  suggest that all bright disks should have substructures formed by giant planets. 
Moreover, the SLR for the substructured disks is independent of the opacity model, but it slightly overpredicts the luminosity for the very large disks (see \autoref{disc:opacity} for more about the opacity).

\subsection{$\mathrm{L_{mm} - M_{\star}}$ relation}
\label{disc:Lmm-Mstar}
\begin{figure}
    \centering
    \includegraphics[width=\linewidth]{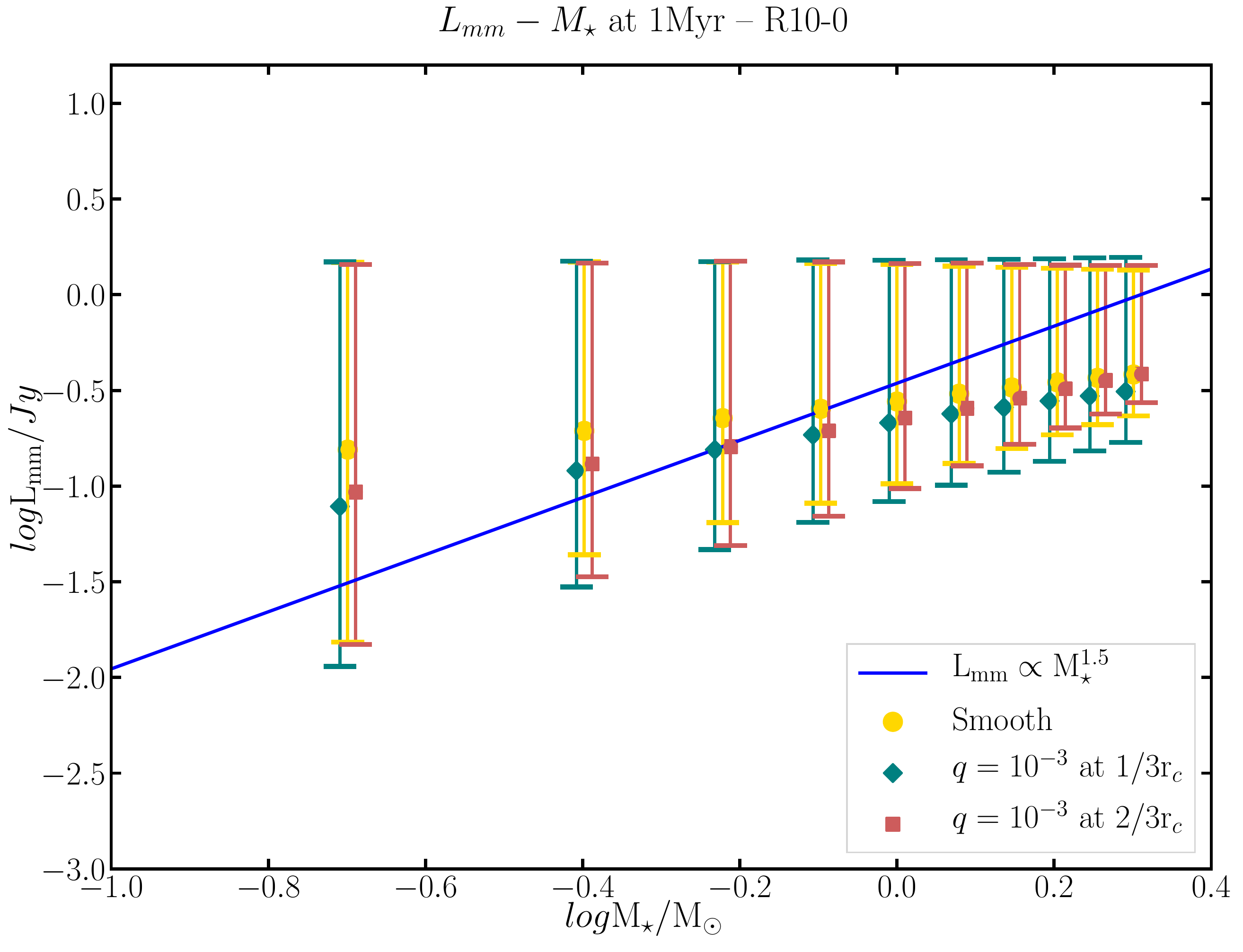}
    \caption[]{$\mathrm{L_{mm}} - M_{\star}$ relation at $\mathrm{1Myr}$ for three different cases for the matching simulations. Smooth case (yellow lines), a planet with planet/star mass ratio $\mathrm{q=10^{-3}}$ at $\mathrm{1/3\rc}$ (green lines) and a planet with the same ratio at $\mathrm{2/3\rc}$ (red lines). The points define the median value of the luminosity at $\mathrm{1Myr}$ and the error bars are the $\mathrm{75\%}$ percentile from the upper and lower value. The blue line is the $\mathrm{L_{mm}\propto M_{\star}^{1.5}}$ correlation from \citet{Andrews2018a}.} 
    \label{fig:L_Mstar}
\end{figure}

In \autoref{subsub:Mstar_tracks} we discuss  that the stellar mass ($\mathrm{M_{\star}}$) is directly correlated with the disks mass ($\mathrm{M_{d}}$) and that the disk temperature is only a weak function of  $\mathrm{L_{\star}}$ (and therefore $\mathrm{M_{\star}}$). The fact that the disk mass scales with the stellar mass implies that the luminosity ($\mathrm{L_{mm}}$) scales with the stellar mass. In \autoref{fig:L_Mstar} the $\mathrm{L_{mm}} - M_{\star}$ relation is shown for three different models for all matching simulations: the smooth case (yellow lines), a planet with planet/star mass ratio $\mathrm{q=10^{-3}}$ at $\mathrm{1/3\rc}$ (green lines), and a planet with the same mass ratio at $\mathrm{2/3\rc}$ (red lines). The markers define the median value of the luminosity at \SI{1}{Myr} and the error bars are the $\mathrm{75\%}$ percentile from the upper and lower value. The blue line is the $\mathrm{L_{mm}\propto M_{\star}^{1.5}}$ correlation from \citet{Andrews2018a}, a correlation that is consistent with those found from previous continuum surveys of comparable size and age \citep[][]{Andrews2013, Ansdell2016ApJ...828...46A, Pascucci2016ApJ...831..125P}. For any of our models the correlation is not as steep as for the \citet{Andrews2018a} model, but the cases with strong substructures have steeper profiles than the smooth cases. The reason is that no correlation between disk size and stellar mass was imposed in the parameter grid. If a size-mass correlation as inferred by \citet{Andrews2018a} were imposed, the mass-luminosity relation is expected to steepen as disks with optically thick substructures would be larger and therefore brighter. However, reproducing the observed mass-luminosity trend will be part of a future population synthesis study.
A similar manifestation of the same trend can be seen in \autoref{fig:corner_13rc}. In the panel $\mathrm{\rc-M_{d}}$, shown as  white dots, is the mean value of the characteristic radius for every disk mass. In order for the correlation to reproduce the observations, more massive disks should have initially been larger. In other words, large low luminosity disks would be expected in the lower right of the SL diagram, but are not observed. Preliminary results indicate that these disks need to have low disk mass $\rm{(M_{d}<10^{-2}M_{\star})}$ and be large in size $(\rm{\rc>\SI{150}{au}})$. Moreover the turbulence parameter should be relatively small $\alpha$ $\rm{\leq10^{-3}}$, otherwise the disk would behave as smooth and woul follow the SLR.

\subsection{Scattering}
Scattering is included in our simulations, as introduced in \autoref{sub:Observables}. Compared to the case where only the absorption opacity is used, the difference is minimal and it can be observed only in a few cases. With the inclusion of scattering, the originally brightest disks (above the SLR) tend to move toward lower luminosity (move down in the SL diagram). This happens for disks that are optically thick, hence for those that contain planets. This effect favors the SLR and allows slightly more disks ($\rm{\sim 2\%}$) to enter  the selected region. 
However, for moderately optically thick disks the emission is larger and a small fraction of disks move up (toward higher luminosity) on the SL diagram (Figure 4 in \citealt{Birnstiel_DSHARP}). This  happens because the derived intensity (\autoref{eq:intensity}) does not saturate to the Planck function, but to a slightly smaller value for a non-zero albedo. This is the well-known effect that scattering makes objects appear cooler than they are in reality.
On the other hand, for small optical depths ($\rm{\tau<<1}$) the effect of scattering is insignificant because the intensity (\autoref{eq:intensity}) approaches $\rm{I_{\nu}^{out}\longrightarrow \epsilon_{\nu}^{eff}B_{\nu}(T_{d})\Delta \tau /\mu}$, which is the identical solution to when $\rm{\kappa_{\nu}^{sca}}$ is set to zero while $\rm{\kappa_{\nu}^{abs}}$ is kept unchanged, as also shown  in \citet{Birnstiel_DSHARP}.

The effect of scattering also depends  on the albedo ($\rm{\eta_{\nu}=1-\epsilon_{\nu}^{eff}}$). For the compositions we use, the maximum effective albedo is $\rm{0.57}$ for R10-0 and $\rm{0.82}$ for DSHARP, while it can reach   $\rm{\sim 0.97}$ for DSHARP-90. For these compositions the effect of scattering is never more than a factor of $\rm{\sim 1.7}$ at a particle size of $\rm{1mm}$. The result of scattering is higher if  we increase the  albedo, but by using a plausible composition the result is effectively negligible.

However, we obtain different results compared to \citet{Zhu2019ApJ...877L..18Z}. In that work the authors  discusse that a completely optically thick disk with high albedo ($\rm{0.9}$) can be constructed, which therefore lies along the SLR with the right normalization (because the high albedo and high optical depth lowers the luminosity). However our findings show that we cannot reach those results from an evolutionary perspective. For smooth disks the dust drifts to the inner part of the disk and the disk is no longer optically thick. On the other hand, disks with substructures create only rings rather than disks that are completely optically thick everywhere.

\subsection{Predictions for longer wavelengths}
A recent study \citep{Tazzari2021MNRAS.506.2804T}, showed a flatter SLR at $\rm{3.1mm}$ ($\rm{Lmm \propto r_{eff}^{1.2}}$), confirming that emission at longer wavelengths becomes increasingly optically thin.

We performed a series of simulations at $\rm{\SI{3.1}{mm}}$ for a comparison  with these results. The disks are fainter and smaller at $\rm{\SI{3.1}{mm}}$. Two effects contribute to this. First, the value of the opacity decreases and the opacity cliff moves to larger particle sizes. This leads the disks to become optically thinner in comparison to the $\rm{\SI{850}{\mu m}}$ case. Second, the intensity at $\rm{\SI{3.1}{mm}}$ is less according to  Plank's spectrum. Therefore, the luminosity will be lower.

In terms of the SLR, the slope for the smooth disks does not change since these disks are never optically thick; therefore, all disks simply move toward lower luminosities and smaller radii.
On the other hand, the substructured disks that cover the SLR do not change in terms of slope, but the large and faint disks (right part of the heat map in \autoref{fig:heatmaps_ricci}) show a larger spread in luminosity compared to the smaller wavelength. Disks that are very optically thick and moderately optically thick have the same luminosity at $\rm{\lambda=850\mu m}$, but at $\rm{\SI{3.1}{mm}}$ because of the decrease in opacity the former category is still optically thick while the latter no longer is, leading to a decrease in luminosity. Therefore the SLR can become flatter if we can take into account these disks that do not belong in the SLR.

With our models the flatter relation from \citet{Tazzari2021MNRAS.506.2804T}, could be explained by substructured and large smooth disks. The heat map in \autoref{app:heatmap} confirms this. In this figure we plot the simulations at $\rm{\SI{3.1}{mm}}$, using the R10-0 opacity model and we overplot the SLR from \citet{Tazzari2021MNRAS.506.2804T}: $\rm{L_{mm} \propto r_{eff}^{1.2}}$. Substructured disks can  explain this relation very well since it is similar to the scaling relation we calculated for disks with strong substructures in \autoref{subsub:scaling_relation}. Small and smooth disks, on the other hand, cannot enter the relation because they are too faint since the particles cannot grow to a size where the opacity cliff is at $\rm{\SI{3.1}{mm}}$.

We should  mention the possibility that the flatter relation can be due to observational bias toward large disks, which tend to be substructured. If small and faint disks are included in the sample, the observed SLR could be steeper and closer to the SLR from \citet{Andrews2018a}. Future observational surveys should  investigate this possibility further.

\subsection{Limitations}
\label{disc:limitations}
\begin{figure*}
    \centering
    \includegraphics[width=\textwidth]{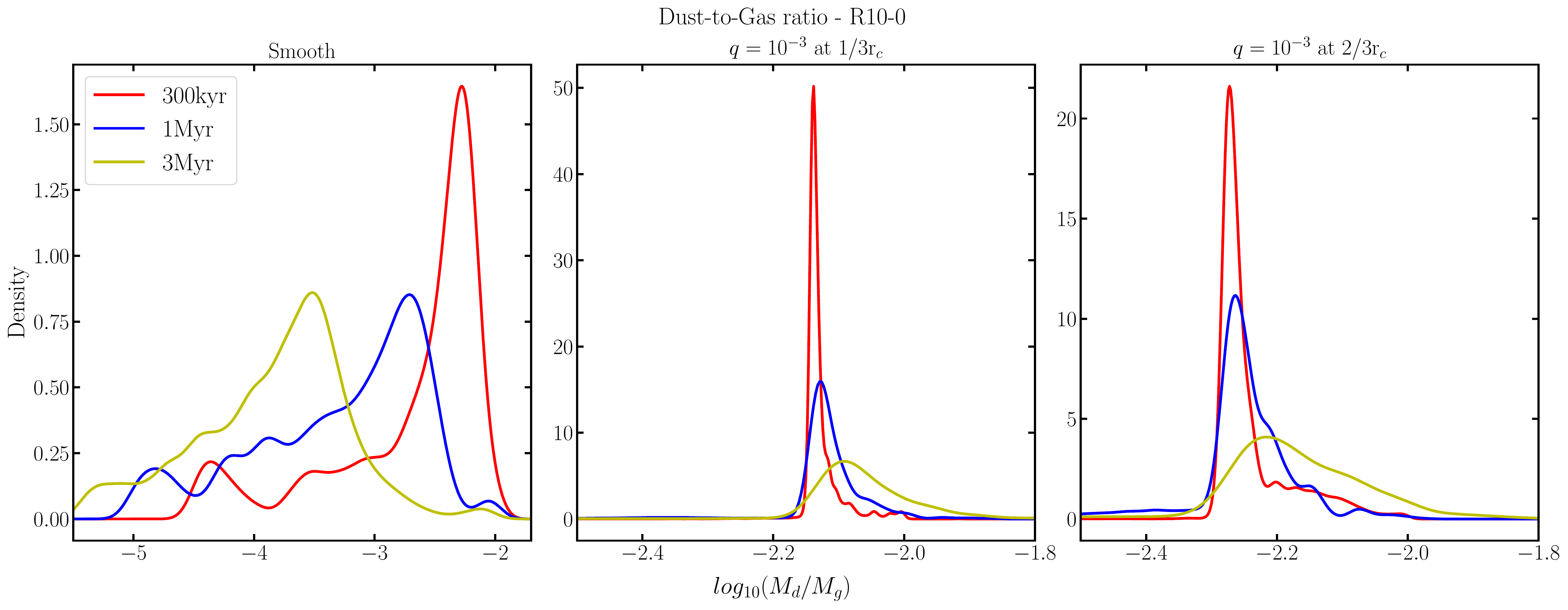}
    \caption[]{Evolution of the global disk dust-to-gas ratio of all matching simulations with the \citet{Ricci_2010} R10-0 opacity model, for three different cases and three different snapshots, from $\mathrm{\SI{300}{kyr}-\SI{3}{Myr}}$. From left to right,  the smooth case, a planet with planet/star mass ratio at $\mathrm{1/3\rc}$, and a planet with the same ratio at $\mathrm{2/3\rc}$. Different limits are used on the x-axis to highlight the evolution of the dust-to-gas ratio. The initial dust-to-gas ratio is $\mathrm{0.01}$.
    For the smooth case the dust-to-gas ratio  decreases by three orders of magnitude up to $\mathrm{\SI{3}{Myr}}$. When a planet is included the disk dust mass is retained and leads to a much higher dust-to-gas ratio. In the case where the planet is the inner part of the disk (middle column), there are cases at $\mathrm{\SI{3}{Myr}}$ where the ratio is higher than $\mathrm{0.01}$. The gas mass moves faster than the dust mass in this case.} 
    \label{fig:d2g_R10}
\end{figure*}

It is important to keep in mind the limitations of this paper. The time span used for the simulations displayed in this paper and the figures is from $\mathrm{\SI{300}{kyr}}$ to $\mathrm{\SI{3}{Myr}}$. This does not exclude the possibility that some disks with high disk mass might evolve a great deal on the SLR diagram for $\mathrm{\SI{10}{Myr}}$-$\mathrm{\SI{20}{Myr}}$. Disk dissipation has not been modeled in this paper, but it will be considered for future work. In \autoref{fig:d2g_R10} we show the kde of the global dust-to-gas ratio\footnote{The dust-to-gas ratio in the disk changes with radius and time and this quantity is simply $\rm{M_{dust}/M_{gas}}$.} for three different snapshots between $\mathrm{\SI{300}{kyr}-\SI{1}{Myr}-\SI{3}{Myr}}$ and for three different cases. Smooth disks lose dust relatively quickly due to radial drift, while disks with planets retain a much higher dust-to-gas ratio because of the the strong trap. In the second panel there are cases where the dust-to-gas ratio increases over the initial $\mathrm{0.01}$. These are substructured disks with intermediate $\mathrm{\alpha}$-values, high fragmentation velocities ($\mathrm{>\SI{1000}{cm/s}}$), and small sizes ($\mathrm{<\SI{60}{au}}$). The gas is removed more quickly than the dust, leading to a higher dust-to-gas ratio. Large values of  $\mathrm{\alpha}$ would lead to less trapping and the dust would drift as usual, while low $\mathrm{\alpha}$ would mean that the disk does not evolve significantly.

As mentioned in \autoref{sec:methods}, the stellar luminosity is not evolved in the simulation. If this were the case the disk luminosity would scale approximately  linearly with the stellar luminosity and would be further modulated by resulting changes in the dust evolution. We therefore expect a general shift of the disks toward lower luminosities, but with the trends that have explored in \autoref{sec:results} remaining the same. Since most of the simulations need to be brighter to remain in the SLR, a change in the luminosity favors higher $\rm{\alpha}$-values and lower fragmentation velocities than the values shown before. An example of a heat map in shown in \autoref{app:stellar_luminosity}.

In our models the planets are already included at the beginning of the simulations and they open a gap in the initially smooth surface density profile relatively quickly. Realistically, the timescales in the outer part of the disk are much longer and the timescale for planet formation changes with the distance to the star \citep{Johansen2017AREPS..45..359J}. Therefore, we would expect that the inner planet should form first and the outer planet later, as  has been suggested (e.g., \citealt{pinilla2015A&A...580A.105P}). Since both planets start at the same time, the inner one might trap more of the total disk mass and the outer bump might be less bright than in our models in reality. The latter will be included in a future work by including the outer planet later in the simulation.

\section{Conclusions}
\label{sec:conclusions}
In this paper we  performed a large population study of 1D models of gas and dust evolution in protoplanetary disks to study how the effective radius and disk continuum emission evolves with time. We varied a range of initial parameters and we included both smooth disks and disks that contain planets. We compared our results with the observed trend between continuum sizes and luminosities from \citet{Andrews2018a} and we managed to constrain the initial conditions. Our findings are as follows:

\begin{enumerate}
    
    \item Disks with strong traps (i.e., massive planets) follow a different SLR than smooth disks. Smooth disks follow the \citet{Andrews2018a} relation, $\mathrm{L_{mm} \propto r_{eff}^{2}}$, as shown by \citet{rosotti2019millimetre}, while disks with massive planets follow $\mathrm{L_{mm} \propto r_{eff}^{5/4}}$. This could mean that not all disks in the \citet{tripathi2017millimeter} and \citet{Andrews2018a} joint sample have a substructure as significant as  HD163296, for example. We explained this result with a simple analytical derivation and we find that if the gas width scales as we measured it from FARGO-3D and if the dust width scales as we expect it from trapping and fragmentation, then theoretically the luminosity scales as $\mathrm{L_{mm} \propto r_{eff}^{5/4}}$.
    
    \item Whether disks   follow the SLR   depends heavily on the opacity model. When the DSHARP \citep{Birnstiel_DSHARP} opacity is used, disks are not as luminous in the first $\rm{\SI{300}{kyr}}$ and the majority of them end up below the SLR. Especially for smooth disks, the DSHARP opacities produce a much lower number of simulations on the SLR compared to models using the \citet{Ricci_2010} R10-0 opacities ($\rm{0.8\%}$ with DSHARP and $\rm{29.6\%}$ with R10-0). Therefore, with this opacity model, only disks with substructures can populate the SLR. On the other hand, R10-0 opacities can reproduce  disks both  with and without substructures since the absolute value of the opacity at $\SI{850}{\mu m}$ is $\rm{\sim 8.5}$ times higher than DSHARP for particle sizes around $\mathrm{\sim \SI{0.1}{mm}}$ (position of the opacity cliff) and the disks become luminous enough to enter the relation.

    \item The SLR is more widely populated when substructures are included in contrast to a tight correlation for smooth disks. Substructured disks cover mostly the upper right part (large and bright disks) of the SL diagram, while the lower left (small and faint) is covered by smooth disks. This is an indication that the SLR can be explained if there is a mixture of both smooth and strong substructured disks.
    
    \item The grain porosity can drastically affect the evolution track of the disk. Throughout our models, relatively compact grains ($\mathrm{<50\%}$ porosity) are preferred for simulations that follow the SLR. If we use slightly porous grains ($\mathrm{10\%}$) by altering the DSHARP opacity, the effect is insignificant, as the shape of the opacity cliff is roughly the same. On the contrary, for semi-porous ($\mathrm{50\%}$) and porous grains ($\mathrm{90\%}$) the opacity cliff flattens out, leading to disks with low luminosity. Only compact grains can explain the SLR for smooth disks, while any porosity can explain it when strong substructures are included.
    
    \item High initial disk mass gives a higher probability for a simulation to follow the SLR. If this applies, the disk starts initially above the SLR (bright) until it reaches a stable state at around $\mathrm{\sim \SI{300}{kyr}}$. By this time it will enter the relation, and depending on the other initial conditions it will either remain there and  be considered  a matching simulation or will leave it.
    
    \item There is a preference toward low $\mathrm{\alpha}$-values (lower than $\mathrm{10^{-3}}$). This result is in line with other more direct methods of determining $\mathrm{\alpha}$ (e.g., \citealt{flaherty_2018ApJ...856..117F}).
    There are multiple reasons for this tendency. For $\mathrm{\alpha}\geq 2.5\cdot 10^{-3}$, disks tend to be more fragmentation dominated, the particle size decreases,  and consequently they are not trapped by the pressure bump (if any) leading them outside the relation. Moreover the diffusivity increases and the peak of the pressure bump smears out, leading to inefficient trapping. On the other hand, if $\mathrm{\alpha}$ is low, the ring that is formed  becomes too narrow and disks tend to have lower luminosity.
    
    \item The location of the planet as a function of the characteristic radius plays a major role in the final outcome. If a planet is included in the inner part of the disk ($\mathrm{1/3 r_{c}}$), the disk has to be significantly larger in order to retain the correct ratio of luminosity to effective radius to stay in the SLR. Instead, when an outer planet ($\mathrm{2/3 r_{c}}$) is included the disk tends to be smaller in size. When two planets are included, the location of the outermost one defines the size of the disk, but a combination of the two defines the luminosity. These results are also affected by the opacity model.
    
    \item We expect a less extended evolution track when substructure is included. The pressure bump halts the dust from drifting farther in, constraining this way the size of the disk and not allowing it to evolve further on the SLR. Furthermore, when two planets are included there is an indication that the inner planet should form first, otherwise there will not be a big enough reservoir of material in order to form.
    
    \item We are not able to construct optically thick disks with high albedo ($\rm{0.9}$) that lie along the SLR with an evolutionary procedure, as opposed to \citet{Zhu2019ApJ...877L..18Z}. Smooth disks are not optically thick due to radial drift, while disks with substructure create only optically thick rings rather than a uniform optically thick distribution.
    
    \item We chose different gap profiles based on \citet{Kanagawa2016} and compared them again to hydrodynamical simulations. We conclude that the depth of the gap does not play an important role to the evolution of the disk on the SLR as long as the planet is big enough to stop the particles from drifting. Instead, the width of the gap is the important parameter  (see \autoref{app:gap_profiles}, where we compare the different profiles for different parameters.)
    
\end{enumerate}

This study shows how the combination of observed and simulated populations allows us to put constraints on crucial unknowns such as the disk turbulence, or the dust opacity. Future work is required to also investigate the effects of disk build-up and dissipation as well as planet migration and planetesimal formation.

\begin{acknowledgements}
 T.B. acknowledges funding from the European Research Council (ERC) under the
 European Union's Horizon 2020 research and innovation programme under grant
 agreement No 714769 and funding by the Deutsche Forschungsgemeinschaft (DFG, German Research Foundation) - 325594231 under Ref no. FOR 2634/1. Furthermore, this research was supported by the Excellence Cluster ORIGINS which is funded by the Deutsche Forschungsgemeinschaft (DFG, German Research Foundation) under Germany's Excelence Strategy - EXC-2094-390783311. G.R. acknowledges support from the Netherlands Organisation for Scientific Research (NWO, program number 016.Veni.192.233) and from an STFC Ernest Rutherford Fellowship (grant number ST/T003855/1)
\end{acknowledgements}

\bibliographystyle{aa}
\bibliography{bibliography}

\begin{appendix}
\section{Derivation of gap width and SLR}
In this section we show the complete derivation of the width versus radius $r$, planet/star mass ratio $q$, and scale  height $h$ relation, as explained in \autoref{subsub:width}, and the derivation of the SLR of disks with strong substructures, as shown in \autoref{subsub:scaling_relation}.

\subsection{Derivation of the gap width}
\label{app:width_slr_derivation}
In addition to our two-pop-py models, we performed 24 hydrodynamical simulations with FARGO-3D \citep{FARG03D_2015ascl.soft09006B} for different planet/star mass ratios, planet locations, and $\mathrm{\alpha}$-values (see \autoref{tabel:gaps_param} in \autoref{app:gap_profiles}). We used these simulations to calculate the width of the outer pressure bump caused by the planet in the gas. The surface density maximum is locally well fitted by a Gaussian, which allows us to measure the width (i.e., the standard deviation) using the curvature at the maximum,

\begin{equation}
    y(x) = \frac{1}{\sqrt{2\pi}\sigma_{g}}e^{-\frac{(x-x_{0})^{2}}{2\sigma^{2}}}\,
,\end{equation}
where $\mathrm{x_{0}}$ is the location of the surface density maximum. We took the logarithmic derivative of the Gaussian,
\begin{equation}
    \frac{\partial lny}{\partial lnx} = \frac{x}{y}\frac{1}{\sqrt{2\pi}\sigma_{g}}\left(-\frac{x-x_{0}}{\sigma^{2}}\right)e^{-\frac{(x-x_{0})^{2}}{2\sigma_{g}^{2}}}\,
,\end{equation}
which gives us
\begin{equation}
    \frac{\partial lny}{\partial lnx} = -x\frac{x-x_{0}}{\sigma_{g}^{2}}
.\end{equation}
If we are close to the pressure maximum, we can approximate that $\mathrm{x\xrightarrow{}x_{0}+\delta x}$, and the equation yields
\begin{equation}
    \frac{dlny}{dlnx} = -\frac{x_{0}\delta x}{\sigma_{g}^{2}}
.\end{equation}
If we take the normal derivative of the above expression, we end up with the following expression: 
\begin{equation}
   \frac{\partial}{\partial \delta x} \frac{\partial lny}{\partial lnx} = -\frac{x_{0}}{\sigma_{g}^{2}}
.\end{equation}
Solving for $\mathrm{\sigma_{g}}$, we measure the width of the maximum by calculating this derivative at the peak location for the azimuthally averaged surface density profile derived from the FARGO3D runs:
\begin{equation}
   \sigma = \sqrt{-\frac{x_{0}}{\frac{\partial}{\partial \delta x} \frac{\partial lny}{\partial lnx}}}
.\end{equation}
We measured all the  widths of our hydrodynamical simulations, and we investigated how  they scale with radius, scale height, and the $\mathrm{\alpha}$-parameter. We fit the width of the pressure bump as multiple power-law as
\begin{equation}
    \sigma_{g} = C \cdot h^{p} \cdot q^{k} \cdot q^{l}\,     
,\end{equation}
where $\mathrm{C}$ is a constant, $h$ is the scale height, $q$ the mass ratio, and $\mathrm{\alpha}$ the turbulence parameter. We find that the width in the measured range scales approximately as
\begin{equation}
\label{eq:app_power_law}
    \sigma_{g} \propto h^{0.81} \cdot q^{0.14} \cdot \alpha^{0.05}\,     
.\end{equation}

\subsection{Derivation of SLR of disks with companions}
\label{app:scaling_relation}
To understand the overall shape of the heat map for the case with strong substructures (i.e., the red lines in \autoref{fig:heatmaps_ricci} and \autoref{fig:heatmaps_dsharp}), we present here the complete derivation of the theoretical estimate from \autoref{subsub:scaling_relation}. This estimate is based on the empirically determined widths (see  \autoref{app:width_slr_derivation}), as explained in \autoref{subsub:width}, and the position of the pressure maximum formed outside the position of the gap-opening planets.

Assuming that all the luminosity of a disk comes from rings that are approximately optically thick, we can approximate
\begin{equation}
    L \simeq A \cdot B_{\nu}\,
,\end{equation}
where $\mathrm{A}$ is the area and $\mathrm{B_{\nu}}$ is the Planck function. If we assume the Rayleigh--Jeans approximation, then we can substitute the Planck function with the temperature $\mathrm{T}$ and the equation becomes
\begin{equation}
    L \propto A \cdot T
.\end{equation}
We make the assumption that the area of the pressure bump scales as $\mathrm{A} \propto r \cdot \sigma_{d}$, where $\mathrm{r}$ is the radius and $\mathrm{\sigma_{d}}$ is the width of the pressure bump in the dust and there is linear scaling of $\mathrm{\sigma_{d}}$ with $\mathrm{h}$. The temperature is $\mathrm{T \propto r^{-1/2}}$ and $\mathrm{h \propto r^{5/4}}$. Combining the above relations, we obtain the theoretical expression for the luminosity:
\begin{equation}
\label{eq:app_luminosity_theory}
    L_{mm} \propto r^{7/4}
.\end{equation}

In the previous section we determined how the width of the pressure maximum depends on the scale height $\mathrm{\sigma_{g} \propto h^{0.81}}$, so we need to find a relation between the $\sigma_{g}$ and $\sigma_{d}$. In the pressure maximum, we assume that the disk is optically thick and that it is fragmentation-limited since there is no radial drift. If the latter applies, then
\begin{equation}
    \sigma_{d} \propto \sigma_{g} \cdot \sqrt{\frac{\alpha}{St}}\,
,\end{equation}
as in \citep{Dullemond2018DSHARP}, where $\mathrm{St}$ is the Stokes number. Moreover,
\begin{equation}
    St \propto \frac{1}{\alpha \cdot c_{s}^{2}}\,
,\end{equation}
where $\mathrm{c_{s}}$ is the sound speed. Combining these two, we obtain that
\begin{equation}
    h_{d} \propto h_{g}\cdot \alpha \cdot c_{s}
.\end{equation}
The sound speed is
\begin{equation}
    c_{s} \propto  r^{-1/4}
.\end{equation}
Using all the above equations we conclude that
\begin{equation}
\label{eq:app_width_r}
    \sigma_{d} \propto \sigma_{g} \cdot r^{-1/4}
,\end{equation}
or, if we express the radius as a function of the scale height,
\begin{equation}
\label{eq:app_width_h}
    \sigma_{d} \propto \sigma_{g} \cdot h^{-1/5}
.\end{equation}
The scale height is $\mathrm{h=c_{s}/\Omega}$, where $\mathrm{\Omega=\sqrt{GM_{\star}/r^{3}}}$ the Keplerian angular velocity, leading to a dependency $\mathrm{h \propto r^{5/4}}$. Therefore, we can use \autoref{eq:app_power_law} and express the calculated $\mathrm{\sigma_{g}}$ as a function of radius: 
\begin{equation}
    \sigma_{g} \propto r^{1.01}
.\end{equation}
Finally, we can calculate the dust width $\mathrm{\sigma_{d}}$ as a function of both the scale height and radius. Substituting into \autoref{eq:app_width_h} we  obtain 
\begin{equation}
    \sigma_{d} \propto  h^{0.61}
,\end{equation}
and into \autoref{eq:app_width_r} we obtain
\begin{equation}
    \sigma_{d} \propto  r^{0.75}
.\end{equation}
If we calculate again how the luminosity scales and replace the scaling of the dust width $\mathrm{\sigma_{d}} \propto r^{0.75}$, we find that
\begin{equation}
\label{eq:app_luminosity_hydro}
    L_{mm} \propto r^{5/4}
,\end{equation}
which is the value that we  fit with the red line in \autoref{fig:heatmaps_ricci}. 

\FloatBarrier

\section{Corner plots}
\label{app:corner}
To search for further correlations between the parameters we visualized the results similar to a corner plot; however, for a grid of parameters instead of the usual sample density. The corner plots are projections of a multi-dimensional parameter space to several two-dimensional planes. In these visualizations every box of the plot is a 2D histogram of a parameter pair, while the color varies according to the number of simulations in every cell. The color bar is given in the upper part of the plot. These visualizations contain only the simulations that match the SLR as the histograms in \autoref{fig:Histograms}.

\begin{figure*}
    \centering
    \includegraphics[width=\linewidth]{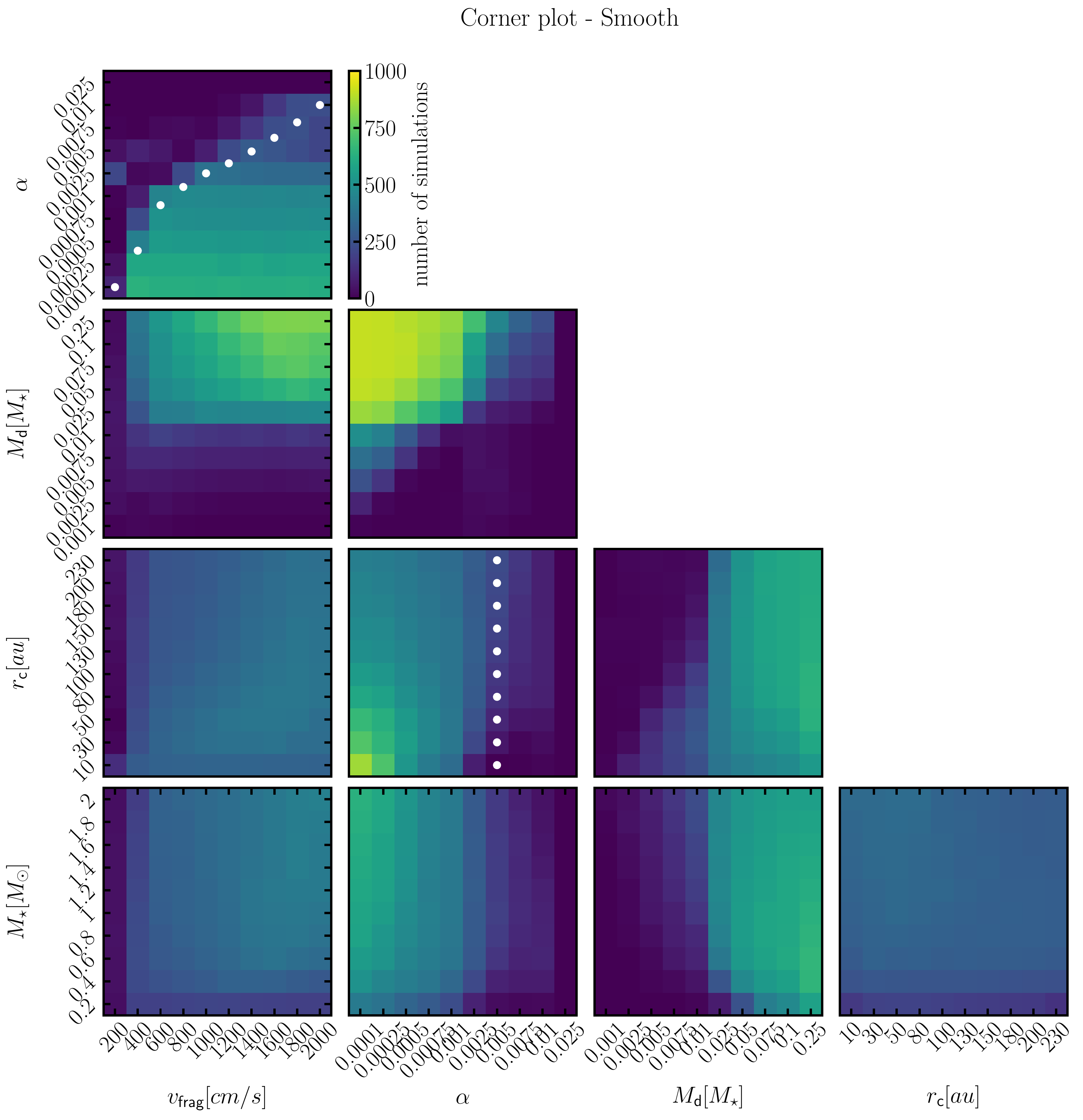}
    \caption[]{Corner plot of the smooth case. From left to right: Fragmentation velocity $\mathrm{v_{frag}}[cm/s]$, turbulence $\mathrm{\alpha}$-value, disk mass $\mathrm{M_{d}}[M_{\star}]$, characteristic radius $\mathrm{r_{c}}[au]$.  
    From top to bottom: $\mathrm{\alpha}$, $\mathrm{M_{d}}[M_{\star}]$, $\mathrm{r_{c}}[au]$, $\mathrm{M_{\star}}[M_{\odot}]$. In the upper left plot the white dots show the track where the distinction between the fragmentation- and the drift-limited regime is. In the middle plot ($\mathrm{\rc-\alpha}$) the white dots show the $\mathrm{\alpha}$-value where the pressure bump cannot hold the dust efficiently anymore. In the panel \rc - $\mathrm{M_{d}}$,  the mean value of the characteristic radius for each  disk mass is shown as a  white
dot.}
    \label{fig:corner_smooth}
\end{figure*}
\begin{figure*}
    \centering
    \includegraphics[width=\linewidth]{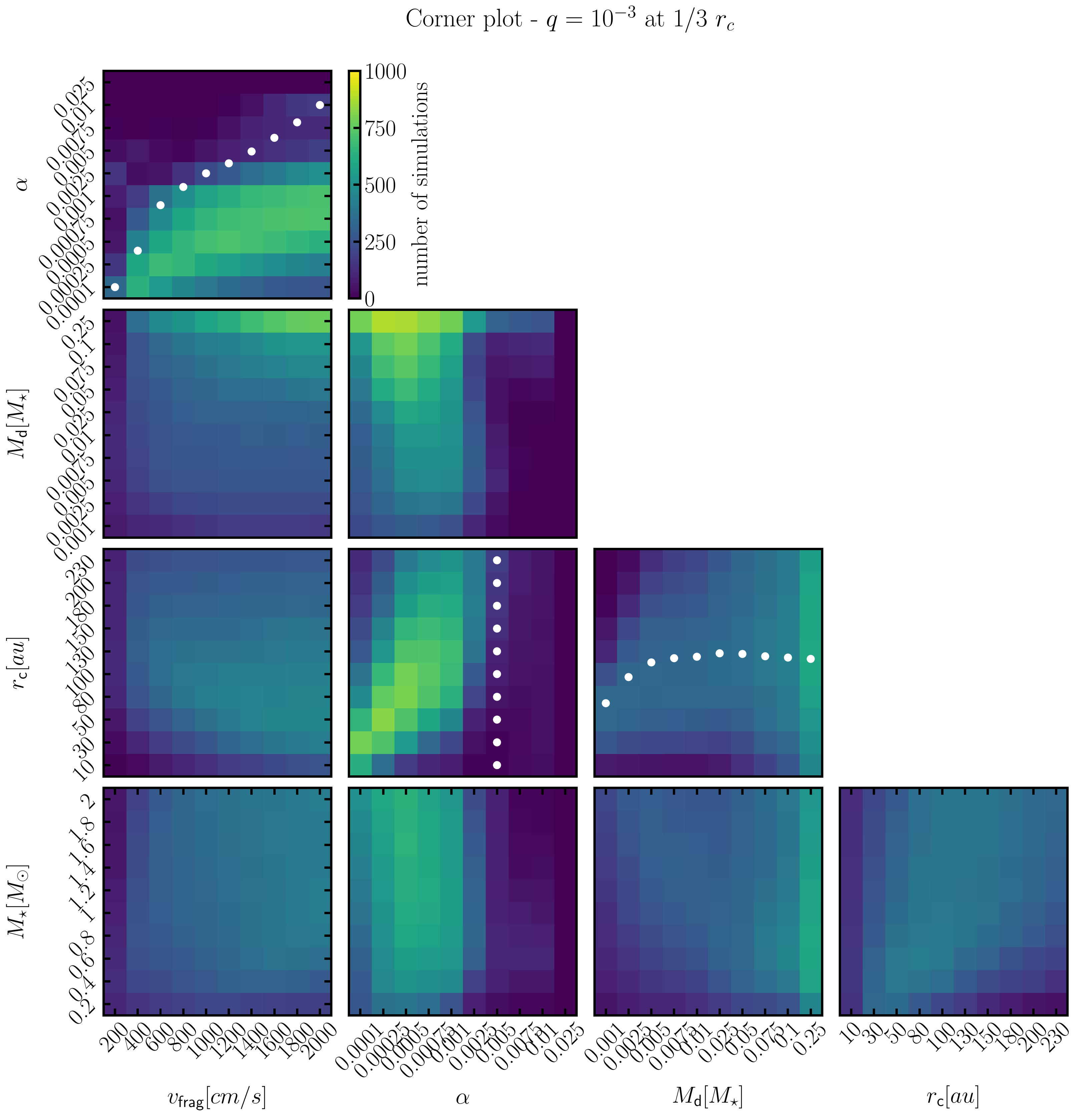}
    \caption[]{Corner plot with a planet with of a planet/star mass ratio $\mathrm{10^{-3}}$ at a location of the 1/3\rc. From left to right: Fragmentation velocity $\mathrm{v_{frag}}[cm/s]$, turbulence $\mathrm{\alpha}$-value, disk mass $\mathrm{M_{d}}[M_{\star}]$, characteristic radius $\mathrm{r_{c}}[au]$.  
    From top to bottom: $\mathrm{\alpha}$, $\mathrm{M_{d}}[M_{\star}]$, $\mathrm{r_{c}}[au]$, $\mathrm{M_{\star}}[M_{\odot}]$. In the upper left plot the white dots show the track where the distinction between the fragmentation and the drift limited regime is. In the middle plot ($\mathrm{\rc-\alpha}$) the white dots show the $\mathrm{\alpha}$-value where the pressure bump cannot hold the dust efficiently anymore. In the panel \rc - $\mathrm{M_{d}}$, shown as  white dots, is the mean value of the characteristic radius for every disk mass.}
    \label{fig:corner_13rc}
\end{figure*}

\begin{figure*}
    \centering
    \includegraphics[width=\linewidth]{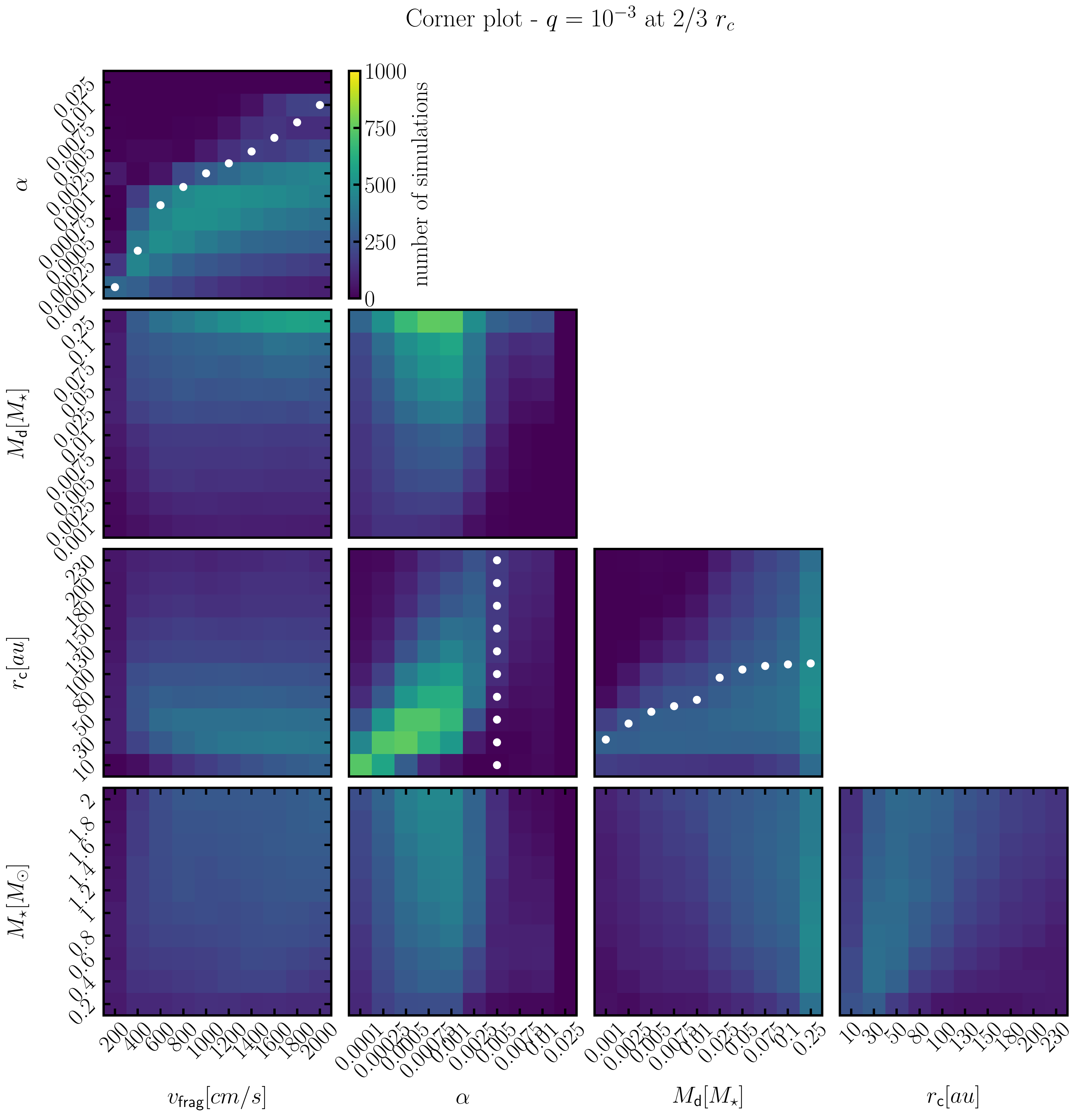}
    \caption[]{Corner plot where we have a Jupiter-mass planet at a location of 2/3\rc. From left to right: Fragmentation velocity $\mathrm{v_{frag}}[cm/s]$, turbulence $\mathrm{\alpha}$-value, disk mass $\mathrm{M_{d}}[M_{\star}]$, characteristic radius $\mathrm{r_{c}}[au]$.     
    From top to bottom: $\mathrm{\alpha}$, $\mathrm{M_{d}}[M_{\star}]$, $\mathrm{r_{c}}[au]$, $\mathrm{M_{\star}}[M_{\odot}]$. In the upper left plot the white dots show the track where the distinction between the fragmentation- and the drift-limited regime is. In the middle plot ($\mathrm{\rc-\alpha}$) the white dots show the $\mathrm{\alpha}$-value where the pressure bump cannot hold the dust efficiently anymore. In the panel \rc - $\mathrm{M_{d}}$, shown as  white dots, is the mean value of the characteristic radius for every disk mass.}
    \label{fig:corner_23rc}
\end{figure*}

In \autoref{fig:corner_smooth} we show the corner plot for the smooth case. The corner plots overall confirm the results discussed in the previous sections. Starting from the upper left part of the plot, we show the correlation between the turbulence parameter $\mathrm{\alpha}$ and the fragmentation velocity. The results show that all these plots are heavily dependent on whether the disks are drift- or fragmentation-limited. The white dots confirm that a scaling $\mathrm{\propto \sqrt{\alpha}}$ nicely explains the correlation: simulations above this relation are strongly fragmentation-limited, and therefore do not follow the SLR  \citep[see][]{rosotti2019time_evolution}. The  bottom left cell (corresponding to $\rm{\alpha}=10^{-4}$ and $\rm{v_{frag}}=$ $\SI{200}{cm/s}$) is empty because it is fragmentation-limited and increasing the $\mathrm{\alpha}$-value only makes the disks more fragmentation-dominated. As the fragmentation velocity increases then the \textit{allowed} $\mathrm{\alpha}$-value increases as well.

In \autoref{fig:corner_13rc} and \autoref{fig:corner_23rc}, we show the corner plots of the cases where a planet with a planet/star mass ratio $\mathrm{q=10^{-3}}$ is included at 1/3 and 2/3 of the \rc, respectively. In both cases we obtain similar results. In the second column a mild covariance can be seen between the turbulence parameter $\mathrm{\alpha}$ and the disk mass. Most of the simulations are toward low $\mathrm{\alpha}$-values and high disk masses, confirming once again the results from the 1D histograms, but showing that these two are the most dominant parameters. 

In the bottom panel a strong correlation between $\mathrm{\alpha}$ and \rc (for a limited $\mathrm{\alpha}$-range) is shown. With vertical white dots we indicate the $\mathrm{\alpha = 5\cdot10^{-3}}$ value, where there is a gradual boundary from matching to discrepant simulations. This relation is visible in the second and third columns of  \autoref{fig:heatmaps_ricci}. In summary, the  leftmost clump of simulations above the correlation is the case where $\mathrm{r_{c}}=$ $\SI{10}{au}$. This clump is mostly above the correlation (unless $\mathrm{\alpha =10^{-4}}$), and it explains the empty row for $\mathrm{r_{c}}=$ $\SI{10}{au}$ in the corner plots of \autoref{fig:corner_13rc} and \autoref{fig:corner_23rc}.

With low $\mathrm{\alpha}$-values there is more trapping in the pressure bump and the emission is optically thick, hence the disks are brighter. Since the disks should not be too bright (because that would locate them above the SLR) or not bright enough (this would locate them below the SLR), this yields only a range of permitted $\mathrm{\alpha}$-values. This range depends on \rc. Moreover, when $\mathrm{\alpha}$ increases, the emission is not optically thick anymore because the dust gets through the trap. More specifically, for $\mathrm{\alpha} \geq 5\cdot10^{-3}$ (as seen in \autoref{fig:mass_acc_1e-3}), the bump can no longer efficiently trap and this is why this transition from matching to discrepant simulations occurs.

Therefore, in the case where a planet is included the strongest correlation is between $\mathrm{\alpha}$-value and the characteristic radius \rc. The area of the parameter space with the highest fraction of matching simulations is where we have small disks (around $\rm{\sim \SI{50}{au}}$) and low $\mathrm{\alpha}$-values. This allows us once again to constrain the disk characteristic radius to a value that is not larger than $\mathrm{\SI{100}{au}}$.

\section{Gap profiles}
\label{app:gap_profiles}
As discussed in \autoref{sec:methods}, we compared the planet profile from \citet{Kanagawa2016} to hydrodynamical simulations using FARGO \citep{FARG03D_2015ascl.soft09006B}. We varied the turbulence parameter $\alpha$, the planet/star mass ratio $q$, and the position of the planetary gap $\rm{r_{p}}$. After obtaining the planet profile from the hydrodynamical simulation, we evolved the disk with two-pop-py \citep{BKE2012} without gas evolution to avoid diffusion of the planetary gap. We obtained the disk continuum luminosity and the effective radius, and we plotted the evolution tracks on the SLR. We evolved the simulations for $\mathrm{\SI{330}{kyr}}$. The parameters that we used for this test are summarized in the following table.
 
\begin{table}[ht!]
\label{tabel:gaps_param}
 \caption{Parameters for the gap comparison}
 \begin{center}
 \centering
 \begin{tabular}{l l l } 
 \toprule
 Parameter & Description & Values \\ [0.3ex]
 \hline
 %\middlerule
 $\alpha$ & viscosity parameter & $\rm{10^{-4}}$, $\rm{10^{-3}}$, $\rm{10^{-3}}$ \\ [0.3ex]
 
 $\rm{r_{p}}$ $\rm{[au]}$ & planet position  & $\rm{30}$, $\rm{60}$, $\rm{100}$  \\ [0.3ex]

 q & planet/star mass ratio & $\rm{3\cdot10^{-4}}$, $\rm{10^{-3}}$, $\rm{3\cdot10^{-3}}$ \\ [0.3ex]
\bottomrule
\end{tabular}
\end{center}

\end{table}

\begin{figure}
    \centering
    \includegraphics[width=\linewidth]{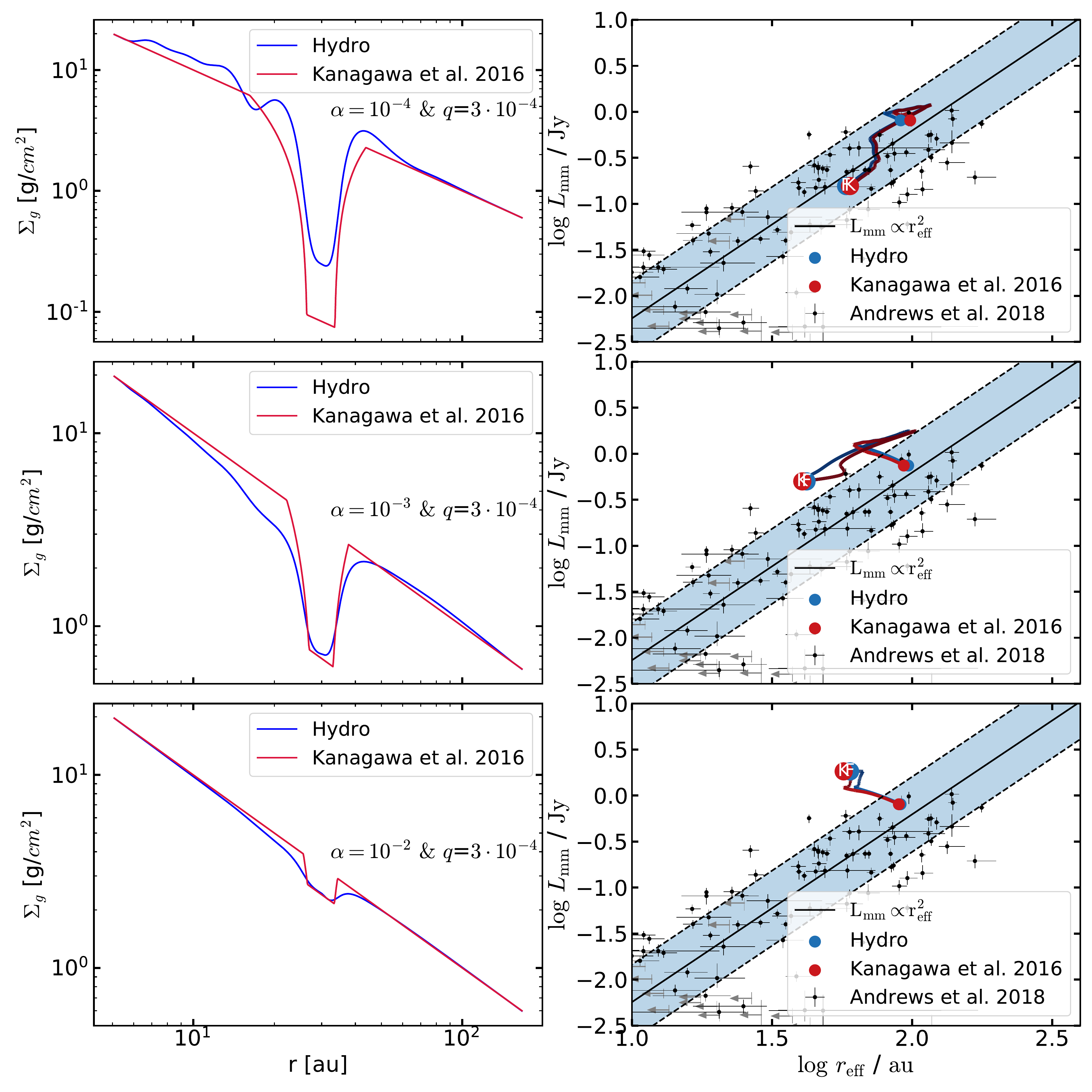}
    \caption[]{Comparison between a hydrodynamical and our dust evolution simulation for different $\mathrm{\alpha}$-values and for a planet/star mass ratio $\mathrm{q=3\cdot10^{-4}}$ at $\mathrm{\SI{30}{au}}$. In the left column are the different gap profiles for different $\mathrm{\alpha}$-values and in the right column the corresponding evolution tracks. Even though the profile from \citet{Kanagawa2016} does not overlapping with that obtained from the hydrodynamical simulation, the tracks in the right column produce similar results.}
    \label{fig:hydro_comp_q3e-4}
\end{figure}

\begin{figure}
    \centering
    \includegraphics[width=\linewidth]{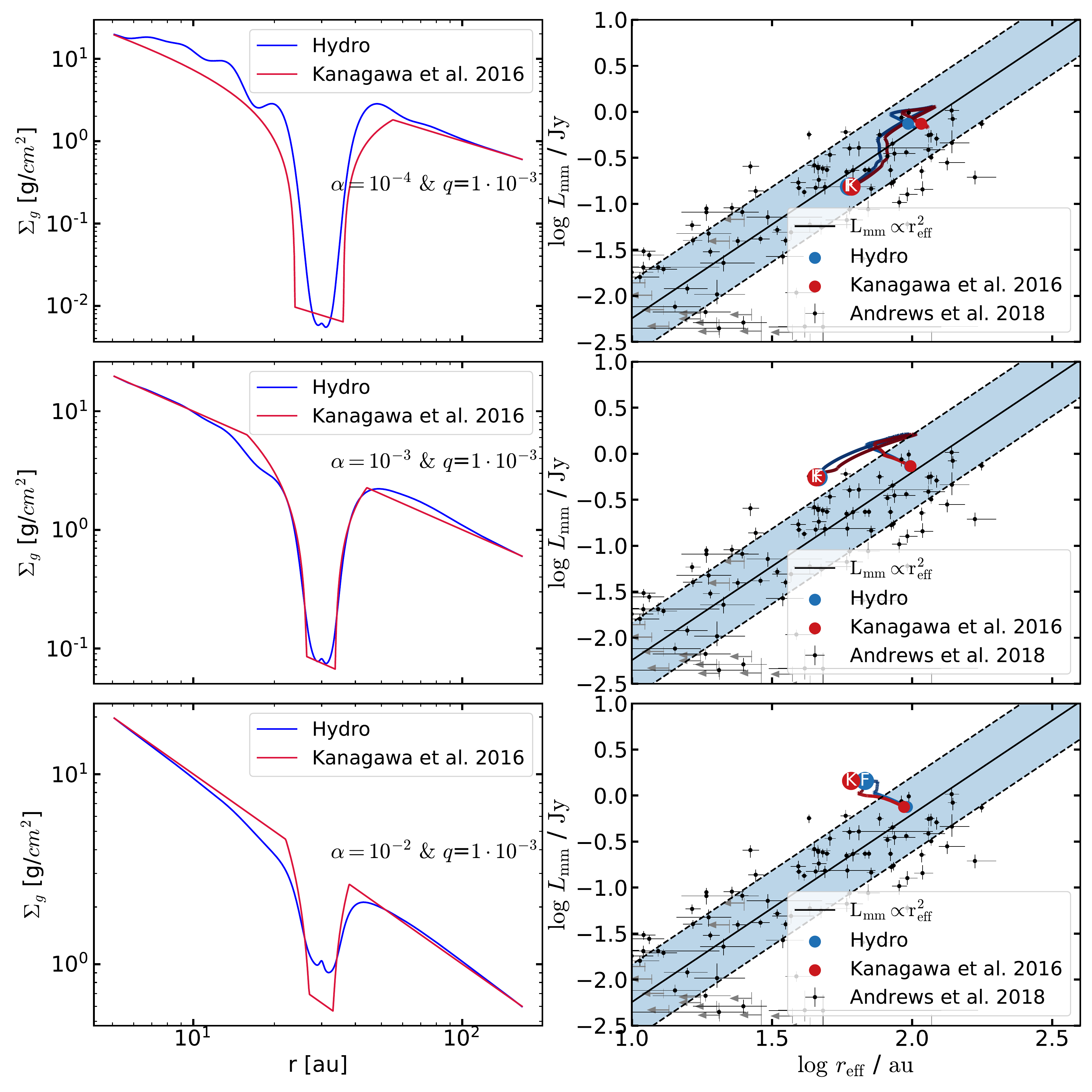}
    \caption[]{Comparison between a hydrodynamical and our dust evolution simulation for different $\mathrm{\alpha}$-values and for a planet/star mass ratio $\mathrm{q=10^{-3}}$ at $\mathrm{\SI{30}{au}}$. In the left column are the different gap profiles for different $\mathrm{\alpha}$-values and in the right column the corresponding evolution tracks. Even though the profile from \citet{Kanagawa2016} does not overlapping with that obtained from the hydrodynamical simulation, the tracks in the right column produce similar results.}
    \label{fig:hydro_comp_q1e-3}
\end{figure}

\begin{figure}
    \centering
    \includegraphics[width=\linewidth]{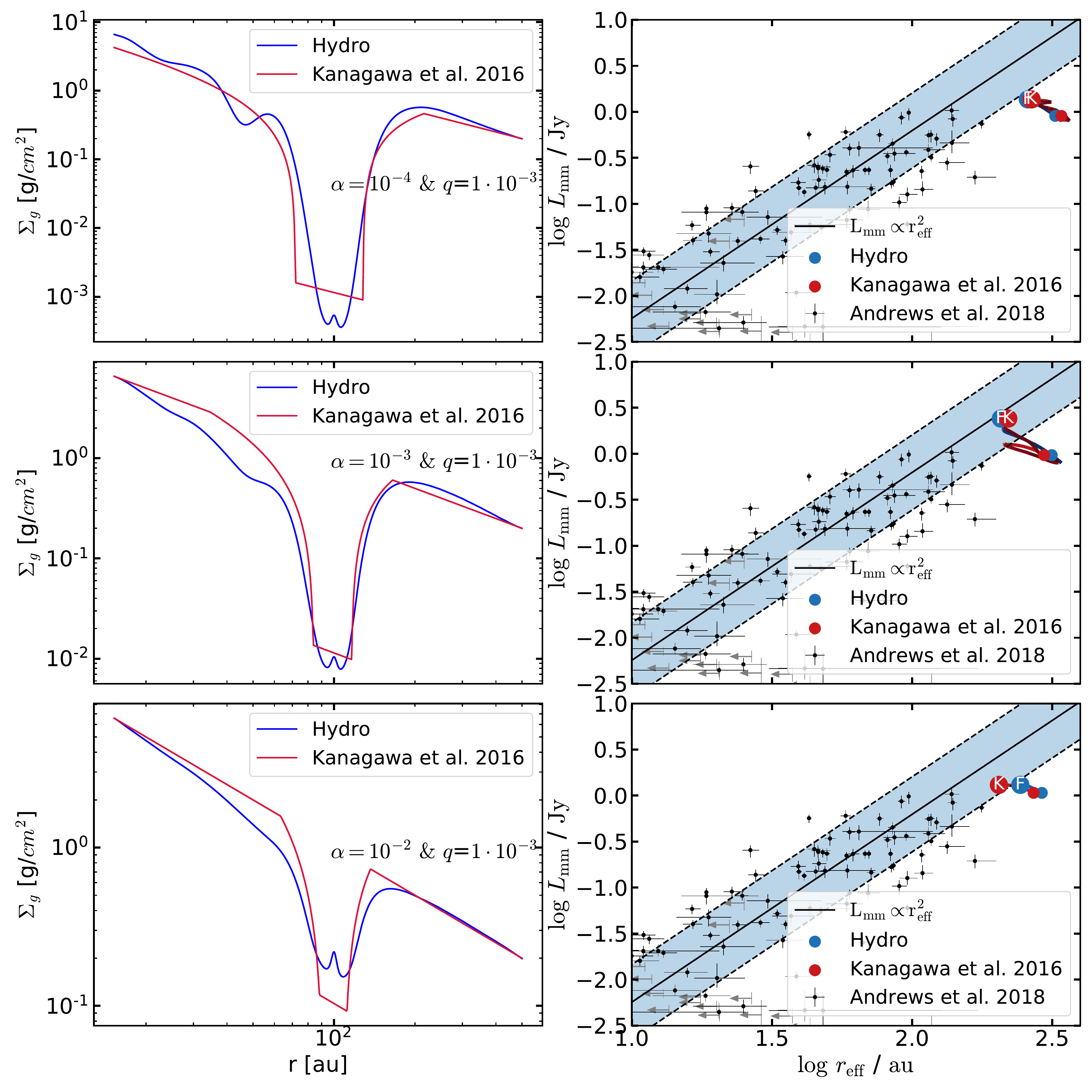}
    \caption[]{Comparison between a hydrodynamical and our dust evolution simulation for different $\mathrm{\alpha}$-values and for a planet/star mass ratio $\mathrm{q=3\cdot10^{-3}}$ at $\mathrm{\SI{100}{au}}$. In the left column are the different gap profiles for different $\mathrm{\alpha}$-values and in the right column the corresponding evolution tracks. Even though the profile from \citet{Kanagawa2016} does not overlapping with that obtained from the hydrodynamical simulation, the tracks in the right column produce similar results.}
    \label{fig:hydro_comp_q3e-3}
\end{figure}

In Figures \ref{fig:hydro_comp_q3e-4}, \ref{fig:hydro_comp_q1e-3}, and \ref{fig:hydro_comp_q3e-3} we show the comparison between the gap that is obtained from the hydrodynamical simulation and the profile from \citet{Kanagawa2016}. In \autoref{fig:hydro_comp_q3e-4} we compare the gap profile for a small planet with planet/star mass ratio of $\mathrm{q=3\cdot10^{-4}}$ for the three different values of the $\mathrm{\alpha}$-parameter. From the low $\mathrm{\alpha}$-value (first row, first column, $\mathrm{\alpha}=10^{-4}$), we observe that the \citet{Kanagawa2016} profile fits  the width of the gap reasonably well, but there is an offset for the depth. As we look at their evolution tracks we do not observe any major difference. For $\mathrm{\alpha}=10^{-3}$ (second row, first column) we have a good fit for the depth and the width in \autoref{fig:hydro_comp_q1e-3} and \ref{fig:hydro_comp_q3e-3}. On the evolution track there is a small fluctuation, but at the end of the simulation they converge to the same point. For the high $\mathrm{\alpha}$-value (third row, first column) the profile nicely fits  the width and the depth of the gap. As a result, the evolution tracks of the simulations (third row, second column) are almost identical. 

Similarly, for the other cases we can observe that even if the profile does not  perfectly fit the bottom of the gap, the difference in the evolution tracks is minimal. The track can be affected  only if the width of the gap is not fitted properly. Therefore, we conclude that our chosen profile will not affect the main point of the paper, which is based on the observable quantities.

\FloatBarrier
\section{Additional heat maps}
\label{app:heatmap}
\begin{figure*}
    \includegraphics[width=\linewidth]{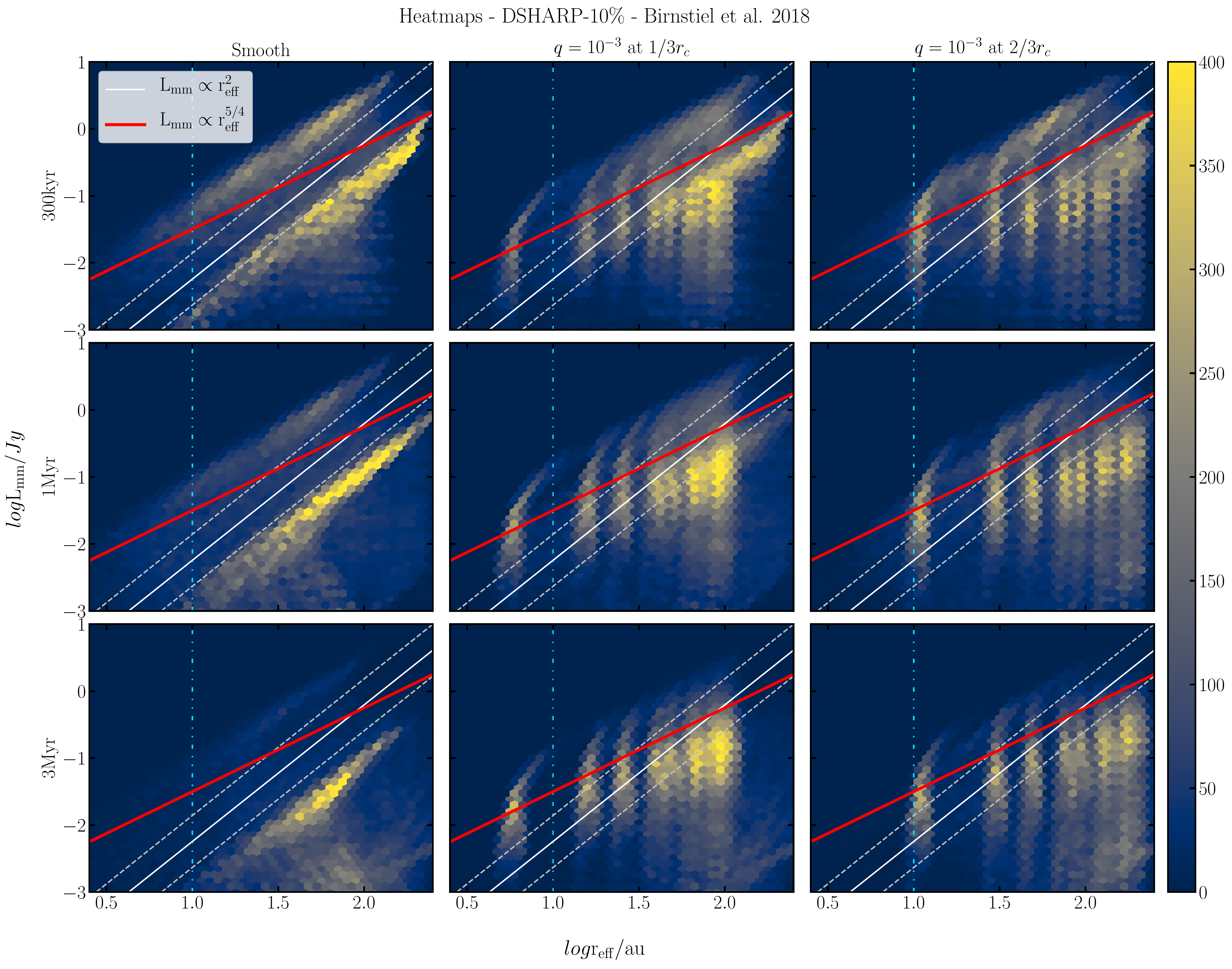}
    \caption[]{Heat maps of representative simulations with the \citet{Birnstiel_DSHARP} D-10 opacities with 10\% porosity. From left to right, the three  columns represent the smooth case, a planet at $\mathrm{1/3r_{c}}$, and a planet at $\mathrm{2/3r_{c}}$. From top to bottom, the rows represent three different snapshots at $\mathrm{\SI{300}{kyr}}$, $\mathrm{\SI{1}{Myr,}}$ and $\mathrm{\SI{3}{Myr}}$. The white solid line is the SLR from \citet{Andrews2018a} and the red solid line our fit for the cases where we include a planet. The color bar shows the number of simulations in a single cell. The blue dash-dotted line shows the minimum limit ($\mathrm{r_{eff} \sim \SI{10}{au}}$) where observational results are available.}
    \label{fig:heatmaps_d10}
\end{figure*}

\begin{figure*}
    \includegraphics[width=\linewidth]{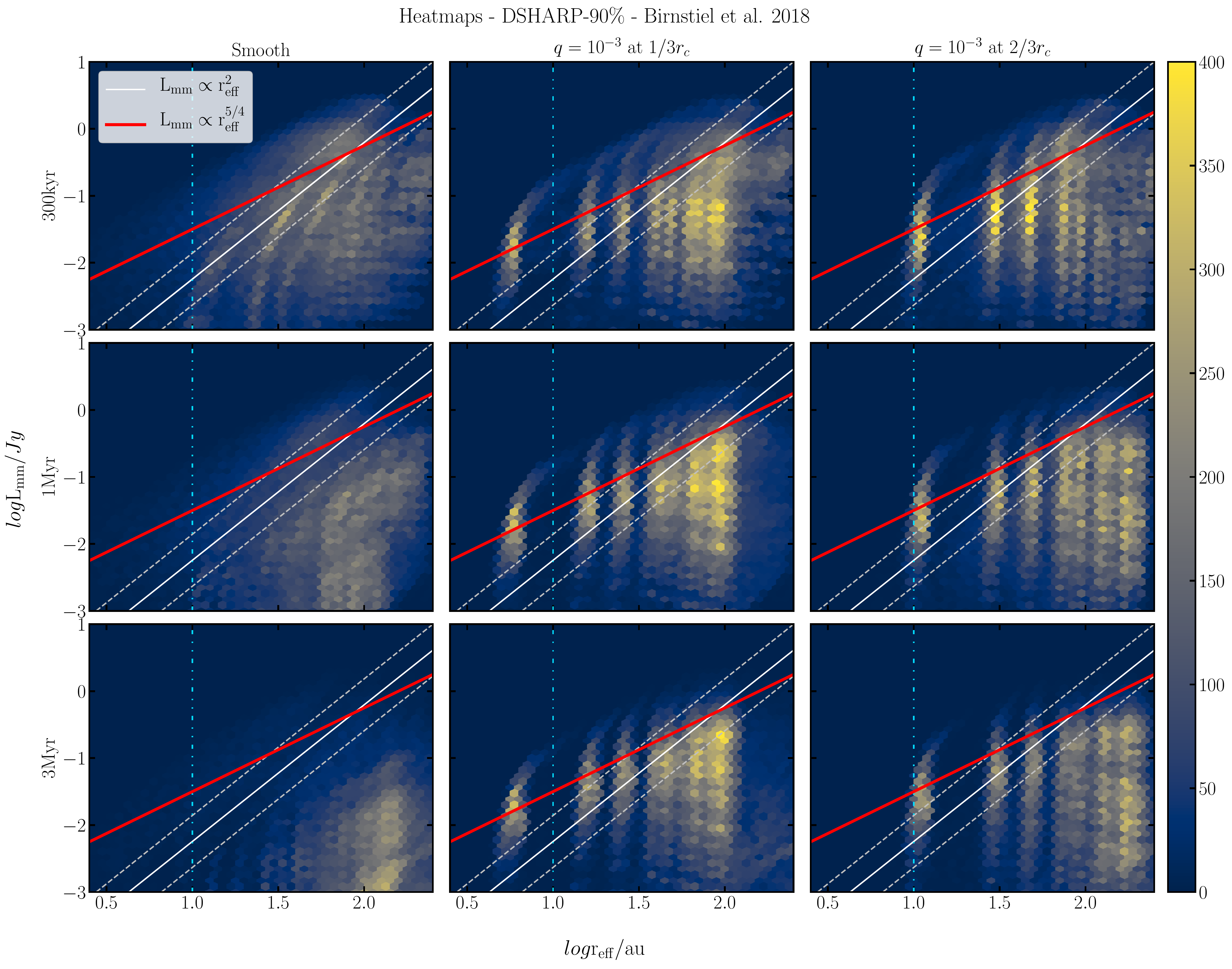}
    \caption[]{Heat maps of representative simulations with the \citet{Birnstiel_DSHARP} D-90 opacities with 90\% porosity. From left to right, the three columns represent the smooth case, a planet at $\mathrm{1/3r_{c}}$, and a planet at $\mathrm{2/3r_{c}}$. From top to bottom, the rows represent three different snapshots at $\mathrm{\SI{300}{kyr}}$, $\mathrm{\SI{1}{Myr,}}$ and $\mathrm{\SI{3}{Myr}}$. The white solid line is the SLR from \citet{Andrews2018a} and the red solid line our fit for the cases where we include a planet. The color bar shows the number of simulations in a single cell. The blue dash-dotted line shows the minimum limit ($\mathrm{r_{eff} \sim \SI{10}{au}}$) where observational results are available.}
    \label{fig:heatmaps_d90}
\end{figure*}

In this section we add the additional heat maps where the $\mathrm{10\%}$ porosity (DSHARP-10) and $\mathrm{90\%}$ porosity (DSHARP-90) opacity is used. We plot the position of every simulation for three different snapshots ($\mathrm{\SI{300}{kyr}}$, $\mathrm{\SI{1}{Myr}}$, $\mathrm{\SI{3}{Myr}}$). For DSHARP-10 (\autoref{fig:heatmaps_d10}) we obtain similar results to the DSHARP case \autoref{fig:heatmaps_dsharp} (with no porosity) as the two opacities are similar to one another (see \autoref{fig:Opacities}). For the DSHARP-90 case (\autoref{fig:heatmaps_d90}), the complete absence of the opacity cliff does not allow a considerable amount of smooth disks to enter the SLR, while the same fraction of substructured disks match, as in the DSHARP-50 \autoref{fig:heatmaps_d50} case.

\begin{figure*}
    \includegraphics[width=\linewidth]{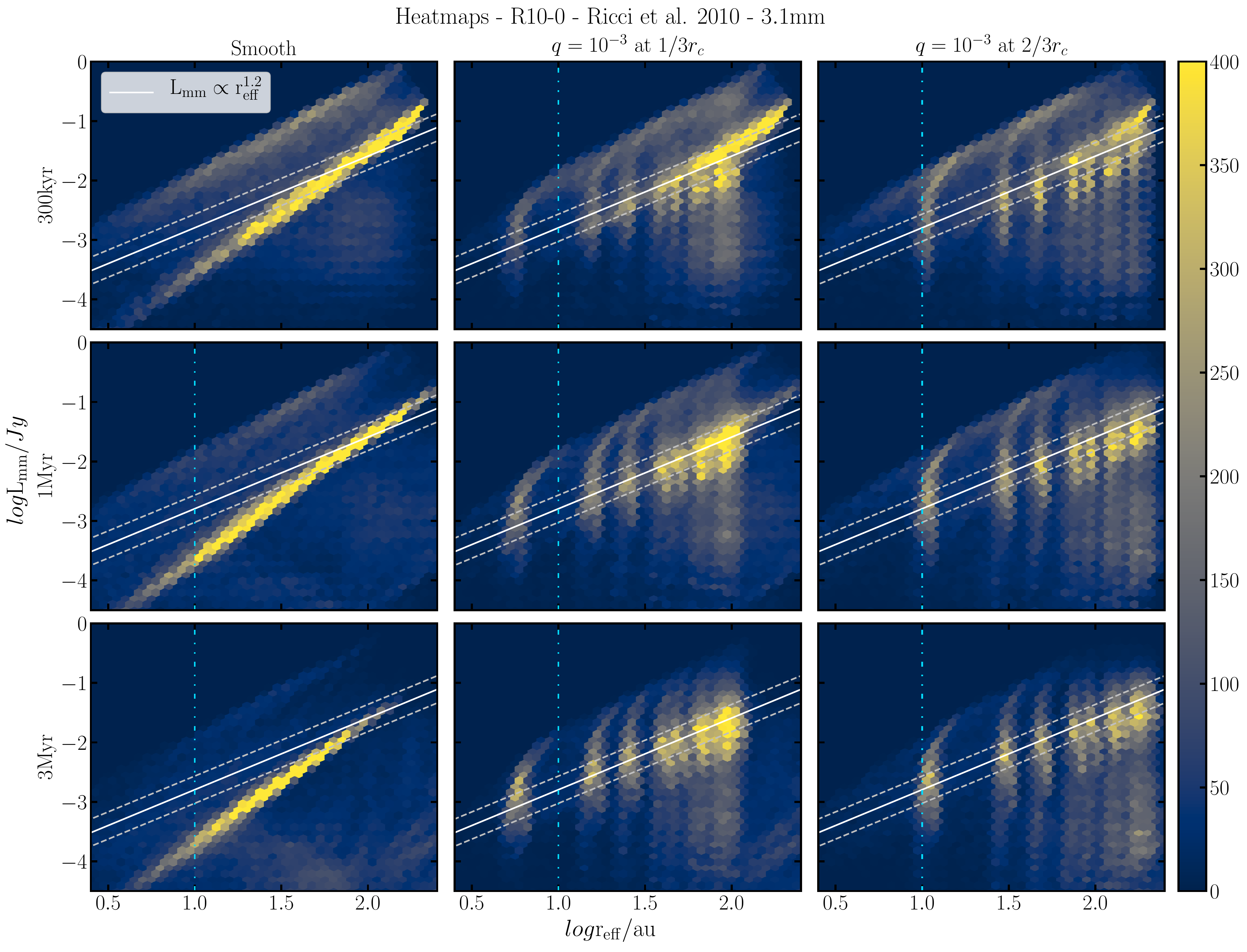}
    \caption[]{Heat maps of representative simulations with the R10-0 opacity at $\SI{3.1}{mm}$ for a direct comparison with \citet{Tazzari2021MNRAS.506.2804T}. From left to right, the three columns represent the smooth case, a planet at $\mathrm{1/3r_{c}}$, and a planet at $\mathrm{2/3r_{c}}$. From top to bottom, the rows represent three different
    snapshots at $\mathrm{\SI{300}{kyr}}$, $\mathrm{\SI{1}{Myr,}}$ and $\mathrm{\SI{3}{Myr}}$. The white solid line is the SLR from \citet{Tazzari2021MNRAS.506.2804T}. The color bar shows the number of simulations in a single cell. The blue dash-dotted line shows the minimum limit ($\mathrm{r_{eff} \sim \SI{10}{au}}$) where observational results are available.}
    \label{fig:heatmaps_3mm}
\end{figure*}

\subsection{Changing the stellar luminosity}
\label{app:stellar_luminosity}
\begin{figure*}
    \includegraphics[width=\linewidth,height=8cm]{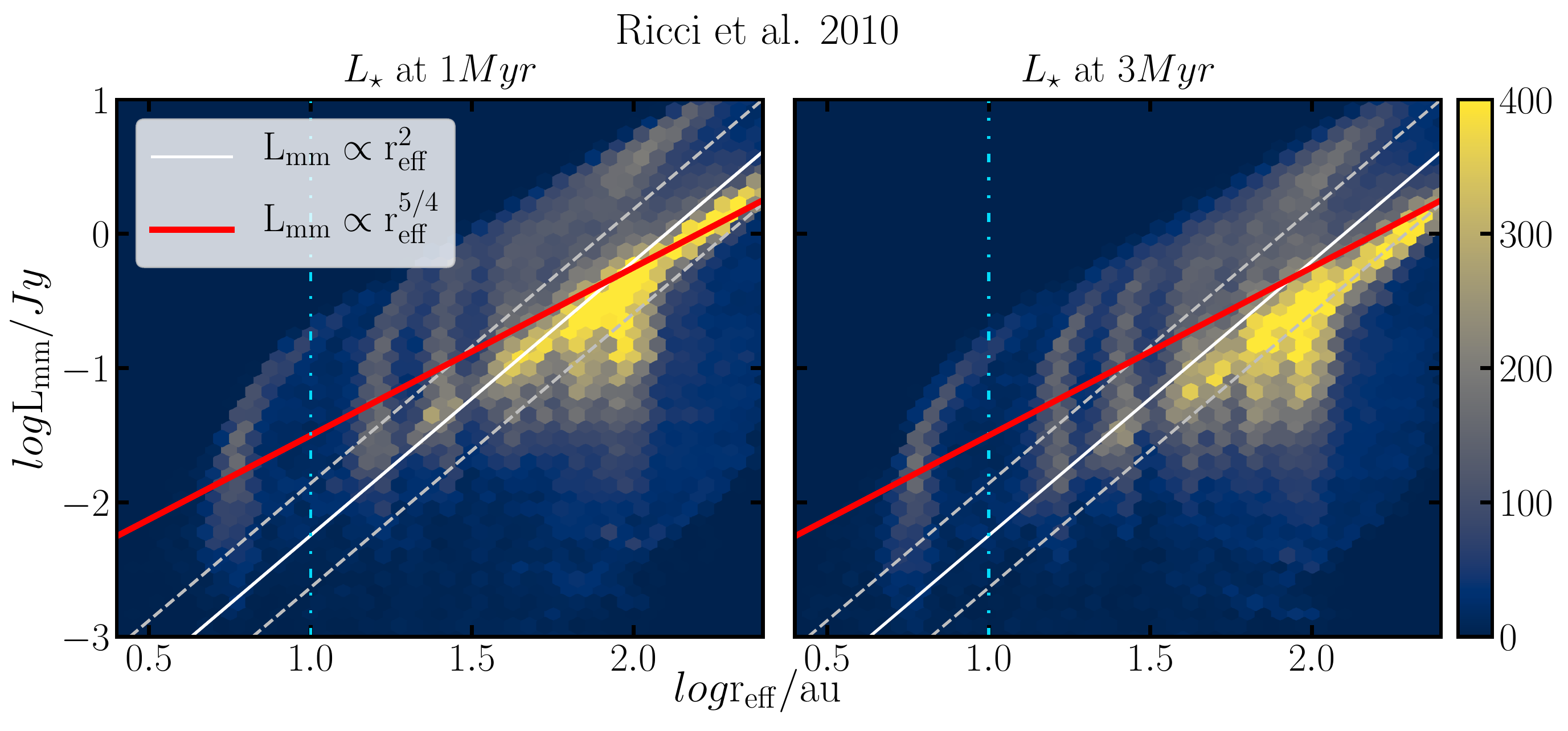}
    \caption[]{Heat map comparison between disks where the stellar luminosity at $\SI{3}{Myr}$ (right panel) is used instead of $\SI{1}{Myr}$ (left panel). There is an overall shift toward lower luminosities, but the trend remains the same.}
    \label{fig:heatmaps_luminosity}
\end{figure*}

In \autoref{disc:limitations} we performed a test where we used the luminosity of the star at $\SI{3}{Myr}$ instead of $\SI{1}{Myr}$, which is used throughout this work. Since the stellar luminosity decreases from $\SI{1}{Myr}$ to $\SI{3}{Myr}$, there is an overall movement of disks to lower luminosities that is shown in \autoref{fig:heatmaps_luminosity} in the right panel. In this example we plot the heat map of a case where there is a planet at $\mathrm{1/3r_{c}}$ at a snapshot of $\mathrm{\SI{1}{Myr}}$.

\end{appendix}
\end{document}